\documentclass[]{aa}

\usepackage{natbib}
\usepackage{breqn}
\usepackage{multirow}
\usepackage{hyperref} 
\usepackage{float}
\usepackage[dvipsnames]{xcolor}
\usepackage{cancel}
\hypersetup{
    colorlinks = true,
    linkbordercolor = {cyan},
    linkcolor={blue},
    citecolor={blue},}

\begin{document}

\def\figureautorefname{Fig.}
\def\equationautorefname~#1\null{Eq. (#1)\null}

\def\sectionautorefname{Sect.}
\def\subsectionautorefname{Sect.}
\def\subsubsectionautorefname{Sect.}

\title{Underestimation of the tidal force and apsidal motion in close binary systems by the perturbative approach: Comparisons with non-perturbative models}
\titlerunning{MoBiDICT}

\author{L. Fellay\inst{1} \and M.-A. Dupret\inst{1} \and S. Rosu\inst{2}}
\institute{STAR Institute, University of Liège, 19C Allée du 6 Août, B$-$4000 Liège, Belgium \and Department of Physics, KTH Royal Institute of Technology, The Oskar Klein Centre, AlbaNova, SE-106 91 Stockholm, Sweden}
\date{May, 2023}

\abstract{Stellar deformations play a significant role in the dynamical evolution of stars in binary systems,  impacting the tidal dissipation and  the outcomes of mass transfer processes.  The prevalent method for modelling the deformations and tidal interactions of celestial bodies solely relies on the perturbative approach, which assumes that stellar deformations are minor perturbations to the spherical symmetry. An observable consequence of stellar deformations is the apsidal motion in eccentric systems,  which has be observationally determined across numerous binary systems.}
{Our objective is to assert the reliability of the perturbative approach when applied to close and strongly deformed binary systems. }
{We have developed a non-perturbative 3D modelling method designed to account for high stellar deformations.  We focus on comparing the properties of perturbatively deformed stellar models with our 3D models, particularly in terms of apsidal motion.  }
{Our research highlights that the perturbative model becomes imprecise and underestimates the tidal force and rate of apsidal motion at a short orbital separation. This discrepancy primarily results from the first-order treatment in the perturbative approach, and cannot be rectified using straightforward mathematical corrections due to the strong non-linearity and numerous parameters of the problem.  We have determined that our methodology affects the modelling of approximately $42\%$ of observed binary systems with measured apsidal motion, introducing a discrepancy greater than $2\%$ when the normalised orbital separation verifies $q^{-1/5}a(1-e^2)/R_1\lesssim 6.5$ ($q$ is the mass ratio of the system,  $a$ is its semi-major axis,  $e$ is its orbital eccentricity and $R_1$ is the radius of the primary star).}
{The perturbative approach underestimates tidal interactions between bodies up to $\sim 40 \%$ for close low-mass binaries.  All the subsequent modelling is impacted by our findings,  in particular,  the tidal dissipation is significantly underestimated.  As a result, all binary stellar models are imprecise when applied to systems with a low orbital separation, and the outcomes of these models are also affected by these inaccuracies.}
\maketitle

\noindent

\section{Introduction}
 
Tidal forces in binary systems result from the gravitational interaction between non-point-like bodies and can be categorised into two components: the equilibrium tides,  corresponding to large-scale circulations resulting from a hydrostatic readjustment (i.e.  stellar deformations) and the dynamical tides corresponding to the excitation of eigenmodes of oscillations from the periodic perturbations from a companion \citep{Zahn1975}. These tidal forces and the subsequent stellar deformations lead to various phenomena such as orbital synchronisation, mass transfer, and tidal dissipation \citep{Jeans1929},   the latter corresponding to the dissipation of orbital energy within the stellar bodies. A direct consequence of the tidal interactions is the apsidal motion in eccentric binaries \citep{Sterne1939}.  It represents the time variation of the argument of periastron $\omega$ or,  in other words, the motion of the apsidal line with time.    Apsidal motion originates from equilibrium tides, dynamical tides \citep{Gimenez1985,Willems2002},  and general relativistic corrections \footnote{In the case of a hierarchical triple system,  a contribution to the apsidal motion arises from the third body orbiting the inner binary \citep{Naoz2013}. In the present study,  we do not account for such an effect as we focus on binary systems.} \citep{Gimenez1985}.  It serves as a valuable observational constraint to understand the structure of deformed stars,  providing insights into the microphysics of stellar models \citep{Rosu2020, Rosu2022a, Rosu2022b, Rosu2023}.  Accurate modelling of tidal and centrifugal deformations is essential to determine the precise stellar structure of each component in binary systems and, consequently, to predict the apsidal motion.
\\~\\When modelling the structure and deformations of binary stars, the perturbative approach is one of the most sophisticated and widely used methods in the literature.  The principle of this methodology is to consider that stellar deformations are small perturbations to the spherical symmetry of stars, only accounting for the leading orders terms in the developments \citep{Sterne1939, Kopal1959, Kopal1978}. This approach allows one to obtain simplified structural deformations from the unperturbed spherically symmetric structure through the well-known stellar structure constant, the $k_2$ coefficient that is obtained from solving the Clairaut-Radau equation \citep{Kopal1959}.  With the perturbative formalism,  the dipolar components of the tidal and centrifugal forces cancel out \citep{Kopal1959,Fitzpatrick2012},  implying that no asymmetry exists in the stellar models, in contradiction with the fundamental Roche geometry.  Moreover,  our previous work \citep{Fellay2023}  already demonstrated that the perturbative approach underestimates the deformations when compared to our non-perturbative models.  Therefore,  it becomes crucial to verify the consequences of the assumptions inherent to the perturbative approach.  Finally,  it is worth mentioning that the tidal distortion and dissipation formalism \citep{Zahn1977,Hut1981} employed by widely recognised and publicly available codes such as MESA \citep{MESA2011,MESA2013,MESA2015, MESA2018,MESA2019} are based on the perturbative approach.  
\\~\\ In this study, we expand on the work done in \cite{Fellay2023} by further investigating the limitations imposed by the assumptions of the perturbative approach, using our non-perturbative method.  Our tool,  called Modelling Binary Deformations Induced by the Centrifugal and Tidal forces (MoBiDICT),  allows us to determine the deformed structure of close binaries in a non-perturbative manner, accounting for the redistribution of mass in the bodies.  By utilising the deformed structure obtained from MoBiDICT, we could calculate the instantaneous tidal acceleration perturbation and its consequence on the apsidal motion of binary systems.  Finally,  MoBiDICT's formalism enabled us to locate and identify the specific spherical order responsible for the  discrepancies seen.
\\~\\In this article, we start, in \autoref{sect_MoBIDICT},  by introducing the physics behind MoBiDICT and the  methodology developed to compute the apsidal motion of binaries in a non-perturbative way.  In \autoref{sect_Theoritical_apsides_pert},  we explain how we developed the perturbative approach formalism usually used to model the apsidal motion of binaries.  \autoref{sect_comparaison_theoritical_models} is dedicated to exploring the discrepancies in models for different types of theoretical stars.  In \autoref{sect_application_observations} we study binary systems for which the apsidal motion has been measured and provided in the literature. 
In \autoref{sec_discussion_population},  we discuss the implication of our findings on the stellar modelling methods used for binary stars. Finally, in the Appendices, we develop all the equations of the perturbative approach starting from a single star's gravitational potential to arrive at the apsidal motion in binaries.

\section{Non-perturbative modelling }\label{sect_MoBIDICT}
\subsection{Set-up of the problem and first models} 
\begin{figure}[h]
\centering
\includegraphics[width=\hsize]{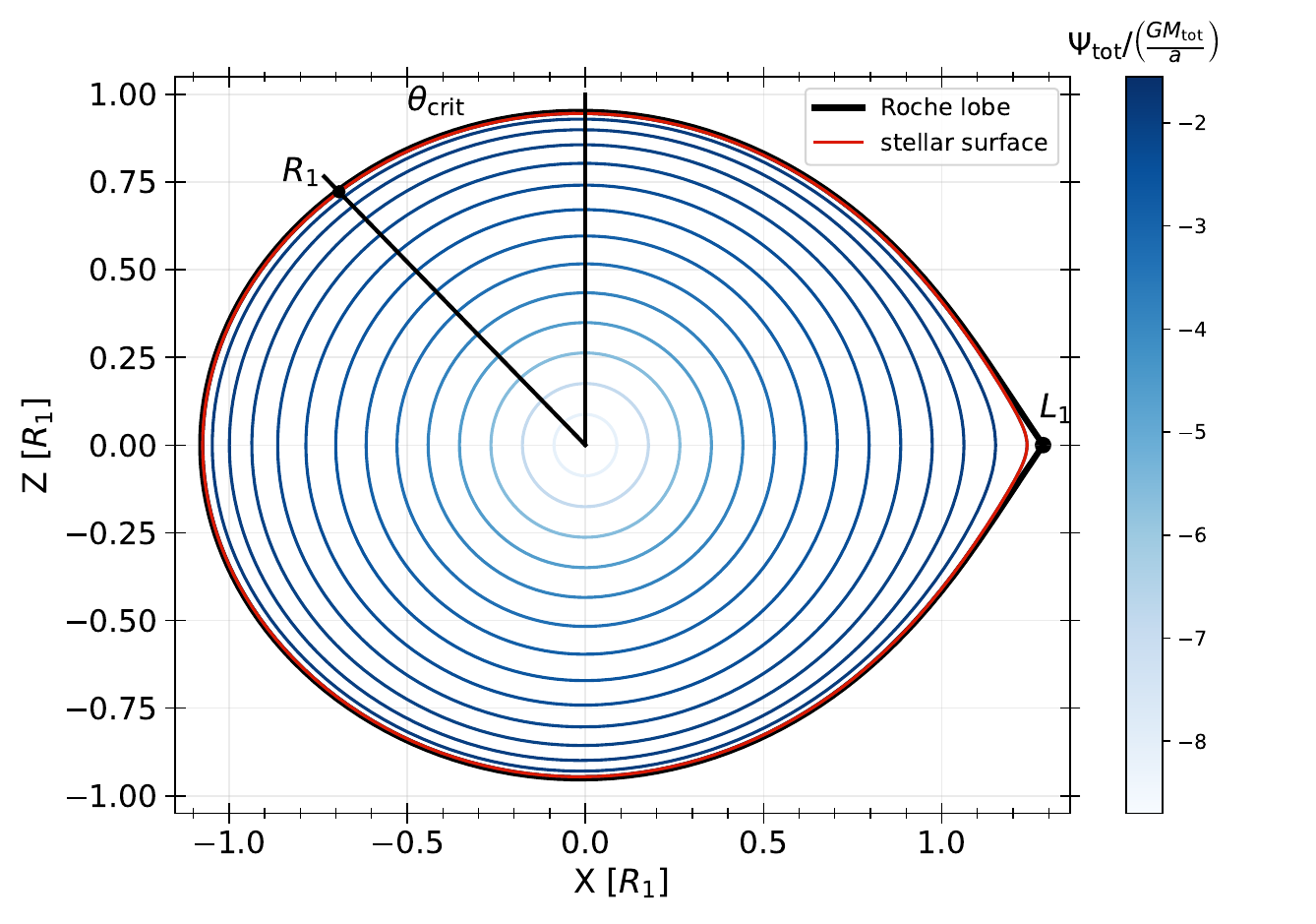}
\caption{View of a $1.0$ $M_\odot$ MS primary star in a binary system, cut in the plane including all the rotations axes of the system.  The companion to this star is a 0.2 $M_\odot$ MS secondary star with an orbital separation of $r=2 R_1$.  The black curve corresponds to the Roche lobe of the primary star.  The black lines respectively correspond to the reference direction for $\theta$ and to the reference direction in witch our model is given by the 1D spherical symmetric model ($\theta=\mu_{\mathrm{crit}}$).  The blue lines are equipotentials inside the star with the colourbar representing the total potential.  On each equipotential the pressure, density, and temperature,  are constant.   $L_1$ is the Lagrangian point and $R_1$ is the radius of the initial 1D model in the reference direction.   }
\label{fig.equipotentials_1M0}
\end{figure}

In this article,  we investigate the deformations in binary systems resulting from three primary sources: the gravitational force, the tidal force arising from a companion, and the centrifugal force originating from the orbital rotation of the system and the individual stellar rotations. Accurate modelling of these phenomena necessitates a non-perturbative treatment of deformations in three dimensions (3D).
\\We considered a binary system consisting of two distinct stars separated by a distance $r$ and with an orbital rotation rate $n$. Assuming the stars ($i = 1,2$) within the system are in hydrostatic equilibrium,  their structures are governed by the equation of hydrostatic equilibrium: 
\begin{equation}\label{eq_hydrostatic_structure}
\frac{\boldsymbol{\nabla} P}{\rho}= -\boldsymbol{\nabla}\left(\Psi_1 +\Psi_2 + \Psi_{\rm{c}}\right)=\textbf{g}_{\rm{eff}},
\end{equation}
and the Poisson equation, 
\begin{equation}\label{eq_Poisson_classique}
\Delta \Psi_i = 4\pi G \rho_i,
\end{equation}
linking the gravitational potential $\Psi_i(r_i, \mu,\phi)$ of a body to its density distribution $\rho_i(r_i, \mu,\phi)$ in 3D.  In \autoref{eq_hydrostatic_structure},  $P$ is the pressure, $\textbf{g}_{\rm{eff}}$ is the effective gravity,  $ \Psi_{\rm{c}}(r_i, \mu,\phi)$ is the centrifugal potential on a 3D grid,  where $r_i$ denotes the radial coordinate of the  star $i$,  $\mu=\cos(\theta)$ its colatitude,  and $\phi$ its azimuthal angle.  The centrifugal force can be solely derived from a potential if the rotations exhibit cylindrical symmetry, with the solid body rotation being a specific case.  Throughout this entire article,  we made the assumption of solid body rotations to simplify the problem (see \autoref{sec_discussion_population} for a discussion on the validity of this assumption).
\\~\\In our recent work \citep{Fellay2023},  we introduced a new tool called MoBiDICT,  specifically designed to precisely model close binaries using a non-perturbative approach to treat the deformations.  Our method iteratively solves the Poisson equation on a 3D structure,  allowing to obtain the gravitational and tidal potential accounting for the redistribution of mass within the bodies.  MoBiDICT takes advantage of the conservative nature of all the forces considered,  resulting in a barotropic model structure (i.e.  the densities are constant on equipotentials).  By assuming that in a given direction ($\mu_{\mathrm{crit}}, \phi_{\mathrm{crit}}$) each star is described by a 1D spherically symmetric input model, we can recompose the entire 3D density profile from the total potential ($\Psi_{\rm{tot}}=\Psi_1 +\Psi_2 + \Psi_{\rm{c}}$) of each star.  This method is thoroughly explained in \cite{Fellay2023}. 
\\~\\Figure~\ref{fig.equipotentials_1M0} illustrates the equipotential lines within a deformed primary star, that were obtained using our non-perturbative modelling method.  In this case,  we applied our tool to a binary system composed of a primary 1 $M_\odot$ main-sequence (MS) star and a 0.2 $M_\odot$ MS companion separated by $r=2 R_1$,  $R_1$ being the radius of the primary given by the 1D spherically symmetric model. 
In the scenario depicted in \autoref{fig.equipotentials_1M0}, we assumed that the rotational velocities of both stars are identical and equal to the orbital rotation rate,  in other terms the system is tidally locked.  However,  when studying eccentric binaries,  pseudo-synchronisation becomes the end result of tidal interactions, making impossible to model the system as entirely synchronised.  Furthermore,  mass transfers between binary components often result in significant transfer of angular momentum \citep{Packet1981},  leaving the accretor star close to its critical rotation while the donor slowly rotates.  Generalising our model to non-synchronised binary systems is necessary to model eccentric binaries and binary evolution.
\subsection{Non-perturbative modelling of non-synchronised binary systems} 
\begin{figure}[h]
\centering
\includegraphics[width=\hsize]{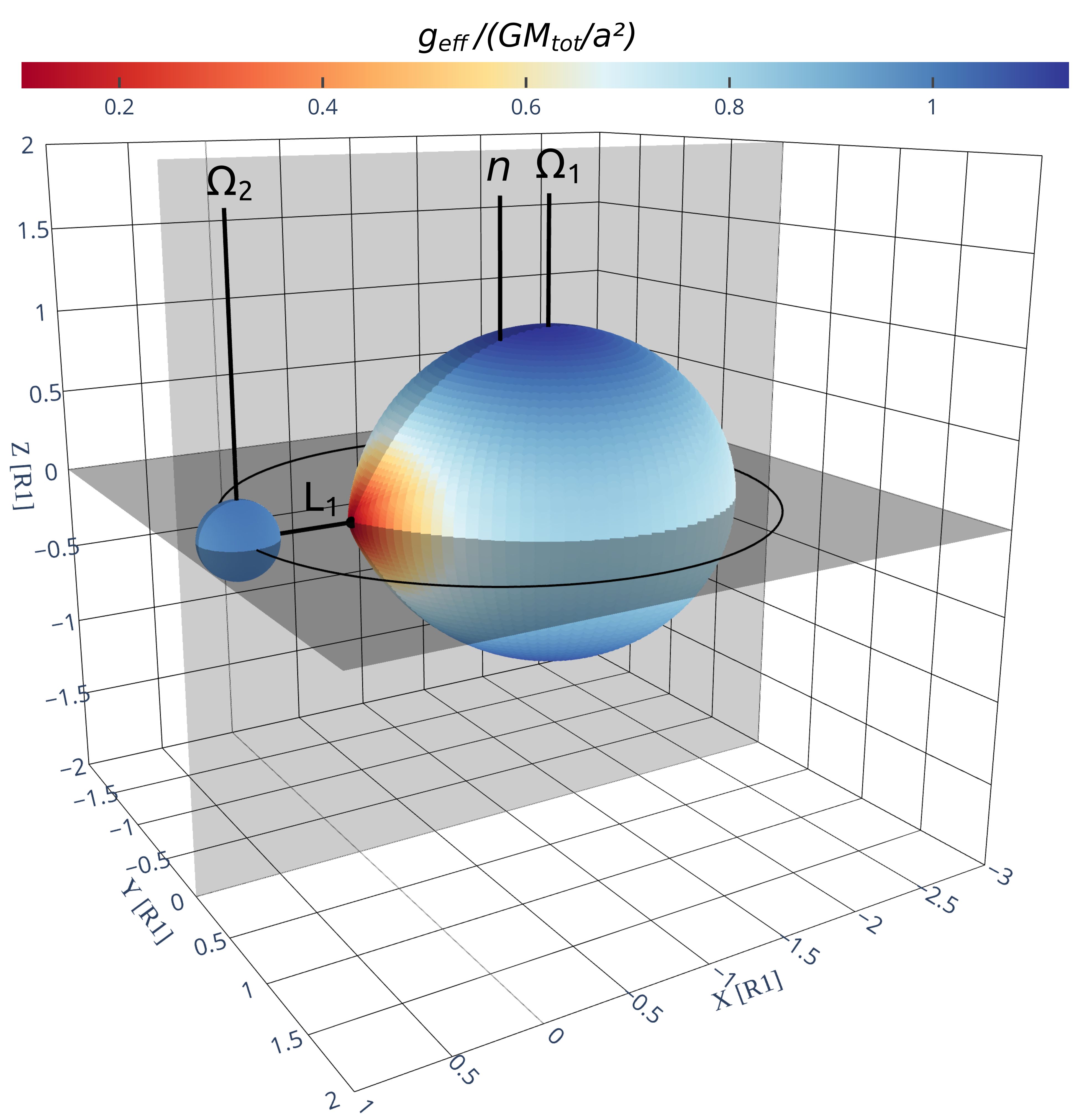}
\caption{Illustration of a binary system configuration with non-synchronised rotations.  The system is composed of 1 $M_\odot$ MS primary star and a 0.2 $M_\odot$ MS companion separated by $r=2 R_1$.  The two grey planes are the planes of symmetry,  each rotation axis is  marked with a black line.  The orbital rotation is originating from the  system centre of mass,  located inside the primary.  The black curve is the secondary orbital path, a black axis is joining the centre of each star with and emphasis on the Lagrangian point L1.  Finally,  the colour code corresponds to the surface effective gravity,  the local flux emitted by a star being a function of this quantity.  }
\label{fig.3D_representation}
\end{figure}

We considered a scenario where each star $i$ composing a system has its individual rotation rate $\Omega_{\star, i }$,  independent from the orbital rotation rate $n$.  In this situation,  assuming alignment of all rotational axes,  two planes of symmetry emerge and are exploited by our method.  The first plane of symmetry is the orbital plane, the second is the plane including all the rotation axes.  A 3D representation of this configuration is shown in \autoref{fig.3D_representation} for the same system as in \autoref{fig.equipotentials_1M0}.
With our non-perturbative method, unsynchronised binary systems can be directly modelled by modifying, and generalising the centrifugal potential.  The previous expression of the centrifugal potential,  which relied on the assumption of solid body synchronised rotation and the circular orbit was
\begin{align}\label{eq_centri_pot_old}
\Psi_{\mathrm{c}}(x,y)=-\frac{n^2}{2} \left[ \left( x_{\rm{CM}}-x\right)^2 +y^2\right] 
\end{align}
and becomes in the non-synchronised cylindrical symetric rotation and circular orbit case:
\begin{equation}\label{eq_centri_pot}
\Psi_{\mathrm{c}}(x,y)=n^2x_{\rm{CM}}x- \int_0^{\sqrt{x^2+y^2}} \Omega_{\star, i }^2(s)s\mathrm{d}s,
\end{equation}
where $s$ is the distance from the stellar rotation axis.  In Eqs.  (\ref{eq_centri_pot_old}) and (\ref{eq_centri_pot}),  we used Cartesian coordinates centred on the star considered.  Here,  $x$ represents the coordinate along the apsidal line (i.e.  the major-axis of the orbit) pointing towards the companion,   $x_{\rm{CM}}$ denotes the distance between the system's centre of mass  and the star's centre,  and $y$  is the coordinate along the perpendicular axis to $x$ included in the orbital plane. 
\\As expected  from the comparison of Eqs.  (\ref{eq_centri_pot_old}) and (\ref{eq_centri_pot}),  for a synchronised system ($\Omega_{\star, i }=n$), these expressions are equivalent up to a constant factor. The effective gravity resulting from the centrifugal force can be obtained by differentiating  \autoref{eq_centri_pot}.  
\\Our model can be generalised to eccentric systems by decomposing the motion of each element as a Keplerian motion around the system's centre of mass plus a cylindrical rotation of the star.   
To obtain the centrifugal potential at a given point of the orbit where the instantaneous distance between the centre of the two binary components is $r$, $n^2$ has to be replaced,  in \autoref{eq_centri_pot},  by $ \tilde{n}^2 $ defined as
\begin{equation}\label{eq_def_n_tilde}
 \tilde{n}^2 = \frac{G M_{\mathrm{tot}}}{r^3},
\end{equation}
so that $\tilde{n}^2r$ is the norm of the Keplerian instantaneous relative acceleration of the bodies in the system.  In \autoref{eq_def_n_tilde}, $M_{\mathrm{tot}}$ is the total mass of the system.
With this decomposition and our assumption of the cylindrical rotation, the acceleration of an element in a star is entirely given by the gradient of a potential. The acceleration induced from our decomposition of the mouvement is the gradient of the centrifugal potential that is expressed for the primary star as
\begin{equation}
\mathbf{\ddot{r_1}}=-\tilde{n}^2 x_{\rm{CM,1}}\mathbf{e_{x_1}} + \Omega_{\star, 1 }^2(s_1)s_1\mathbf{e_{s_1}},
\end{equation}
where the bold symbols denote vectorial quantities, $\mathbf{\hat{e}_{x_1}}$ is the unit vector linked to the $x$-axis of the primary and $\mathbf{\hat{e}_{s_1}}$ is the unit vector linked to the gradient of $s$. In all this article, we symplify further the problem by only consiering solid body rotations.

\subsection{Force perturbation}\label{subsect_Perturbating force}
In classical celestial mechanics theory,  the relative acceleration of two bodies in a binary system is entirely given by its radial component expressed as
\begin{equation}\label{eq_relative_motion}
\mathbf{\ddot{r}}=\mathbf{\ddot{r_2}}-\mathbf{\ddot{r_1}}=\left[ \frac{F_{21}}{M_1}-\frac{F_{12}}{M_2} \right]  \mathbf{\hat{e}_{r}}=  \frac{F_{21}}{m_{\mu}} \mathbf{\hat{e}_{r}},
\end{equation}
where the bold symbols denote vectorial quantities, $F_{21}$ is the force exerted by the secondary star on the primary (noting $F_{2\rightarrow 1} \equiv F_{21}$), $F_{12}$ its opposite,  $\mathbf{\hat{e}_{r}}$ is the radial unit vector centred on the secondary and pointing towards the primary, $M_i$ are the individual masses of the bodies and $m_{\mu}$ is the reduced mass of the system defined as
\begin{equation}
m_{\mu}=\frac{M_1M_2}{M_1+M_2}.
\end{equation} 
The last step of \autoref{eq_relative_motion} was deduced from the third Newton Law ($F_{12}=-F_{21}$). The total force per unit of mass exerted by the secondary on its primary can be decomposed as
\begin{equation}\label{eq_perturbed_acceleration}
 \frac{F_{21}}{m_{\mu}}= -\frac{G M_{\mathrm{tot}}}{r^2}+ F_{\mathrm{R}}, 
\end{equation}
In \autoref{eq_perturbed_acceleration}, $F_{\mathrm{R}}$ is the radial perturbed acceleration resulting from considering stars as non-spherical extended interacting bodies. This perturbation is responsible for all the equilibrium tidal-induced dynamical effects of binary evolution,  including orbital migration (i.e.  variation of systems semi-major axis),  eccentricity dissipation (i.e.  reduction of orbital eccentricty),  and  apsidal motion (i.e.  time variation of the argument of the periastron). 
\\In the framework of our modelling, the force exerted by the secondary on the primary  can be expressed as 
\begin{align}
 F_{21}&=-\int \rho_{1}F_{2} \mathrm{d} V_1 = -\int \rho_{1}\frac{\partial \Psi_{2}}{\partial x}\mathrm{d} V_1,
\end{align}
where $F_{2}$ is the gravitational force from the secondary inside the primary.
By decomposing the densities and the forces in unperturbed spherical symmetric terms,  respectively $\rho_{10}$ and $F_{20}$,  and non-spherical perturbations,  respectively $\rho_{1}^{\prime}$ and $F_{2}^{\prime}$,  the force exerted by the secondary on the primary  can be rewritten as 
\begin{align}
F_{21}=-\int (\rho_{10}+\rho_1^{\prime})(F_{20}+F_2^{\prime})\ \mathrm{d} V_1.
\end{align}
The spherically symmetric component of the force can be simply identified as being 
\begin{equation}
-\frac{1}{m_{\mu}}\int \rho_{10}F_{20} \mathrm{d} V_1= -\frac{G M_{\mathrm{tot}}}{r^2}.
\end{equation}
Therefore, the perturbed acceleration is expressed as
\begin{align}
F_{\mathrm{R}}= -\frac{1}{m_{\mu}}\int \left(\rho_{10}F_2^{\prime}+\rho_1^{\prime}F_{20} + \rho_1^{\prime}F_2^{\prime}\right)\ \mathrm{d} V_1.
\end{align}
In MoBiDICT, all quantities are projected on spherical harmonics to facilitate the solving of the Poisson equation (\autoref{eq_Poisson_classique}), for example the densities and potentials are decomposed as 
\begin{equation}
\rho_i(r_i,\mu, \phi)=\sum^L_{\ell=0} \sum^{(\ell-p)/2}_{k=0}   \rho_{i, \ell}^m (r_i) Y_{\ell, m} (\mu, \phi),
\end{equation}
and
\begin{equation}
\Psi_i(r_i,\mu, \phi)=\sum^L_{\ell=0} \sum^{(\ell-p)/2}_{k=0}   \Psi_{i, \ell}^m (r_i) Y_{\ell, m} (\mu, \phi),
\end{equation}
where $L$ is the free parameter of our modelling defining the maximum degree of the spherical harmonics to be accounted for, $p=0$ if $\ell$ is even, $p=1$ otherwise, and $m=2k+p$. In this formalism, the decomposition in spherically symmetric and non-symmetric terms is straightforward. The $\ell=0, m=0$ term is the spherically symmetric component of each quantity while all other spherical orders are non-symmetric terms.  In our formalism the perturbed acceleration can thus be obtained with 
\begin{align}\label{eq_R_MoBiDICT}
& F_{\mathrm{R}}= -\frac{1}{m_{\mu}}\int \rho_{1,0}^0Y_{0, 0} \left( \sum^L_{\ell=1} \sum^{(\ell-p)/2}_{k=0}  F_{2,\ell}^m \right)  \mathrm{d} V_1\\ &- \frac{1}{m_{\mu}}\int  \left(\sum^L_{\ell=1} \sum^{(\ell-p)/2}_{k=0}  \rho_{1,\ell}^m Y_{\ell, m}\right)F_{2,0}^0  \nonumber \mathrm{d} V_1 \\ &- \frac{1}{m_{\mu}}\int  \left(\sum^L_{\ell=1} \sum^{(\ell-p)/2}_{k=0}  \rho_{1,\ell}^m Y_{\ell, m}\right)\left(\sum^L_{\ell=1} \sum^{(\ell-p)/2}_{k=0}  F_{2,\ell}^m \right)  \mathrm{d} V_1.\nonumber
\end{align}
Finally, each force term projected on the spherical harmonic basis $F_{2,\ell}^m$ can be decomposed and re-expressed in a spherical coordinate frame (to be integrated) with
\begin{align}\label{eq_decompositionF_lm_npert}
F_{2,\ell}^m = \frac{\partial r_{1}}{\partial x_1}    \frac{\partial \Psi_{2, \ell}^m Y_{\ell, m}}{\partial r_1} + \frac{\partial \theta_{1}}{\partial x_1}    \frac{\partial \Psi_{2, \ell}^m Y_{\ell, m}}{\partial \theta_1}   +\frac{\partial \phi_{1}}{\partial x_1}    \frac{\partial \Psi_{2, \ell}^mY_{\ell, m}}{\partial \phi_1}.
\end{align}
The spherical harmonics $Y_{\ell, m}$ in \autoref{eq_decompositionF_lm_npert} are functions of $\mu_2$ and $\phi_2$. The expressions of the $\partial \Psi_{2, \ell}^mY_{\ell, m}/\partial r_1, \partial \Psi_{2, \ell}^mY_{\ell, m}/\partial \theta_1 $,  and $\partial \Psi_{2, \ell}^mY_{\ell, m}/\partial \phi_1$ terms can be found in \cite{Fellay2023}.  More than having a non-perturbative description of the resulting force,  the main advantage of our formalism is the possibility of naturally compare the contributions of the different spherical harmonic orders by selecting the desired orders in the sums. 
\subsection{Non-perturbative apsidal motion computation}\label{subsect_apsidal_motion_mobi}
\begin{figure}[h]
\centering
\includegraphics[width=\hsize]{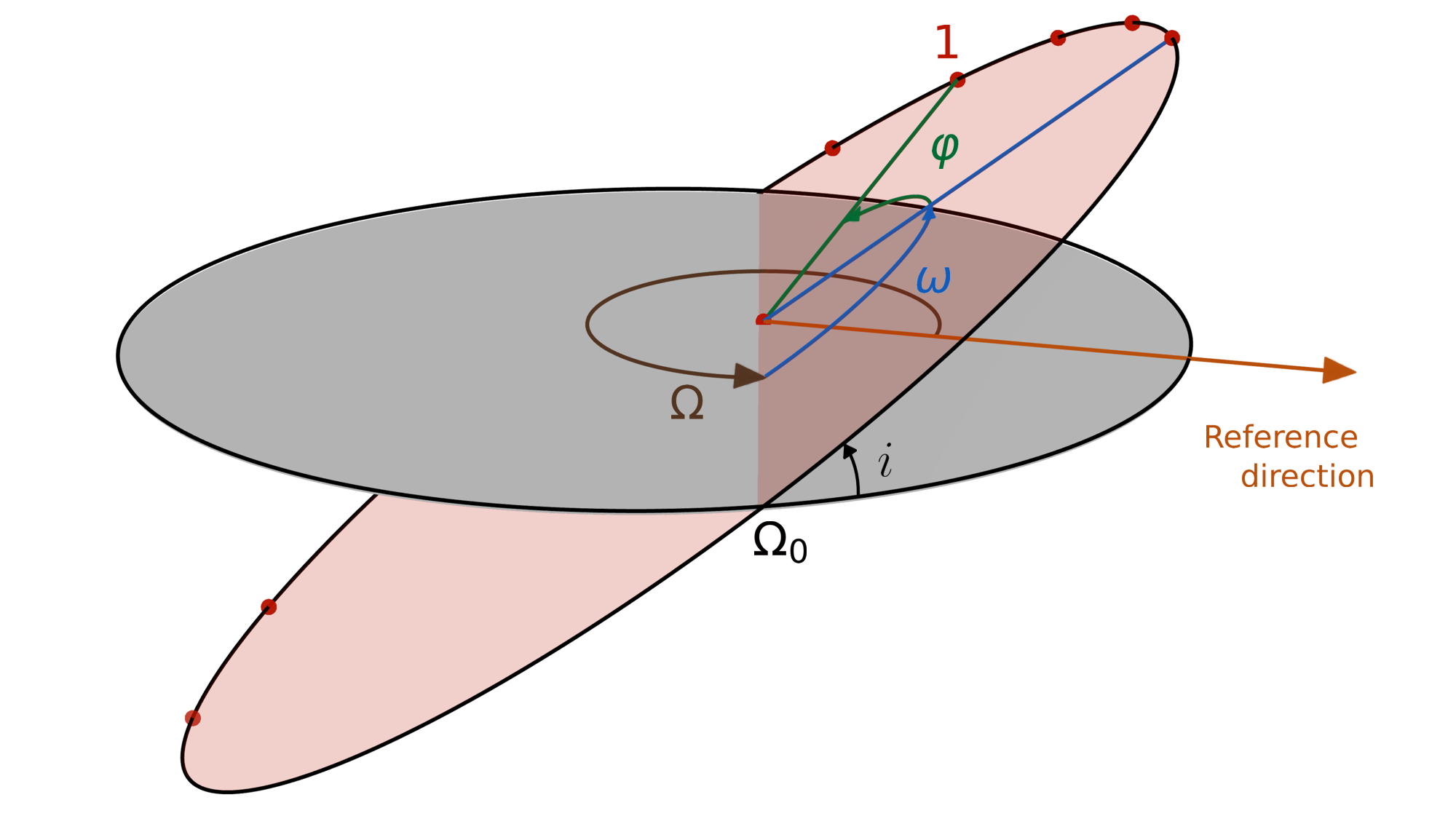}
\caption{Orbital trajectory of a star in a binary system with an eccentricity of $0.2$.  The orbital plane is in red,  the grey plane is a the reference plane (i.e. the plane of the sky), $i$ is the inclination of the system compared to the reference plane,  $\Omega$ is the longitude of the ascending node,  $\Omega_0$ is the ascending node,  $\omega$ is the argument of the periastron,  and $\varphi$ is the true anomaly.  The reds dots on the trajectory represent the different points at which the snapshots are taken to compute the apsidal motion. }
\label{fig.2D_scheme_orbits}
\end{figure}

In this section, we developed the formalism used to model the apsidal motion with our non-perturbative method.  \autoref{fig.2D_scheme_orbits} illustrates the setup of an eccentric binary system, the apsidal motion of a system corresponding to $\mathrm{d}\omega/\mathrm{d}t$.
In eccentric binary systems, aspidal motion is a direct observational constraint on the deformations of distorted bodies. Recent investigations \citep{Rosu2020,Rosu2022a,Rosu2022b} highlight its utility in constraining physical processes occurring during stellar evolution, such as internal mixing.  
\\The apsidal motion is induced by the equilibrium tidal perturbed acceleration $F_{\mathrm{R}}$, the general relativity and dynamical tidal perturbed acceleration.  The general relativistic component of the apsidal motion can be expressed as 
\begin{equation}
\frac{\mathrm{d}\omega_{\mathrm{rel}}}{\mathrm{d}t}=\frac{2 \pi}{P_{\mathrm{orb}}} \frac{3G M_{\mathrm{tot}}}{c^2 a (1-e^2)},
\end{equation}
where $a$ is the system's semi-major axis, $e$ its eccentricity, $P_{\mathrm{orb}}$ its orbital period, $c$ is the celerity,  and $G$ the gravitational constant. 
In this article,  we studied the equilibrium tides component of the apsidal motion and neglected the dynamical tides component, the latter will be the focus of a forthcoming article.  
From the Gauss planetary equations for a perturbed acceleration along the apsidal line (see e.g.  \cite{Kopal1959}),  the apsidal motion of a binary system is classically given by 
\begin{equation}\label{eq_domega_base}
\dot{\omega}_N\equiv\frac{\mathrm{d}\omega_N}{\mathrm{d}t}= -\dfrac{\sqrt{1-e^2}}{e} \sqrt{\dfrac{a}{GM_{\mathrm{tot}}}}\langle F_{\mathrm{R}} \cos{\varphi}\rangle,
\end{equation}
where $\varphi$ is the true anomaly and  $\langle F_{\mathrm{R}} \cos{\varphi}\rangle$ denotes the time average of the perturbed acceleration expressed as 
\begin{equation}\label{eq_R_basic}
\langle F_{\mathrm{R}} \cos{\varphi}\rangle=\dfrac{\int F_{\mathrm{R}} \cos{\varphi} \mathrm{d}t }{\int \mathrm{d}t}.
\end{equation}
To obtain the apsidal motion of a system through a non-perturbative approach,  the sole quantity to be computed by our non-perturbative modelling is the time average perturbed acceleration.  To achieve this,  we took advantage of the fast computing capabilities of our method,  we captured snapshots of stellar deformations at various points along the binary orbits, as depicted in \autoref{fig.2D_scheme_orbits}.  Using these snapshots,  we can derive the time-averaged perturbed acceleration by reformulating its expression with the second Kepler's law:
\begin{equation}\label{eq_dt_dphi}
\dfrac{\mathrm{d}t}{\mathrm{d}\varphi} =\frac{r^2}{n a^2 \sqrt{1-e^2}},
\end{equation}
where $r$ is the separation between the stars expressed,  following the first Kepler's Law,  as
\begin{equation}\label{eq_separation}
r=\dfrac{a(1-e^2)}{1+e\cos(\varphi +\omega)}.
\end{equation}
By combining Eqs. (\ref{eq_domega_base}),  (\ref{eq_R_basic}),  (\ref{eq_dt_dphi}) and (\ref{eq_separation}),  the apsidal motion obtained with our non-perturbative modelling is given by
\begin{equation}
\frac{\mathrm{d}\omega_N}{\mathrm{d}t}= \dfrac{1}{\pi} \dfrac{\left(1-e^2\right)^2}{e}   \sqrt{\dfrac{a}{GM_{\mathrm{tot}}}}    \int_0^\pi  \dfrac{F_{\mathrm{R}} \cos{\varphi} }{\left(1+e\cos(\varphi)\right)^2 }    \mathrm{d}\varphi,
\end{equation}
where the integrals are only taken over half of an orbit by symmetry.  In practice,  with our method's implementation,  merely fifteen equally spaced values of  $\varphi$ over half of an orbit are sufficient to reach a relative precision of $10^{-3}-10^{-4}$ on the apsidal motion in the vast majority of close binaries.

\section{Perturbative model}\label{sect_Theoritical_apsides_pert}
\subsection{The Clairaut-Radau equation}\label{subsect_Clairaut}
Perturbative models are obtained by assuming that the deformations of the stellar models are small (more precisely, it is assumed that $|r_\ell^m| \ll 1$ in \autoref{eq_perturbative_1}),  retaining only first order terms in derivations. Similar to our modelling approach,  the gravitational potential  of each star $i$ is obtained through the solution of the Poisson equation projected on a spherical harmonics basis, 
\begin{align}\label{eq_solution_poisson}
\Psi_{i,\ell}^m(r_i)\ =\ &- \frac{4\pi G}{(2\ell+1) r_i^{\ell+1}} \int_0^{r_i}  \rho_{i,\ell}^m(r_i^\prime) r_i^{\prime \ell+2} \textrm{d}r_i^\prime \\ \ &-   \frac{4\pi G r_i^\ell}{2\ell+1}   \int_{r_i}^{\infty} \rho_{i,\ell}^m(r_i^\prime) r_i^{\prime 1-\ell}   \textrm{d}r_i^\prime.\nonumber
\end{align}
The main distinction in this modelling is to treat stars as lightly perturbed spherically symmetric bodies. This assumption greatly simplifies the problem and allows us to relate the deformations of a body directly to its unperturbed structure and system properties.  In this framework,  the deformations of each body are linked to its structural coefficients $\eta_{\ell}$ (see \autoref{eq_perturbative_1} and \autoref{eq_def_eta_ell}).  These coefficients can be obtained by solving the Clairaut-Radau first-order differential equation (\cite{Kopal1959},  and derived in \autoref{annexe_Clairaut-Radau}) expressed as
\begin{equation}\label{eq_clairaut-radau}
\overline{r}\frac{\mathrm{d} \eta_{\ell}}{\mathrm{d}\overline{r}}+6\frac{\rho(\overline{r})}{\overline{\rho}(\overline{r})}(\eta_{\ell}+1)+\eta_{\ell}(\eta_{\ell}-1)=\ell(\ell+1),
\end{equation}
using the boundary conditions
\begin{equation}
\eta_\ell(0)=\ell-2.
\end{equation}
In the Clairaut-Radau equation,  $\overline{r}$ is the average radius of a chosen isobar while $\overline{\rho}$ is the average density under the chosen isobar. The relationship between the deformed structure and the $\eta_\ell$ coefficients is explained and derived in \autoref{annexe_Clairaut-Radau}.  

\subsection{Perturbation of the gravitational potential and force by a companion}\label{subsect_perturbation_gravity}
We considered the perturbation induced  by a point-mass secondary on the primary. As developed in \autoref{annexe_perturbation_surface}, the potential perturbation caused by the secondary at the surface of the primary is given by 
\begin{align}\label{eq_pertubation_surface_potential}
\Psi_{1\ell}^m(R_{1})&=\frac{1+\ell-\eta_{1\ell}^{m}(R_{1})}{\eta_{1\ell}^{m}(R_{1})+\ell}\left(\Psi_{2,\ell}^m(R_{1}) +\Psi_{\textrm{c},\ell}^m(R_{1}) \right) \\ &= 2 k_{\ell, 1}\left(\Psi_{2,\ell}^m(R_{1}) +\Psi_{\textrm{c},\ell}^m(R_{1}) \right),\nonumber
\end{align}
where $\Psi_{2,\ell}^m(R_{\textrm{1}})$ is the ($\ell,  m$) component of the gravitational potential generated by the secondary evaluated at the surface of the primary,  and the centrifugal components of the potential are noted  $\Psi_{\textrm{c},\ell}^m$.  The first terms of each of these quantities can be found in \autoref{annexe_first_termes}. The coefficients $k_{\ell, 1}$ also known as Love numbers,  characterise the surface non-spheroidal deformation response of a body subjected to a perturbative potential. Mathematically, these coefficients are defined as:
\begin{equation}\label{eq_k2}
k_{\ell, 1}=\frac{1+\ell-\eta_{1\ell}^{m}(R_{1})}{2(\eta_{1\ell}^{m}(R_{1})+\ell)}.
\end{equation}
\autoref{eq_pertubation_surface_potential} provides a direct comparison between the models, where the quantity $\Psi_{1,\ell}^m(R_{1})$ is equivalent to $\Psi_{1, \ell}^m(R_{1})$ of our non-perturbative iterative method.  One particular application to \autoref{eq_pertubation_surface_potential} concerns the dominant contribution to deformations and perturbations, the quadrupolar terms ($\ell=2$ terms).  For those terms,  the sum of the centrifugal and tidal potentials generated at the surface of the primary by the secondary treated as a point-mass body is expressed, in the non-synchronised case,  as 
\begin{equation}\label{eq_pertubation_surface_potential_l2_m0}
\Psi_{\textrm{c},2}^0(R_{\textrm{1}})+\Psi_{2,2}^0(R_{\textrm{1}})= \left( \dfrac{R_{1}}{r} \right)^2\left(\dfrac{\Omega_{\star,1}^2}{3} r^2 +  \dfrac{1}{2}\dfrac{GM_2}{r}\right),
\end{equation}
and
\begin{equation}\label{eq_pertubation_surface_potential_l2_m2}
\Psi_{\textrm{c},2}^2(R_{\textrm{1}})+\Psi_{2,2}^2(R_{\textrm{1}})= -\dfrac{1}{4}\dfrac{GM_2}{R_{\textrm{1}}} \left(\dfrac{R_{\textrm{1}}}{r} \right)^{3}.
\end{equation}
In both synchronised and non-synchronised systems, the dipolar component of the total potential  is null ($\Psi_{\textrm{c},1}^1(R_{\textrm{1}})+\Psi_{2,1}^1(R_{\textrm{1}})$) when the secondary is treated as a point-mass,  as gravitational and centrifugal forces cancel out according to the third Kepler law.  As highlighted earlier,  when computing the apsidal motion of a system,  it is imperative to account for the tidally induced acceleration perturbation $F_{\mathrm{R}}$.  By only considering the spherical order lower than $\ell=3$, the perturbed acceleration can be decomposed as follows:
\begin{equation}\label{eq_R_base_pert}
F_{\mathrm{R}}= -\frac{1}{m_{\mu}} \left[  F_{2^{\prime}10} +F_{201^{\prime}} +\cancel{F_{2^{\prime}1^{\prime}}}     		\right],
\end{equation}
where $F_{2^{\prime}10}$ is the force exerted by the quadrupolar deformation of the secondary on the unperturbed primary, $F_{201^{\prime}}$ is the force exerted by the unperturbed secondary on the already perturbed primary, and $F_{2^{\prime}1^{\prime}}$ is the force exerted by the perturbed secondary on the already perturbed primary.  The last term, $F_{2^{\prime}1^{\prime}}$,  is a second order term,  thus neglected in the pertubative approach.  In accordance with Newton's third law,  the force exerted by the unperturbed primary on the already perturbed secondary is the opposite of the force exerted by the perturbed secondary on the unperturbed primary, mathematically $F_{201^{\prime}}=-F_{1^{\prime}20}$. As shown in \autoref{annexe_perturbation of the force}, the force exerted by the perturbed  secondary on the unperturbed primary is given by
\begin{equation}\label{eq_F210_pert}
F_{2^{\prime}10}=2\frac{GM_1^2}{r^2} k_{2,2} \left( \dfrac{R_2}{r}\right)^5\left[ 3+\frac{1}{2} \frac{M_{\mathrm{tot}}}{M_1} \left( \dfrac{\Omega_{\star, 1}}{\tilde{n}}\right)^2 \right],
\end{equation}
similarly,  $F_{1^{\prime}20}$ can be obtained by permuting the primary and secondary in  \autoref{eq_F210_pert}.
Finally, combining Eqs.  (\ref{eq_R_base_pert}) and  (\ref{eq_F210_pert}) the perturbed acceleration is expressed as 
\begin{align}\label{eq_R_pert}
F_{\mathrm{R}}&=-2\frac{G M_{\mathrm{tot}}}{r^2}\left[ k_{2,2}\dfrac{1}{q} \left( \dfrac{R_2}{r}\right)^5\left(3+\frac{q+1}{2} \left( \dfrac{\Omega_{\star, 2}}{\tilde{n}}\right)^2\right) \right.  \\ &+ \left. k_{2,1} q \left( \dfrac{R_1}{r}\right)^5\left(3+\frac{1}{2} \frac{q+1}{q} \left( \dfrac{\Omega_{\star, 1}}{\tilde{n}}\right)^2 \right)       \right],\nonumber
\end{align}
with $q=M_2/M_1$.
\subsection{Perturbative apsidal motion computation}\label{subsec_pert_apsidal_motion_theory}
In the case of an eccentric orbit,  the mean orbital rotation rate $n$ is distinct from the instantaneous orbital rotation rate $\tilde{n}$ experienced by each individual binary component.  Using this formalism, the perturbed acceleration can be rewritten by grouping the tidal and centrifugal terms as 
\begin{align}\label{eq_F_tides_total}
F_{\mathrm{R}}&=-2\frac{G M_{\mathrm{tot}}}{a^2}\left[ \left( 3k_{2,2}\dfrac{1}{q} \left( \dfrac{R_2}{a}\right)^5+3k_{2,1} q \left( \dfrac{R_1}{a}\right)^5\right)\left( \dfrac{a}{r}\right)^7 \right. \\ &+\left( k_{2,2}\dfrac{q+1}{2q} \left( \dfrac{R_2}{a}\right)^5 \left( \dfrac{P_{\mathrm{orb}}}{P_{\star, 2}} \right)^2\right. \nonumber\\ &+ \left.\left. k_{2,1}\frac{q+1}{2}\left( \dfrac{R_1}{a}\right)^5\left( \dfrac{P_{\mathrm{orb}}}{P_{\star, 1}}\right)^2\right)\left( \dfrac{a}{r}\right)^4\right]\nonumber \\ &= -2\frac{G M_{\mathrm{tot}}}{a^2}\left[ R_{\mathrm{tides}}\left( \dfrac{a}{r}\right)^7+R_{\mathrm{c}}\left( \dfrac{a}{r}\right)^4\right],\nonumber
\end{align}
where $P_{\star, 1}$ and $P_{\star, 2}$ are respectively the rotation period of the primary and secondary,  and $R_{\mathrm{tides}}$ and $R_{\mathrm{c}}$ can be identified in \autoref{eq_F_tides_total}.  The apsidal motion of a system is dependent on the time average of the perturbed acceleration over an orbit $\langle F_{\mathrm{R}} \cos{\varphi}\rangle$. Using \autoref{eq_R_basic}, the perturbed acceleration  averaged over an orbit is given by 
\begin{align}\label{eq_R_avrage_integrals}
\langle F_{\mathrm{R}} \cos{\varphi}\rangle&=-\frac{2}{P_{\mathrm{orb}}}\frac{G M_{\mathrm{tot}}}{a^2}\left[ R_{\mathrm{tides}} \int_0^{P_{\mathrm{orb}}} \left( \dfrac{a}{r}\right)^7 \cos{\varphi}\ \mathrm{d}t\right. \\ &+ \left.R_{\mathrm{c}}\int_0^{P_{\mathrm{orb}}}\left( \dfrac{a}{r}\right)^4\cos{\varphi}\ \mathrm{d}t \right].\nonumber
\end{align}
The integrals in \autoref{eq_R_avrage_integrals} can be re-expressed with the zero-order Hansen coefficients $X_0^{n,m}(e)$ defined for every positive $m$  as
\begin{equation}
X_0^{n,m}(e)\equiv\frac{1}{P_{\mathrm{orb}}}\int_0^{P_{\mathrm{orb}}} \left( \dfrac{r}{a}\right)^n \cos(m\varphi)\ \mathrm{d}t.
\end{equation}
Without entering  into the detailed expressions of these coefficients,  the two orders that are of interest in our case are the $n=-4, m=1$ and $n=-7, m=1$ that are given by 
\begin{equation}
X_0^{-4,1}(e)=e(1-e^2)^{-5/2},
\end{equation}
and,
\begin{equation}
X_0^{-7,1}(e)=\dfrac{5}{2}e(1-e^2)^{-11/2}\left(1+\dfrac{3}{2}e^2+\dfrac{1}{8}e^4\right).
\end{equation}
The average perturbed acceleration is finally given by
\begin{equation}\label{eq_R_average}
\langle F_{\mathrm{R}} \cos{\varphi}\rangle=-2\frac{G M_{\mathrm{tot}}}{a^2}\left[ R_{\mathrm{tides}} X_0^{-4,1}(e)+ R_{\mathrm{c}}X_0^{-7,1}(e) \right],
\end{equation}
and can be inserted in \autoref{eq_domega_base} to obtain the apsidal motion of a binary system with the perturbative approach:
\begin{align}\label{eq_apsidal_motion_perturbative}
& \frac{\mathrm{d}\omega_N}{\mathrm{d}t}=\frac{2\pi}{P_{\mathrm{orb}}}\left[15 f(e) \left(\left( \frac{R_1}{a}\right)^5 q k_{2,1}+\left( \frac{R_2}{a}\right)^5 \frac{k_{2,2}}{q} \right) \right. \\ &+ \left.  g(e) \left(k_{2,1} (q+1)  \left( \dfrac{P_{\mathrm{orb}}}{P_{\star, 1}}\right)^2\left( \frac{R_1}{a}\right)^5 \right.\right. \nonumber\\ &+\left.\left. k_{2,2} \frac{q+1}{q}   \left( \dfrac{P_{\mathrm{orb}}}{P_{\star, 2}}\right)^2 \left( \frac{R_2}{a}\right)^5 \right)\  \right],\nonumber
\end{align}
with,
\begin{equation}
f(e)=\dfrac{1+\dfrac{3}{2}e^2+\dfrac{1}{8}e^4}{(1-e^2)^5},
\end{equation}
and,
\begin{equation}
g(e)=\frac{1}{(1-e^2)^2}.
\end{equation}

\section{General model comparisons}\label{sect_comparaison_theoritical_models}
This section is dedicated to a theoretical comparison of different types of stars encompassing a range of potential binary systems across different mass regimes.  Here,  we present the input stellar models used to compare the perturbative and non-perturbative modelling approaches.  We chose four models to represent a broad spectrum of stars and structural characteristics. The physical ingredients of the chosen models are detailed in \autoref{table_stella_models}, those models are the same models compared in \cite{Fellay2023}.
\begin{table}[h]
\centering
\caption{Summary of the stellar properties of the 1D input models used in this work.}\label{table_stella_models}
\resizebox{\hsize}{!}{\begin{tabular}{llllllll}
\hline\hline
\multirow{2}{*}{Stellar parameters} & \multirow{2}{*}{$0.2\ M_\odot$} &  \multirow{2}{*}{$1.0\ M_\odot$} &  \multirow{2}{*}{$1.5\ M_\odot$} &  \multirow{2}{*}{$20\ M_\odot$} \\
                                    &                                                          &                                                          &                                                         &                       \\ 
\hline
Mass {[}$M_\odot${]}                & $0.2$                              & $1.0$  & $1.5$ & $20.0$          \\
Radius {[}$R_\odot${]}              & $0.22$                               & $1.03$   & $11.1$ & $6.01 $          \\
Age {[}Gyr{]}                     & $2.0$                               & $2.0 $ & $2.0  $ & $0.002  $      \\
Evolutionary stage        & MS & MS  & RGB                         & MS         \\
Effective temperature {[}$K${]}     & $3342 $                                   & $6080$     & $4621  $    & $35\ 859 $            \\
Luminosity {[}$L_\odot${]}          & $0.005$                              & $1.31$  & $51.1$ & $ 53\ 921 $         \\
Initial Hydrogen $X_0$           & $0.72 $                 & $0.72  $             & $0.72  $ & $0.72  $         \\
Initial Metallicity $Z_0$           & $0.010 $                 & $0.010 $             & $0.010 $ & $0.010$         \\
Core Hydrogen $X_{\rm{c}}$           & $0.693$                             & $0.469 $  & $0 $ & $0.582 $         \\
$k_2$           & $0.151$                             & $0.0104 $ & $0.0593  $ & $0.0127 $         \\
$\Omega_{\star, i}/n  $         & $1.0$                             & $1.0$ & $1.0$ & $1.0$       \\
\hline
\end{tabular}
}\end{table}
Table 1 in \cite{Fellay2023} has only been extended to include the $k_2$ value of each model.  As shown by \autoref{eq_pertubation_surface_potential}, to a higher $k_2$ corresponds a greater surface response from a body when subjected to a potential perturbation to its surface.  Moreover, $k_2$ can be used as an indicator of a body's deformability.  In addition,  our results from \cite{Fellay2023}, showed that the low-mass and red giant branch (RGB) stars are the most distorted by a companion while the massive and solar type stars are expected to be less deformed  and  impacted by our non-perturbative method for a given $a/R_1$ ratio.  In the following sections, we compared  the surface deformations and apsidal motion obtained with our non-perturbative modelling method and the perturbative modelling for twin binary systems ($q=1$) composed of the stars presented in \autoref{table_stella_models}.  Twin binary systems being systems composed of two identical stars (i.e.  stars having the same mass, radius, effective temperature, and rotational period).   
\subsection{Potential and forces discrepancy}\label{subsect_potential_force_discrepancy}
\begin{figure}[h]
\centering
\includegraphics[width=\hsize]{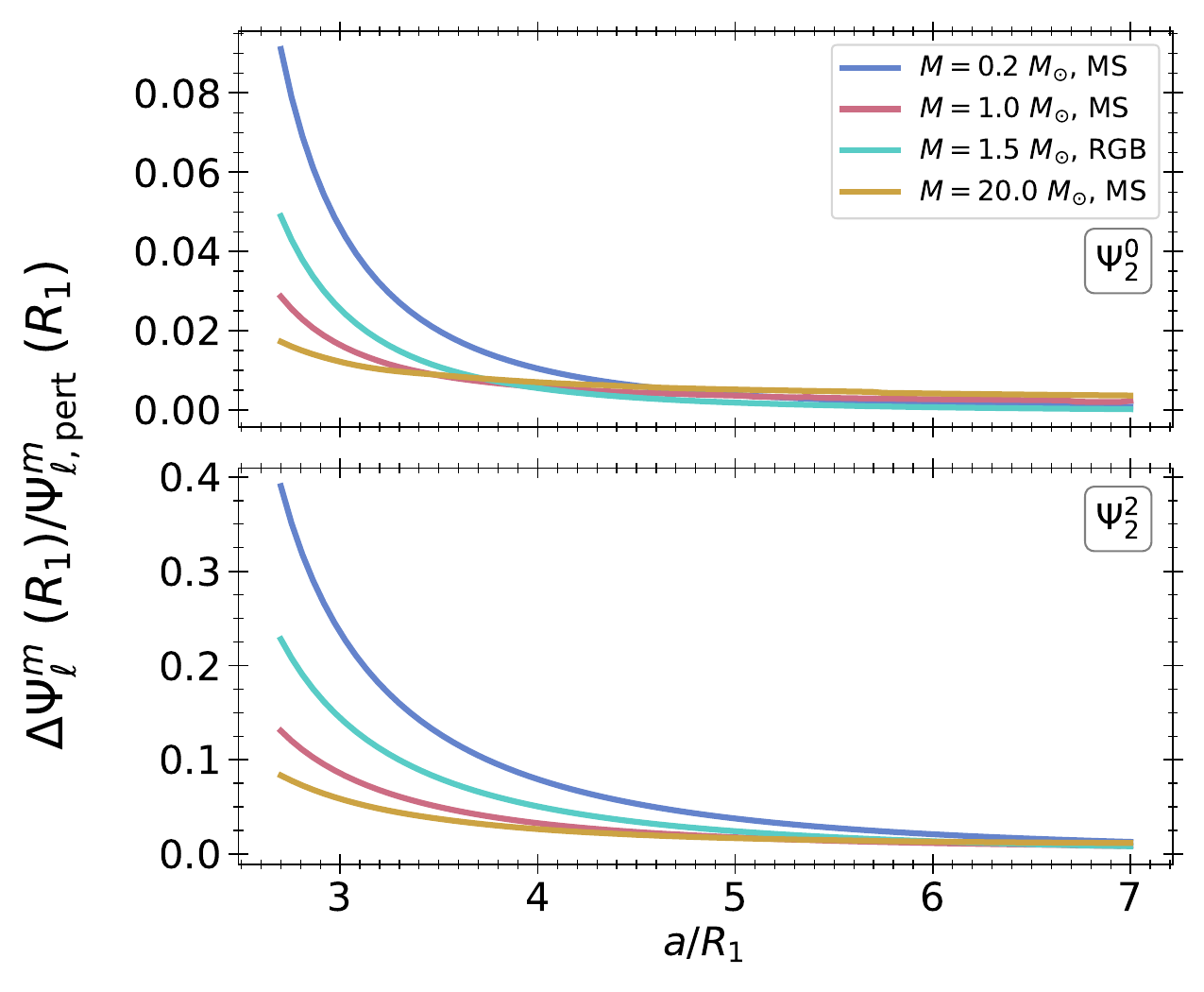}
\caption{Surface potential discrepancy between perturbative and non-perturbative modelling approaches as a function of the orbital separation normalised by the stellar radii for the different binary components presented in \autoref{table_stella_models}. The upper panel corresponds to the $\ell=2$, $m=0$ component of the potential while the lower panel is the $\ell=2$, $m=2$ term.}
\label{fig.diff_pot_surface_all_models}
\end{figure}

The perturbation of the surface potential is a reliable indicator of the deformations undergone by a star. Within either our formalism or the perturbative formalism, the perturbation of the surface potential is given by all spectral terms of the surface potential $\Psi_{i\ell}^m(R_{i})$ with $\ell>0$. In the perturbative method, this quantity is given by \autoref{eq_pertubation_surface_potential}, which is applied to the $\ell=2$ terms in Eqs.  (\ref{eq_pertubation_surface_potential_l2_m0}) and (\ref{eq_pertubation_surface_potential_l2_m2}). Within our non-perturbative method, $\Psi_{i\ell}^m(R_{i})$ are obtained in an iterative process, solving Poisson's equation at each step and for each spherical order.  Consequently, there is no analytical expressions available for this quantity in our method.  To compare the surface deformations with different modelling procedures we introduce the  discrepancy of surface potential,  denoted$\Delta\Psi_{\ell}^m(R_1)$ and defined as:
\begin{equation}
\Delta\Psi_{\ell}^m(R_1)=\Psi_{\ell, \mathrm{MoBi}}^m(R_1)-\Psi_{\ell, \mathrm{pert}}^m(R),
\end{equation}
where $\Psi_{\ell, \mathrm{MoBi}}^m$ is the spectral surface potential obtained with our non-perturbative method and $\Psi_{\ell, \mathrm{pert}}^m$ is the same quantity obtained with the perturbative approach. In \autoref{fig.diff_pot_surface_all_models}, we show this surface potential discrepancy as a function of the orbital separation for twin binary systems composed of the stars presented in \autoref{table_stella_models}.  Our focus in this analysis is on the dominant terms, the $\ell=2$ terms.
\\The surface potential discrepancy,  varies significantly depending on the stellar type and the orbital separation. For both $\ell=2$ components, stars with the higher $k_2$ exhibit more pronounced effects from our non-perturbative treatment,  resulting in differences of up  to $\sim 40 \%$ in the $\ell=2, m=2$ term for low mass stars.  This implies that the higher the deformations, the more substantial the underestimation of the surface deformations by the perturbative approach.  Furthermore,  orbital separation plays a critical role,   with closer star configurations yielding stronger deformations and more notable disparities in the surface potential.  
\\~\\By comparing the two panels in \autoref{fig.diff_pot_surface_all_models},  we observe a significantly smaller difference in surface potential for  the $\ell=2$, $m=0$ term, regardless of the model.  The $\ell=2$, $m=0$ component arises from both centrifugal and tidal deformations, while the $\ell=2$, $m=2$ term is solely originating from the tidal deformation. Both our modelling and perturbative modelling are sharing the same assumptions on the centrifugal potential, therefore, the difference originating from tidal contribution in the $\ell=2$, $m=0$ term is diluted by the centrifugal contribution.  The major difference between the perturbative and non-perturbative modelling approaches is thus appearing in the treatment of the tidal deformation. 
\subsection{Apsidal motion}\label{subsect_apidal_motion_theoritical_model}
\begin{figure}[h]
\centering
\includegraphics[width=\hsize]{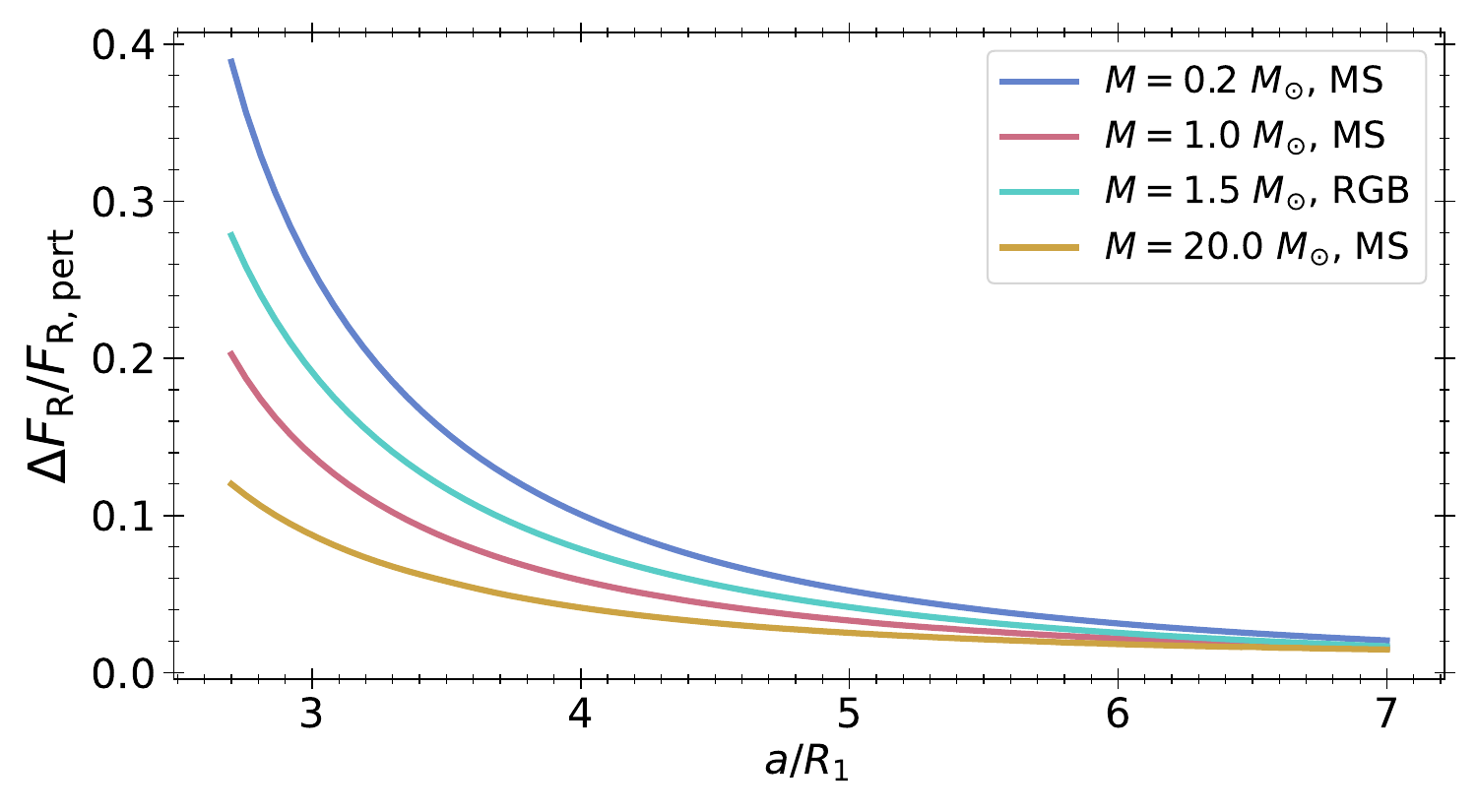}
\caption{Perturbed acceleration discrepancy as a function of the separation normalised by the stellar radii for the twin binary systems composed of the stars presented in \autoref{table_stella_models}. The colour code denotes the different twin binary systems. }
\label{fig._Delta_R_general}
\end{figure}

\begin{figure}[h]
\centering
\includegraphics[width=\hsize]{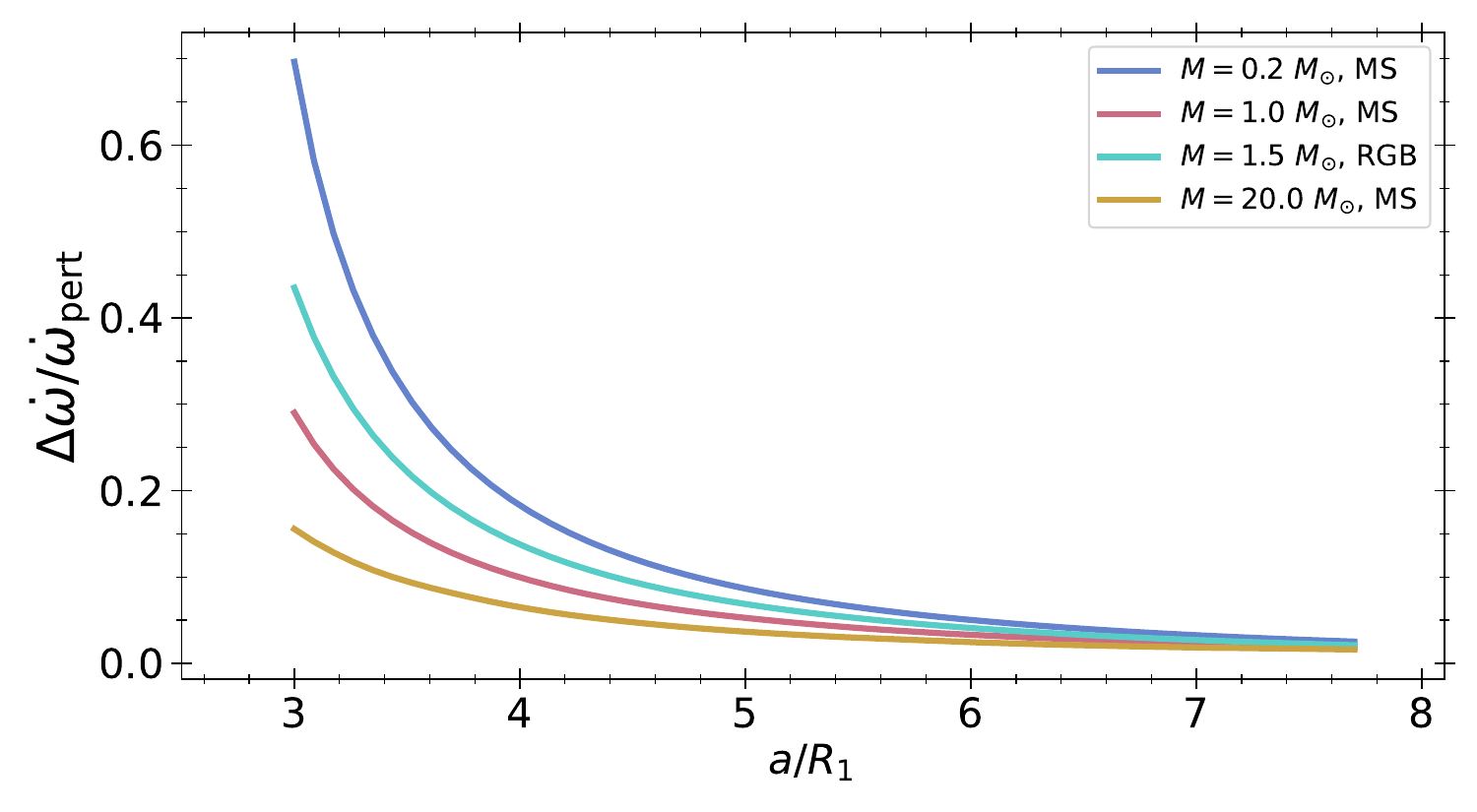}
\caption{Evolution of the apsidal motion relative difference as a function of the orbital separation normalised by the stellar radii.  The eccentricity of the orbits were set to $e=0.1$ and the colour code denotes the different twin binary systems. }
\label{fig._delta_apside_general}
\end{figure}

In \autoref{subsect_apsidal_motion_mobi} we showed that the apsidal motion of an eccentric binary system is proportional to the time average of the perturbed acceleration.  As detailed in \autoref{subsect_potential_force_discrepancy},  the perturbed acceleration denotes the tidal acceleration originating from a nearby companion.  A modification of the surface $\ell=2$ potential  as illustrated in \autoref{fig.diff_pot_surface_all_models}  introduces variations in the perturbed acceleration.  In this section, we investigated the perturbed acceleration discrepancy resulting from the potential differences seen and the consequent impact on the apsidal motion of eccentric binaries.  
\\~\\In the classical  perturbative model,  the perturbed acceleration $F_{\mathrm{R}}$ is solely originating from the quadrupolar component of the deformation. The detailed expression of $F_{\mathrm{R}}$ is given in \autoref{eq_R_pert} and derived in \autoref{annexe_perturbation of the force}.  In opposition,  our methodology  incorporates a non-perturbative treatment of all spherical orders spanning  $\ell=1,2,...L$.  The exact expression of $F_{\mathrm{R}}$ is given by \autoref{eq_R_MoBiDICT}.  To quantify the disparities between perturbative and non-perturbative perturbed accelerations, we define their difference $\Delta F_{\mathrm{R}}$ as follows:
\begin{equation}
\Delta F_{\mathrm{R}}=F_{\mathrm{R, MoBi}}-F_{\mathrm{R, pert}}.
\end{equation}
The difference of perturbed accelerations as a function of the separation between the twin binary systems components presented in \autoref{table_stella_models} is illustrated in \autoref{fig._Delta_R_general}.
\\The perturbed acceleration discrepancy observed in \autoref{fig._Delta_R_general} is in agreement with the results showed in  \autoref{fig.diff_pot_surface_all_models}.  Models exhibiting the greatest surface potential discrepancies are also exhibiting notable perturbed acceleration differences.  Moreover,  models with higher $k_2$,  namely the low mass and RGB stars, are the most impacted by our modelling.  Solar-like stars and massive stars are comparatively less affected,  even if in close orbit $F_{\mathrm{R}}$ discrepancies can still reach up to, respectively,   20$\%$ and $12\%$.
\\~\\To assess the discrepancy of apsidal motion resulting from the perturbed acceleration differences,  we need to consider binaries with eccentric orbits.  To limit the free parameters of the system we  compare binary systems with synchronised rotations and a fixed eccentricity,  $e=0.1$. The impact of the eccentricity on the apsidal motion is explored for systems in \autoref{subsect_orbital_ecc}.  We introduce the apsidal motion difference: 
\begin{equation}
\Delta \dot{\omega}= \dot{\omega}_{\mathrm{MoBi}}-\dot{\omega}_{\mathrm{pert}},
\end{equation}
to compare the apsidal motion originating from the non-perturbative and perturbative approaches.  In the perturbative approach, the apsidal motion is given by \autoref{eq_apsidal_motion_perturbative}. With our non-perturbative method we required to capture snapshots of the binary system along half of an orbit to estimate the perturbed acceleration time average and hence the apsidal motion (see \autoref{subsect_apsidal_motion_mobi}).  For each binary system composed of the stars presented in \autoref{table_stella_models},  \autoref{fig._delta_apside_general} illustrates $\Delta \dot{\omega}$ as the function of the binaries orbital separation normalised by the stellar radii. 
The apsidal motion discrepancy is a direct result of the perturbed acceleration illustrated in \autoref{fig._Delta_R_general}.  Binary systems with greater $F_{\mathrm{R}}$ discrepancis exhibit the highest apsidal motion differences.  For the low mass and RGB stars,  the apsidal motion discrepancy can reach up to respectively  $70\%$ and $45\%$ when the stars are close to contact,  in their periastron.  Similarly, for the solar type and massive stars we found an apsidal motion discrepancy up to respectively  $30\%$ and $15\%$. These discrepancies vary depending on the exact structure of the stars or the architecture and the orbital parameters of the system. 
\subsection{Dependency on the orbital eccentricticty}\label{subsect_orbital_ecc}
\begin{figure}[h]
\centering
\includegraphics[width=\hsize]{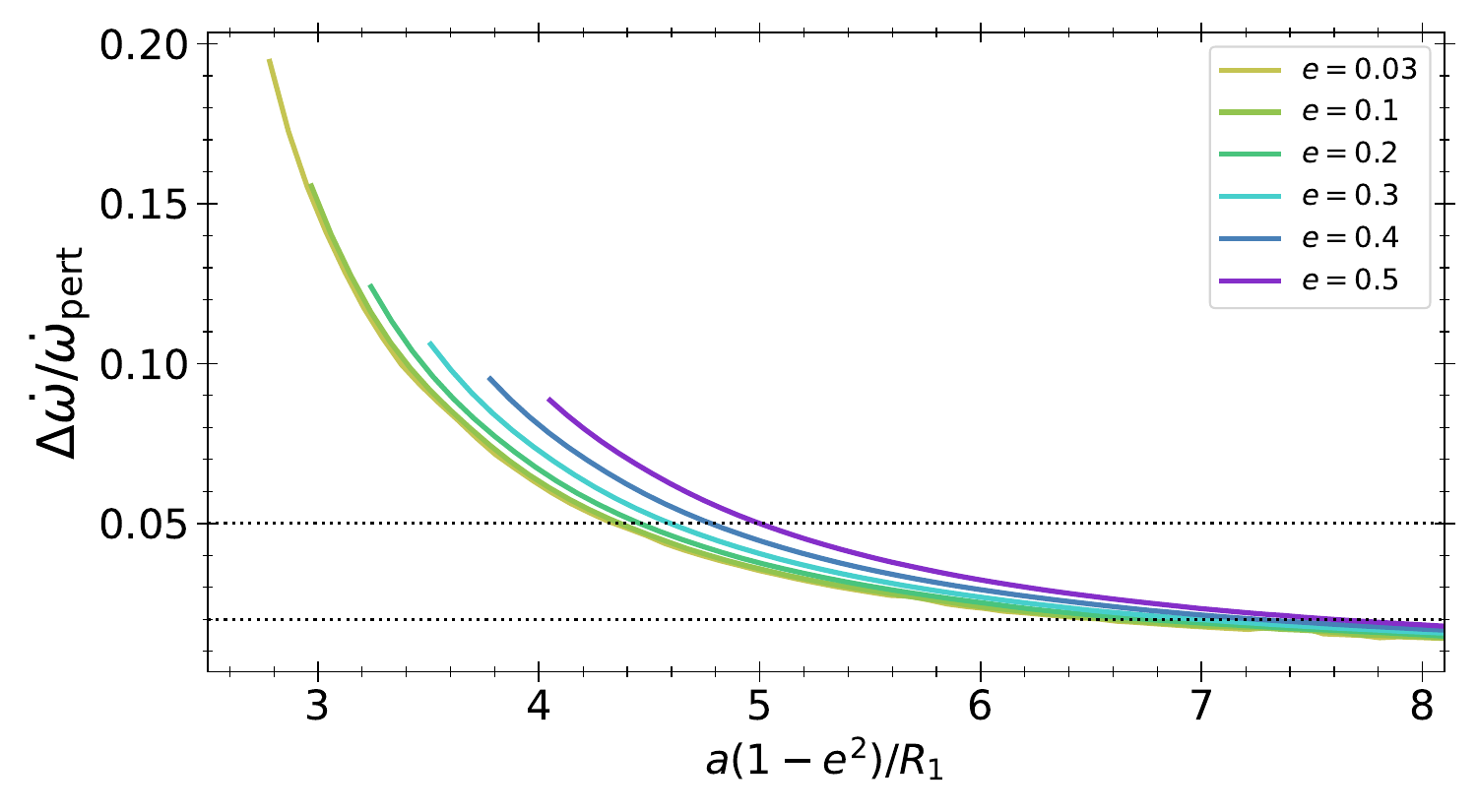}
\caption{Apsidal motion relative different for a twin binary system composed of two 20 $M_\odot$ stars presented in \autoref{table_stella_models}.  Each colour correspond to the variation of discrepancy as a function of the orbital separation scaled by the eccentricity and stellar radii  for different values of the orbital eccentricity.  The black lines are lines of constant model relative difference,  corresponding to $5\%$ and $2\%$ discrepancy. }
\label{fig.apsidal_motion_diff_ecc}
\end{figure}

In the previous section,  we fixed the eccentricity of the systems to infer the modelled apsidal motion discrepancy.  In this section,  we study the impact of varying the orbital eccentricity on the apsidal motion discrepancy.  We expect that binaries with higher eccentricities are closer in their periastron, consequently more deformed and more impacted by our methodology.  We focussed on the twin binary system composed of 20 $M_\odot$ stars as this system is less impacted by our modelling and will consequently give a lower limit to the modelling discrepancies for a given semi-major axis.  We computed a grid of models with different eccentricities and orbital separations,  looking at the apsidal motion relative difference. The results from this computation are illustrated in \autoref{fig.apsidal_motion_diff_ecc}\footnote{ In \autoref{fig.apsidal_motion_diff_ecc},  we have chosen the empirical parameter $a(1-e^2)/R_1$ for the $x$-axis such as the curves are the most superposed at high orbital separation.  }.
The eccentricity is indeed a crucial parameter directly impacting the modelling discrepancies.  As expected, binary systems with higher eccentricities are more significantly impacted by our modelling method for a given orbital separation.   For our least deformed model, we found that the modelling discrepancy reaches at least $5\%$ when $a(1-e^2)/R_1\lesssim 4.5$ and $2\%$ when $a(1-e^2)/R_1\lesssim 6.5$.  For all our other theoretical models these thresholds are higher,  by applying the same methodology to the twin system composed of $0.2 M_\odot$ stars,  we found that the modelling discrepancy thresholds of $5\%$ and $2\%$ are respectively located at $a(1-e^2)/R_1\lesssim 6.5$ and $a(1-e^2)/R_1\lesssim 4.5$.
\\In addition to this work,  we verified the precision of our method at higher orbital separation and found that for $a(1-e^2)/R_1=20$ the discrepancies between the models reach $0.1\%$ indicating a slow decrease of the apsidal motion discrepancy at high orbital separation and a sufficient precision to impose the thresholds given in this section.

\subsection{Dependency on $k_2$ and the orbital separation}\label{subsect_theoritical_dependancy_k2}
\begin{figure}[h]
\centering
\includegraphics[width=\hsize]{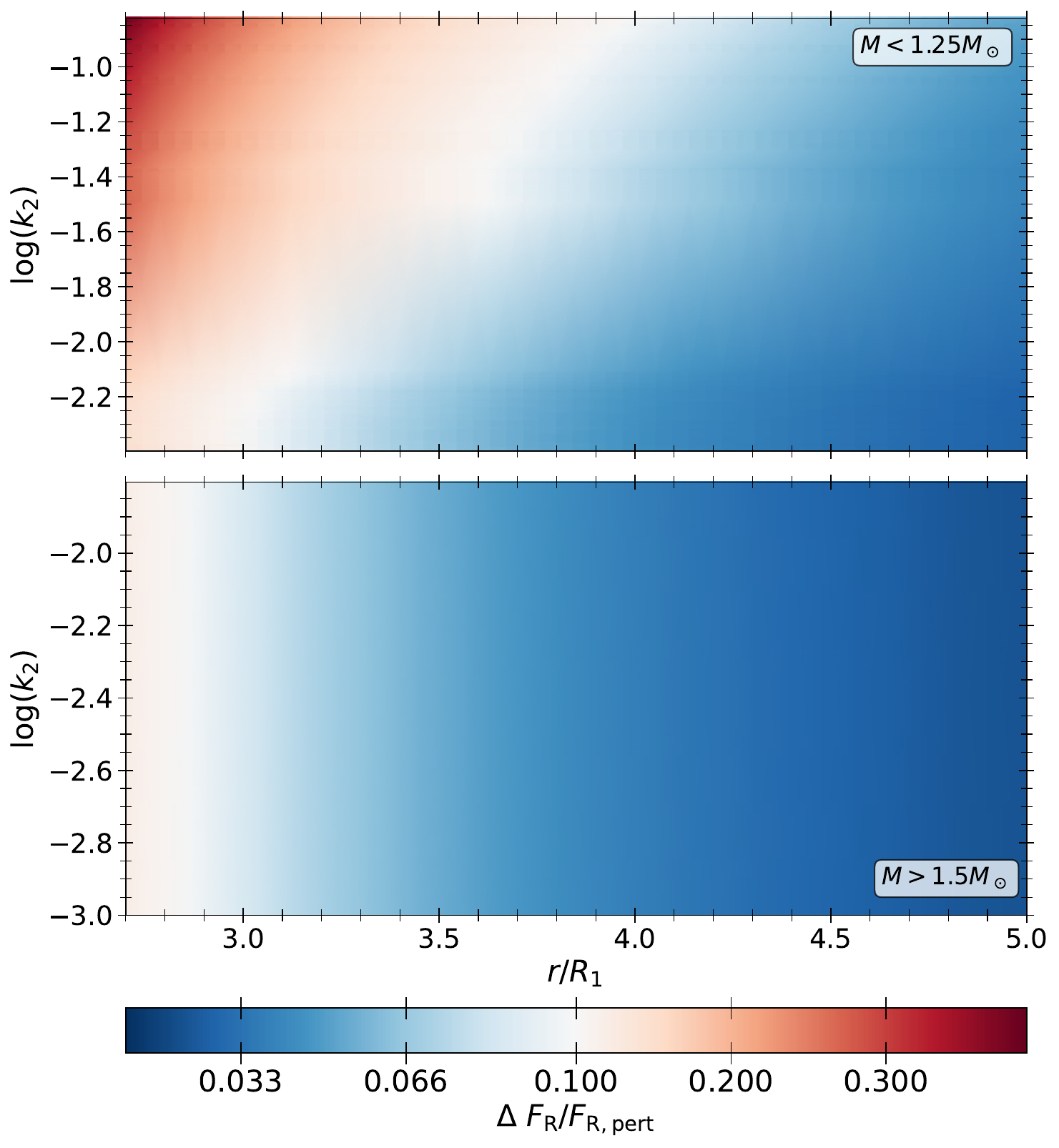}
\caption{Mapping of the acceleration perturbation discrepancy for our grid of low-mass stars and intermediate-high mass stars.  The colour code denotes the difference of perturbed acceleration between the perturbative and our non-peturbative approach as a function of the models $k_2$ and orbital separation.}
\label{fig.grid_perturbed_acceleration}
\end{figure}

In the previous sections,  we saw that a hierarchy exists between the perturbed acceleration of different stellar models and the resulting apsidal motion discrepancy.  We observed that stellar models with higher $k_2$ values were more affected by our modelling approach.  Another crucial quantity impacting the discrepancies is the separation of the binary components.  In this section,  we aim to comprehensively assess modelling discrepancies across a diverse range of stellar models and orbital separation.
\\~\\To explore this,  we constructed a grid of models of MS stars with masses ranging from $0.2 M_\odot$ to $30 M_\odot$ using  the Code Liégeois d'Evolution Stellaire \citep[CLES,][]{Scuflaire2008a}.  For each mass in our grid we took several models along the MS to have diverse models and stellar structures with similar $k_2$ to explore whether discrepancies are solely dependent on $k_2$ and the orbital separation.  
 \\~\\ \autoref{fig.grid_perturbed_acceleration},  illustrates the evolution of the acceleration perturbation relative difference $\Delta F_{\mathrm{R}}/F_{\mathrm{R,pert}}$ between our modelling and the perturbative approach.  The top panel of the figure represents a group of stars that are either fully convective or composed of a radiative core and convective envelope,  specifically stars with masses below $1.25 M_\odot$.  The lower panel corresponds to stars with a convective core,  which includes stars with masses greater than $1.5 M_\odot$.
\\~\\\autoref{fig.grid_perturbed_acceleration} reveals a distinct behaviour between the two groups of stars.   While the lower $k_2$ limit of the low mass stars and the upper limit of the intermediate-high mass stars have similar $k_2$,  their acceleration perturbation discrepancy is not similar. 
This result indicates that discrepancies between the models are not only a function of $k_2$ but  also involve a component related to the intrinsic stellar structure.  From our model grid,  we observed that stars with different structures, particularly those with radiative or convective cores, displayed significant variations in discrepancy despite similar $k_2$ values.  This phenomenon is probably related to the differences in density profiles between stars having or not a convective core.  If looking for a possible parametrisation solely dependent on $k_2$ and orbital separation to correct the modelling discrepancies in the perturbative approach,  the non-linear behaviours of the models greatly reduces its feasibility.  We attempted an MCMC analysis using linear combinations of power laws of the orbital separation,  for both the intermediate-high mass stars and low-mass stars.  However, the results were unsatisfactory to propose an empirical parametrisation.  As detailed in \autoref{subsect_low_order_spherical_apsides}, the modelling discrepancies indeed arise  from a combination of discrepancies in the $\ell=1, 2, 3$ terms.  Non-perturbative modelling is necessary to recover the discrepancies seen in this section.  However,  \autoref{fig.grid_perturbed_acceleration} can be used to get an order of magnitude of the underestimation of tidal forces introduced by the perturbative modelling.  The updated grid of $k_2$ values provided by \cite{Claret2023} could for example be used to easily locate observed stellar systems in this diagram. 
\subsection{Dependency on the mass ratio $q$}\label{subsect_theoritical_dependancy_q}
Our analysis so far focussed on the dependency on $R_1/a$ as this latter parameter has the most important impact on the results. However, the perturbative apsidal motion has a direct dependency on $8$ parameters: ($n,(n/\Omega_1)^2,(n/\Omega_2)^2 ,(R_1/a)^5, (R_2/a)^5, q, e, k_{21},k_{22}$), see \autoref{eq_apsidal_motion_perturbative}. With our non-pertubative modelling the dependencies on $k_{21},k_{22}$ become dependencies on the entire density profile of each star. Characterising the dependency of our result on mass ratio $q$ is useful as this quantity can strongly impact the results and is a direct sub-product of the observations of binaries. To study and modify the $q$ parameter in our modelling, different stellar models with the same age and initial chemical properties have to be used for the secondary star, impacting $4$ of the parameters of our modelling for a given orbital separation:  ($n, (R_2/a)^5, q, k_{22}$ or $\rho_2(r)$). Consequently the effect of a change of $q$ will be seen indirectly, through, in particular $(R_2/a)^5$. For a given $q$ parameter different than one, the behaviour of the apsidal motion discrepancy as a function of the orbital separation is dependent on the regime of $(R_2/R_1)^5$, as discussed hereafter. 
\subsubsection{$R_1^5\gg R_2^5$}
\begin{figure}[h]
\centering
\includegraphics[width=\hsize]{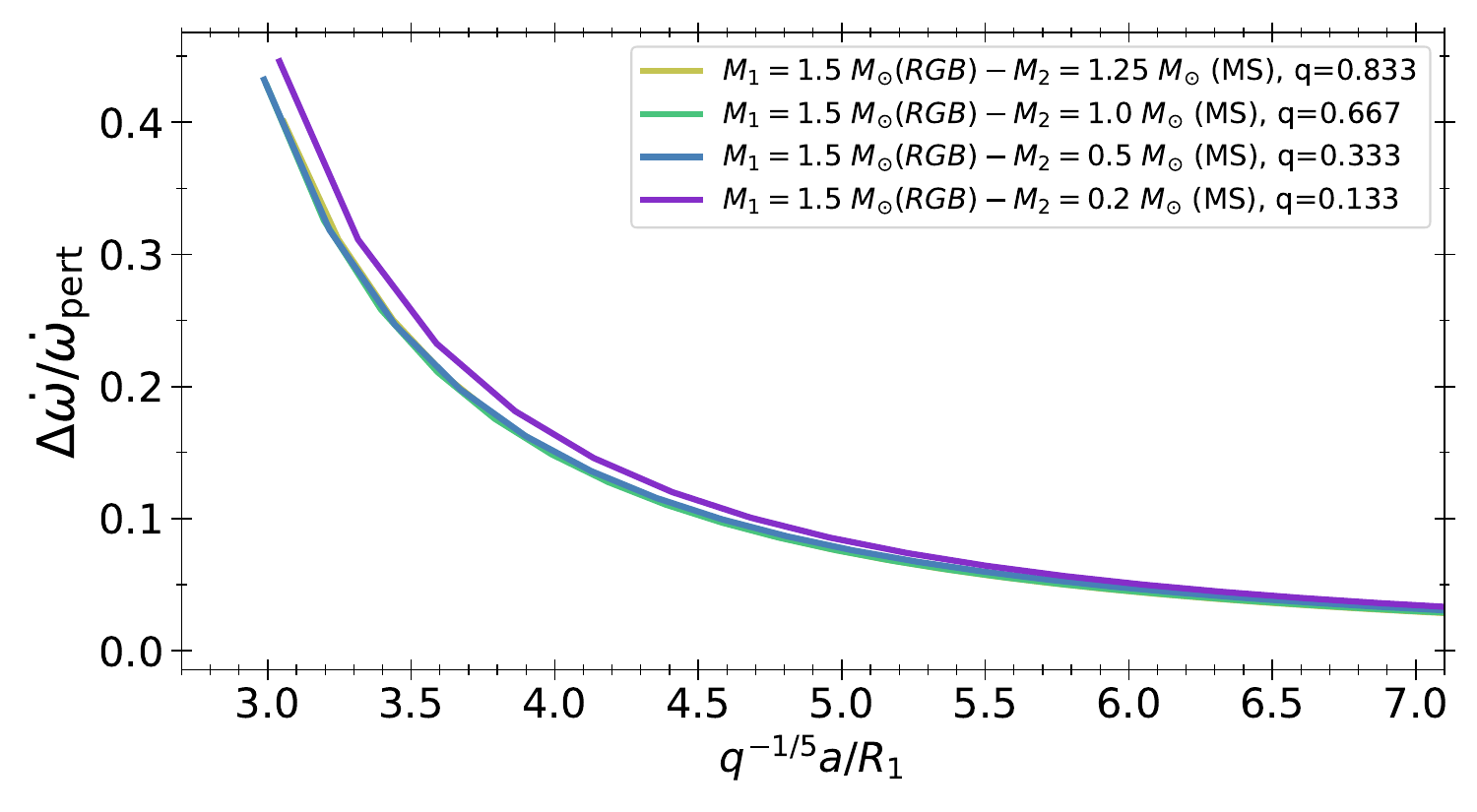}
\caption{Apsidal motion discrepancy as a function of the orbital separation scaled with the $q$ parameter.  Each system is composed of a primary $1.5 M_\odot$ RGB star presented in \autoref{table_stella_models},  and a secondary star with a lower mass. The stars composing the systems have the same age ($2$ Byr) and initial chemical composition ($X_0=0.72$, $Z_0=0.010$). Each colour corresponds to a secondary with a different mass. }
\label{fig.dependency_q_RGB}
\end{figure}

This regime describes the case of a system composed of either a star plus a compact object or an evolved star plus a MS star. To simulate this situation,  we chose to model the primary star as being the $1.5$ $M_\odot$ RGB star described in \autoref{table_stella_models}. The secondary is a MS star with a lower mass,  ranging between 0.2 and 1.25  $M_\odot$,  the same age ($2$ Byr) and initial chemical composition ($X_0=0.72$, $Z_0=0.010$).  
In \autoref{fig.dependency_q_RGB}, we show the evolution of the apsidal motion discrepancy for the different systems considered.  Based on \autoref{eq_apsidal_motion_perturbative}, we chose the $x$-axis of \autoref{fig.dependency_q_RGB} to be $q^{-1/5} R_1/a$ to maintain the scale between $(R_1/a)^5$ and $q$.
We can see that all the systems are superposed in \autoref{fig.dependency_q_RGB}  which indicates a clear proportionality of the apsidal motion discrepancy on $q$.  When $R_1^5\gg R_2^5$, as seen in \autoref{eq_apsidal_motion_perturbative}, the terms with $(R_1/a)^5$ are dominating for a given orbital separation, and the contribution from the secondary is negligible. In this situation, only the deformations of the primary impact the apsidal motion. As both the deformations and the apsidal motion are proportional to $q$ for a given orbital separation, the apsidal motion discrepancy is also proportional to $q$. 
\subsubsection{$R_1^5\simeq R_2^5$}
\begin{figure}[h]
\centering
\includegraphics[width=\hsize]{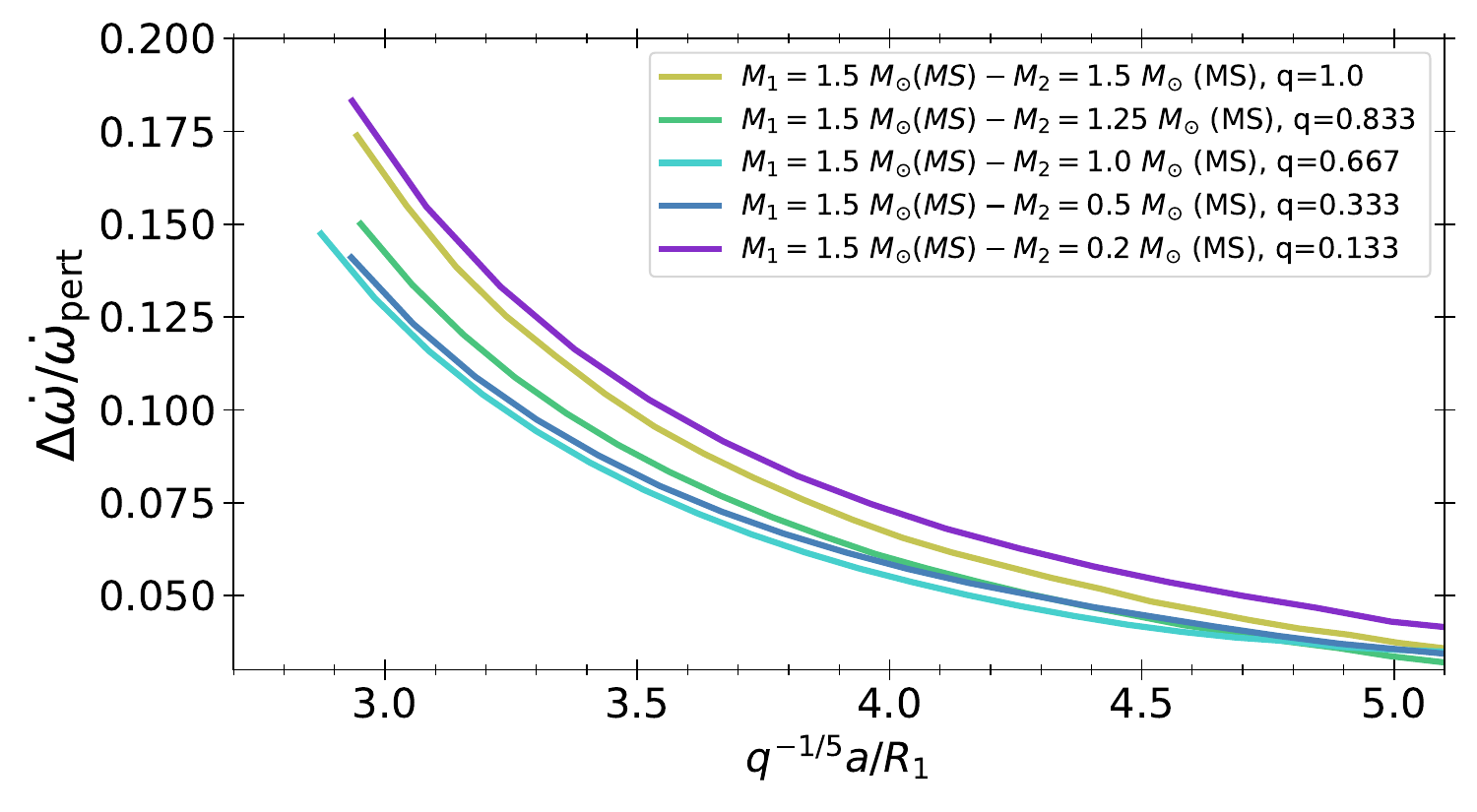}
\caption{Apsidal motion discrepancy as a function of the orbital separation scaled with the $q$ parameter.  Each system is composed of a primary $1.5 M_\odot$ MS star,  and a secondary star with a lower or equal mass. The stars composing the systems have the same age ($1.5$ Byr) and initial chemical composition ($X_0=0.72$, $Z_0=0.010$).  Each colour corresponds to a secondary with a different mass. }
\label{fig.dependency_q_MS}
\end{figure}

This regime describes systems composed of stars in similar evolutionary stages and structural properties.  To simulate this situation we combined a primary MS star with a mass of $1.5$ $M_\odot$, an age of $1.5$ Byr,  an initial chemical composition of $X_0=0.72$, $Z_0=0.010$, with a secondary star with the same properties but a lower or equal mass.  In \autoref{fig.dependency_q_MS}, we explore the apsidal motion discrepancy as a function of $q^{-1/5} R_1/a$ for these binary systems. 
In the case where the primary and secondary have similar radii despite different $q$, the properties of the secondary have an impact on the apsidal motion discrepancy, as illustrated by comparing Figs.  \ref{fig.dependency_q_RGB} and \ref{fig.dependency_q_MS}. The apsidal motion discrepancy is still dominated by the contribution from the primary, consequently a partial  proportionality to $q$ can be seen. However, additional contributions from the other parameters of the secondary, in particular its density profile, have to be accounted for. 
\\~\\In our determination of the limit after which the modelling discrepancy reaches $2\%$ and $5\%$ (see \autoref{subsect_orbital_ecc}), we assumed that $q=1$. In practice, observed systems often have $q<1$, and,  to include the dependency of these limits on $q$ we modified them. As for both cases  presented above the dependency on $q$ dominates the apsidal motion discrepancy,  we need that $q^{-1/5}a(1-e^2)/R_1\lesssim 4.5$ to have at least $5\%$ modelling discrepancy and $q^{-1/5}a(1-e^2)/R_1\lesssim 6.5$ to have at least $2\%$ modelling discrepancy. By adopting these modifications we introduced a little bias in the regime where $R_1^5\simeq R_2^5$ however, as showed before, the dependency on $q$ still dominates. 
\section{Application to observed systems}\label{sect_application_observations}
\subsection{Modelling of observed binaries and quantification of the extra mixing}\label{subsect_application_observations}
In this section,  we apply our non-perturbative modelling technique to existing twin binary systems, comparing the stellar parameters obtained with our approach to those from the perturbative theory.  As we study twin binary systems,  modelling one star is sufficient to describe the entire system. We integrated MoBiDICT to a minimisation method following a Levenberg-Marquardt algorithm to compute on the fly the non-perturbative apsidal motion resulting from twin binaries which unperturbed structure are given by our 1D single star stellar evolution code,  CLES.  By using this methodology we followed the procedure of \cite{Rosu2020, Rosu2022a, Rosu2022b, Rosu2023},  neglecting the interactions of the binaries during their evolution.  To avoid such assumptions, our modelling method would have to be directly coupled to binary stellar evolution codes, this work will be conducted in the future with the code BINSTAR \citep{Siess2013} and will be presented in a forthcoming article.  
\\~\\We selected four well-known close observed twin binary systems from the literature: PV Cas \citep{Torres2010, Claret2021, Marcussen2022}, IM Per  \citep{Lacy2015, Claret2021}, Y Cyg \citep{Gimenez1987, Harmanec2014, Claret2021, Marcussen2022}, and  HD152248 \citep{Rosu2020,Rosu2020b}.  The first binary system is composed of MS intermediate mass stars,  the second one of sub-giant stars,  and the last two ones of massive stars. The properties of these systems are given in the second and fifth line blocks of \autoref{table_results_stellar_modelling}. For each of these binary systems,  an apsidal motion was determined in the literature  and is used in this section as a constraint on the stellar structure.  The systems were selected to be twin systems with accurate determination of the orbital and stellar parameters,  and an apsidal motion primarily driven by tidal forces.  
\\~\\ For this modelling we were inspired by the findings of \cite{Rosu2020, Rosu2022a, Rosu2022b, Rosu2023},  who demonstrated that modelling twin binary systems using 1D stellar evolution codes with apsidal motion as a constraint necessitates significant additional mixing during stellar evolution.  In our modelling,  we used the same constraints for all observed systems:  the masses, radii, effective temperatures,  and apsidal motions.  The constraints for each system are listed in the second line block of \autoref{table_results_stellar_modelling}.   The objective of our modelling technique is to generate an evolutionary track such as the end model is fitting all the observed constraints of the system.  As we have four constraints, we used four free parameters for our stars, their initial mass $M_{\mathrm{0}}$,  initial hydrogen mass fraction $X_{\mathrm{0}}$,  and age, as well as the overshooting parameter $\alpha_{\mathrm{over}}$.  In our modelling,  we used the overshooting to control the additional mixing necessary to reproduce the observed constraints as a strong degeneracy is existing between the overshooting and the mixing (see e.g. \cite{Rosu2020,Rosu2022a}).  For all the models, we used the AGSS09 abundances \citep{Asplund2009}, the FreeEOS equation of state \citep{Irwin2012}, the OPAL opacities \citep{Iglesias1996}, the $T(\tau)$ relation from Model-C of \citet{Atmosphere1981} for the atmosphere,  the mixing length theory of convection implemented as in \citet[][]{Cox1968}, and the nuclear reaction rates of \cite{Reaction2011}. For the massive stars,  the mass loss rate was computed, accounting for the metallicity of the models,  with the prescription of \cite{Vink2001} assuming that the scaling factor for the mass loss rate equals $\xi=1$. The equivalent averaged mass loss along the entire evolution is given in the fourth line block of \autoref{table_results_stellar_modelling}.
\\~\\In our modelling procedure,  we started by finding models reproducing the observations with the perturbative approach.  We then fixed  the initial mass and hydrogen mass fraction,  only using the constraints on the radii and apsidal motion to model the systems with MoBiDICT.  This approach aims to directly evaluate the impact of our non-perturbative modelling on the required additional mixing and age.  The objective of our modelling is not to provide an accurate estimation of the stellar properties of each system but rather to compare the stellar properties obtained when modelling the apsidal motion with the perturbative and non-perturbative approaches.  The results from our modelling procedure are given both for the non-perturbative and perturbative modelling approaches in the third line block of \autoref{table_results_stellar_modelling}.
\begin{table*}[h]
\centering
\caption{Observed and model properties of the systems studied.  The first line block corresponds to the observational constraints used in our modelling, the second line block gives the values of the free parameters obtained,  the third line block gives the assumed parameters of the models, and the fourth, provides the assumed parameters of the system originating from observations.  All the models were compared using $\chi^2$ defined as the sum of the squared relatives differences between the observed constraints and the models.}\label{table_results_stellar_modelling}
\resizebox{\hsize}{!}{
\begin{tabular}{lllllllllllll}
\hline\hline
\multirow{2}{*}{Targets}    &                                     & \multicolumn{2}{c}{\multirow{2}{*}{PV Cas$^{1,2,3}$}}       &  & \multicolumn{2}{c}{\multirow{2}{*}{IM Per$^{2,4}$}}        &  & \multicolumn{2}{c}{\multirow{2}{*}{Y Cyg$^{5,2,3}$}}        &  & \multicolumn{2}{c}{\multirow{2}{*}{HD152248$^{6,7}$}}      \\
                            &                                     & \multicolumn{2}{c}{}                              &  & \multicolumn{2}{c}{}                               &  & \multicolumn{2}{c}{}                              &  & \multicolumn{2}{c}{}                               \\ \cline{3-4} \cline{6-7} \cline{9-10} \cline{12-13} 
                            &                                     & \multirow{2}{*}{Pert.} & \multirow{2}{*}{Non-Pert.} &  & \multirow{2}{*}{Pert.} & \multirow{2}{*}{Non-Pert.} &  & \multirow{2}{*}{Pert.} & \multirow{2}{*}{Non-Pert.} &  & \multirow{2}{*}{Pert.} & \multirow{2}{*}{Non-Pert.} \\
                            &                                     &                        &                          &  &                        &                           &  &                        &                          &  &                        &                           \\ \hline
                            &                                     &                        &                          &  &                        &                           &  &                        &                          &  &                        &                           \\
\multicolumn{1}{l|}{}       & M{[}$M_\odot${]}                    & \multicolumn{2}{c}{2.78(8)}                       &  & \multicolumn{2}{c}{1.78(1)}                        &  & \multicolumn{2}{c}{$17.72(30)$}                   &  & \multicolumn{2}{c}{$29.5(5)$}                      \\
\multicolumn{1}{l|}{Obs.}   & R{[}$R_\odot${]}                    & \multicolumn{2}{c}{2.28(4)}                       &  & \multicolumn{2}{c}{2.38(4)}                        &  & \multicolumn{2}{c}{$5.8(1)$}                      &  & \multicolumn{2}{c}{$15.07(12)$}                    \\
\multicolumn{1}{l|}{Const.} & $T_{\mathrm{eff}}${[}$K${]}         & \multicolumn{2}{c}{10200(250)}                    &  & \multicolumn{2}{c}{7570(160)}                      &  & \multicolumn{2}{c}{$33200(500)$}                  &  & \multicolumn{2}{c}{$34000(1000)$}                  \\
\multicolumn{1}{l|}{}       & $\dot{\omega}$ {[}$^\circ$cycle$^{-1}${]}      & \multicolumn{2}{c}{0.0212(2)}                     &  & \multicolumn{2}{c}{0.0146(4)}                      &  & \multicolumn{2}{c}{$0.06186(30)$}                 &  & \multicolumn{2}{c}{$0.0293(13)$}                   \\
                            &                                     &                        &                          &  &                        &                           &  &                        &                          &  &                        &                           \\ \hline
                            &                                     &                        &                          &  &                        &                           &  &                        &                          &  &                        &                           \\
\multicolumn{1}{l|}{}       & $M_0${[}$M_\odot${]}                & 2.780                   & 2.780                     &  & 1.78                   & 1.78                      &  & 20.4          & 20.4                 &  & 33.16                 & 33.16                      \\
\multicolumn{1}{l|}{Free}   & $X_0$                               &0.750                    & 0.750                       &  & 0.730  & 0.730                                        &  & 0.743 & 0.743                   &  & 0.726                & 0.726                   \\
\multicolumn{1}{l|}{Param.} & $\alpha_{\mathrm{over}}$            & 0.923                    & 0.951                     &  & 0.306                     & 0.333                &  & 1.01            & 1.05                   &  & 1.29                  & 1.28                   \\
\multicolumn{1}{l|}{}       & Age {[}Gyr{]}                       & 3.07                   & 3.09                    &  & 1.33& 1.35                                      &  & 0.00347 & 0.00353                  &  & 0.00521                  & 0.00520                    \\
                            &                                     &                        &                          &  &                        &                           &  &                        &                          &  &                        &                           \\ \hline
                            &                                     &                        &                          &  &                        &                           &  &                        &                          &  &                        &                           \\
\multicolumn{1}{l|}{Models}       & $Z_0$                               & \multicolumn{2}{c}{0.0245}                         &  & \multicolumn{2}{c}{0.012}                          &  & \multicolumn{2}{c}{0.012}                         &  & \multicolumn{2}{c}{0.017}                          \\
\multicolumn{1}{l|}{Fixed}       & $\alpha_{\mathrm{MLT}}$             & \multicolumn{2}{c}{1.95}                          &  & \multicolumn{2}{c}{1.95}                           &  & \multicolumn{2}{c}{1.80}                          &  & \multicolumn{2}{c}{1.80}                           \\
\multicolumn{1}{l|}{Param.}      & $\dot{M} ${[}$M_\odot$yr$^{-1}${]} & \multicolumn{2}{c}{0}                             &  & \multicolumn{2}{c}{0}                              &  & \multicolumn{2}{c}{0.8$\cdot 10^{-6}$}                           &  & \multicolumn{2}{c}{1.18$\cdot 10^{-6}$}                           \\
                            &                                     &                        &                          &  &                        &                           &  &                        &                          &  &                        &                           \\ \hline
                            &                                     &                        &                          &  &                        &                           &  &                        &                          &  &                        &                           \\
\multicolumn{1}{l|}{} & $P_{\mathrm{orb}}$ {[}days{]}                    & \multicolumn{2}{c}{1.75047}                       &  & \multicolumn{2}{c}{2.25422}                        &  & \multicolumn{2}{c}{2.996321}                      &  & \multicolumn{2}{c}{5.816498}                       \\
\multicolumn{1}{l|}{Obs.}       & $i$ {[}$^\circ${]}                       & \multicolumn{2}{c}{85.8}                          &  & \multicolumn{2}{c}{84.4}                           &  & \multicolumn{2}{c}{86.47}                         &  & \multicolumn{2}{c}{67.6}                           \\
\multicolumn{1}{l|}{Param.}       & $e$                                 & \multicolumn{2}{c}{0.0325}                        &  & \multicolumn{2}{c}{0.0491}                         &  & \multicolumn{2}{c}{0.145}                         &  & \multicolumn{2}{c}{0.13}                           \\
\multicolumn{1}{l|}{}       & $P_\star$/$P_{\mathrm{orb}}$                     & \multicolumn{2}{c}{0.998}                         &  & \multicolumn{2}{c}{0.933}                          &  & \multicolumn{2}{c}{0.688}                         &  & \multicolumn{2}{c}{0.88}                           \\
                            &                                     &                        &                          &  &                        &                           &  &                        &                          &  &                        &                           \\ \hline
\multirow{2}{*}{$\chi^2 $}  &                                     & \multirow{2}{*}{0.11}    & \multirow{2}{*}{0.18}      &  & \multirow{2}{*}{0.11}    & \multirow{2}{*}{0.20}      &  & \multirow{2}{*}{0.04}  & \multirow{2}{*}{0.23}    &  & \multirow{2}{*}{0.05}  & \multirow{2}{*}{0.03}      \\
                            &                                     &                        &                          &  &                        &                           &  &                        &                          &  &                        &                           \\ \hline
\end{tabular}}
\begin{flushleft}
\footnotesize{\textbf{References.} (1) \cite{Torres2010};  (2) \cite{Claret2021}; (3) \cite{Marcussen2022}; (4) \cite{Lacy2015}; (5) \cite{Harmanec2014};  (6) \cite{Rosu2020}; (7) \cite{Rosu2020b}.}
\end{flushleft}
\end{table*}
\\~\\For the modelling of each system, the parameters presented in the fourth line block of  \autoref{table_results_stellar_modelling} were empirically selected to facilitate the convergence of our Levenberg-Marquardt method.  For the first system,  $Z_0$ was chosen from a model grid to minimise the $\chi^2$ and be able to reproduce the observations as we could not achieve a coherent fit with a solar metallicity.  For the other systems, a solar metallicity was adopted as no constraints were given from the observations.  
\\\autoref{table_results_stellar_modelling} shows that both the perturbative and non-perturbative approaches can accurately reproduce all observed system constraints.  For the four systems, we find that extremely large extra-mixing  is required to reproduce the observational properties of stars.  Comparing the stellar parameters obtained with MoBiDICT and the perturbative approach,  our non-perturbative technique necessitates even more extra mixing for the first three systems.  While for the three first systems,  the leading contribution to the $\chi^2$ is the reproduction of the apsidal motion and the radius of the stars, for HD152248,  the major difficulty was to reproduce the observed effective temperature.  This can partly explain the smaller impact of the non-perturbative treatment on the overshooting parameter found in this system.  Nonetheless,  our approach does not reduce the extremely high extra-mixing required to reproduce the apsidal motion motion inferred by \cite{Rosu2020} but rather exacerbate it.  This additional mixing could stem from various sources, including rotation or dynamical tides.  We do not think that it should be concluded from our study that such large overshooting is present in these stars. Other physical processes neglected in our study such as dynamical tides are known to significantly affect the apsidal motion \citep{Willems2002}. 


\subsection{Apsidal motion  contribution from the different sources}\label{subsect_low_order_spherical_apsides}
\begin{figure}[h]
\centering
\includegraphics[width=\hsize]{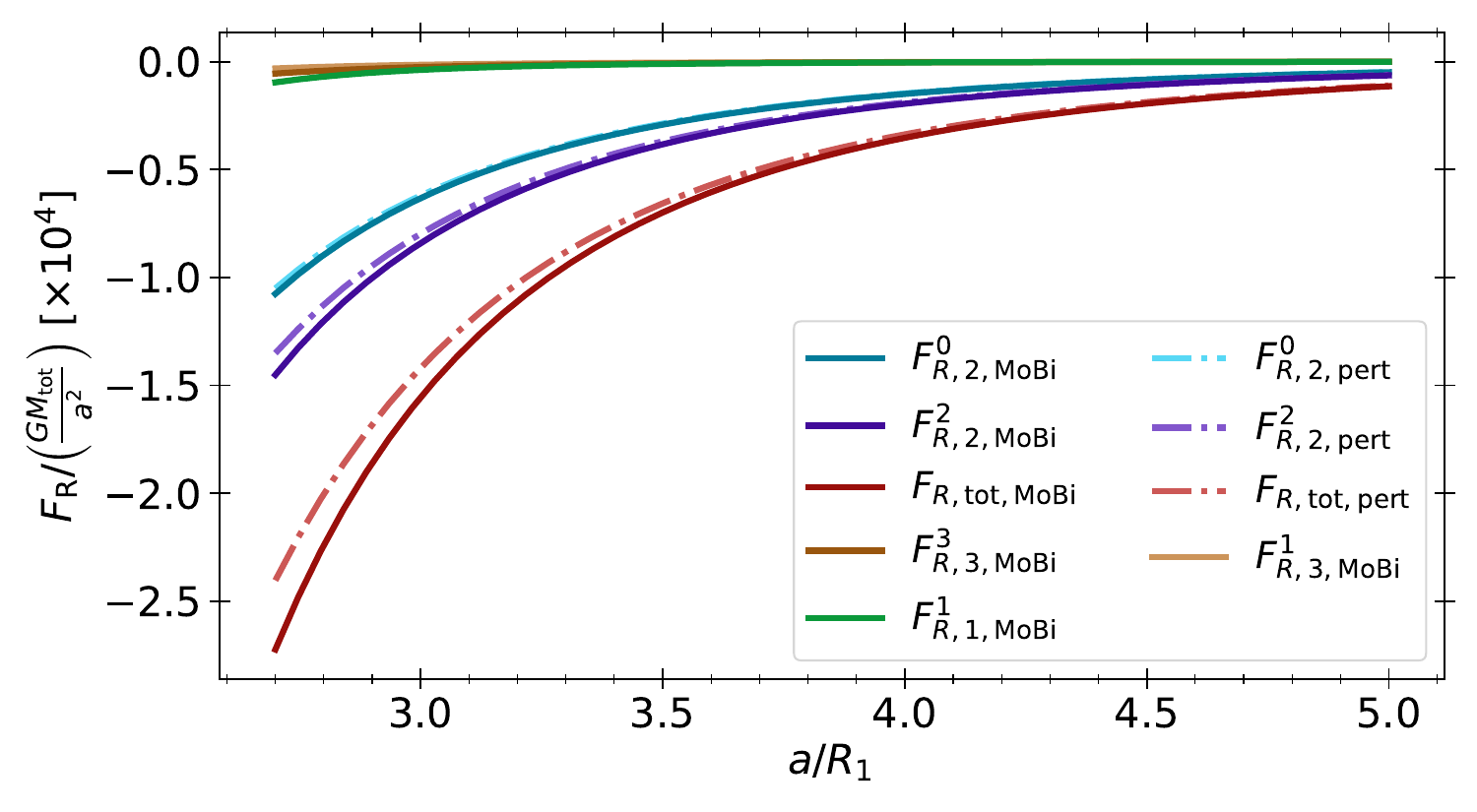}
\caption{Decomposition of the perturbed acceleration for the IM-Per system with the perturbative and non-perturbative models as a function of the system orbital separation normalised by the stellar radii.   Each colour represents one spherical order: the lighter curves correspond to the perturbative model while the darker curves correspond to the with non-perturbative approach. The red curves are the total of all the spherical orders. }
\label{fig.decomposition_R_IM_Per}
\end{figure}

In this section,  we focus on exploring the origin of the apsidal motion discrepancy seen for the observed targets presented in \autoref{sect_application_observations}.  We reused the structures from the modelling done in \autoref{table_results_stellar_modelling} and looked at the different contributions to the apsidal motion for each  binary system.   \autoref{table_apside_low_order_decomposition} provides a decomposition of the apsidal motion for each observed binary system using the models obtained with our non-perturbative method.  In the right part of \autoref{table_apside_low_order_decomposition}, we also decomposed the difference between the perturbative and non-perturbative approaches based on the considered spherical orders,  revealing the key contributors to the apsidal motion discrepancies.
\begin{table*}[h] \centering
\caption{Decomposition of the apsidal motion of the modelled observed systems.  The relativistic component of the apsidal motion is noted $\dot{\omega}_{\mathrm{rel.}}$,  $\dot{\omega}_{\mathrm{pert.}}$ is the contribution from the pertubative approach and $\dot{\omega}_{\mathrm{non-pert.}}$ is the correction induced by the non-perturbative approach.  The latest contribution in then decomposed as spherical harmonics in the right region of the table noted $\dot{\omega}_{\mathrm{non-pert.}}$ contributions.  }\label{table_apside_low_order_decomposition}
\resizebox{\hsize}{!}{
\begin{tabular}{lllllllllllll}
\hline\hline
                         &                                              &                                        &                                                                                                                              &                                                                                                                               &                                                                                                                                &                                                                                                                                    &  & \multicolumn{5}{c}{\multirow{2}{*}{$\dot{\omega}_{N,  \mathrm{non-pert.}}$ contributions {[}$\%${]}}}                                                          \\
                         &                                              &                                        &                                                                                                                              &                                                                                                                               &                                                                                                                                &                                                                                                                                    &  & \multicolumn{5}{c}{}                                                                                                                                       \\ \cline{9-13} 
\multirow{2}{*}{Targets} & \multicolumn{1}{c}{\multirow{2}{*}{$a$/$R_1$}} & \multicolumn{1}{c}{\multirow{2}{*}{$e$}} & \multicolumn{1}{c}{\multirow{2}{*}{\begin{tabular}[c]{@{}c@{}}$\dot{\omega}_{\mathrm{obs}}$\\ {[}$^\circ$cycle$^{-1}${]}\end{tabular}}} & \multicolumn{1}{c}{\multirow{2}{*}{\begin{tabular}[c]{@{}c@{}}$\dot{\omega}_{\mathrm{rel.}}$\\ {[}$^\circ$cycle$^{-1}${]}\end{tabular}}} & \multicolumn{1}{c}{\multirow{2}{*}{\begin{tabular}[c]{@{}c@{}}$\dot{\omega}_{N,\mathrm{pert.}}$\\ {[}$^\circ$cycle$^{-1}${]}\end{tabular}}} & \multicolumn{1}{c}{\multirow{2}{*}{\begin{tabular}[c]{@{}c@{}}$\dot{\omega}_{N,\mathrm{non-pert.}}$\\ {[}$^\circ$cycle$^{-1}${]}\end{tabular}}} &  & \multirow{2}{*}{$\ell=1$} & \multirow{2}{*}{$\ell=2$} & \multirow{2}{*}{$\ell=3$} & \multirow{2}{*}{$\ell>3$} & \multirow{2}{*}{$F^{\prime}\rho^{\prime}$} \\
                         & \multicolumn{1}{c}{}                         & \multicolumn{1}{c}{}                   & \multicolumn{1}{c}{}                                                                                                         & \multicolumn{1}{c}{}                                                                                                          & \multicolumn{1}{c}{}                                                                                                           & \multicolumn{1}{c}{}                                                                                                               &  &                           &                           &                           &                           &                                            \\ \cline{1-7} \cline{9-13} 
                         &                                              &                                        &                                                                                                                              &                                                                                                                               &                                                                                                                                &                                                                                                                                    &  &                           &                           &                           &                           &                                            \\
PV Cas$^{1,2,3}$                  &              4.80                                & 0.0325                                 & 0.0212(2)                                                                                                                    &              0.0012 &                0.0192            &  0.0008 &   &               23         &          38                 &         38                  &             2              &                $\sim0$                              \\
IM Per$^{2,4}$                   &       4.60                                       & 0.0491                                 & 0.0146(4)                                                                                                                    &       0.0007                                                                                                                        &                                                                                                                               0.0131 &           0.0008                                                                                                                         &  &                          25 &     37                      &     37                      &     2                      &                    $\sim0$                        \\
Y Cyg$^{5,2,3}$                   &                     4.95                         & 0.145                                  & 0.0618(3)                                                                                                                   &                                                   0.0029                                                                            &0.0565 &                                                                                                                                   0.0025 &  &        22                   &                          45 &            32              &           2                &                 $\sim0$  \\
HD152248$^{6,7}$                 &                  3.47                            & 0.13                                   & 0.0293(13)                                                                                                                   &          0.0026                                                                                                                     &                                                                                                                               0.0250 &                                0.0018                                                                                                    &  &                          23 &                      42     &       32                    &         2                  &     $\sim0$                                        \\
                         &                                              &                                        &                                                                                                                              &                                                                                                                               &                                                                                                                                &                                                                                                                                    &  &                           &                           &                           &                           &                                            \\ \hline
\end{tabular}}
\begin{flushleft}
\footnotesize{\textbf{References.} (1) \cite{Torres2010};  (2) \cite{Claret2021}; (3) \cite{Marcussen2022}; (4) \cite{Lacy2015}; (5) \cite{Harmanec2014};  (6) \cite{Rosu2020}; (7) \cite{Rosu2020b}.}
\end{flushleft}
\end{table*}
The analysis of  \autoref{table_apside_low_order_decomposition}  reveals that, for all chosen targets, the contributions from our non-perturbative modelling approach surpass the uncertainties in the observational apsidal motion determinations.  Generally, the corrections provided by MoBiDICT are comparable in magnitude to the general relativistic corrections. Then, going into more details,  the perturbed acceleration discrepancy between the non-perturbative and perturbative methods can be attributed to several main components:
\begin{itemize}
\item $\ell=1$: Constituting $\sim23\%$ of the modelling discrepancy,  this component cannot be recovered by the perturbative model  due to its inherent formalism and assumptions;
\item $\ell=2$:  Representing  $\sim40\%$ of the modelling discrepancy,  this component cannot either be recovered by the perturbative approach.  The differences arise from the first-order assumption of the perturbative approach.  We tried taking into account the effect from the quadrupolar deformation of the primary star on the quadrupolar component of its companion and the reverse impact of this modification on the quadrupolar deformation of the primary star.  However this effect is negligible as each of such additional effect considered has an increase  dependency to $(R/a)^3$ of a power three; 
\item $\ell=3$: Accounting for $\sim35\%$ of the modelling discrepancy,  this component can be at least partially recovered by the perturbative model by going at higher spherical orders than usually considered.  In \autoref{subsect_higher_order_spherical_apsides}, we expand the computation of the pertubative apsidal motion to the $\ell=3$ assessing the extent to which these discrepancies can be rectified;
\item $\ell>3$: The contribution of these components only represents $2\%$ of the discrepancies. Our analysis suggests that computing spherical orders higher than three is unnecessary, in particular when comparing these discrepancies with the contributions from other spherical orders;
\item $F^{\prime}\rho^{\prime}$: This term corresponds to the assumption made to neglect the third contribution to the apsidal motion in \autoref{eq_R_base_pert}. With our modelling method we see that this assumption is totally valid even in the most distorted cases.
\end{itemize}
For the case of IM Per,  we examine the individual contributions to this discrepancy. \autoref{fig.decomposition_R_IM_Per} illustrates the evolution of the components of the perturbed acceleration with orbital separation normalised by the stellar radii  for the perturbative and non-perturbative models.
For the $\ell=2$ terms, \autoref{fig.decomposition_R_IM_Per} shows that most of the discrepancies are originating from the $\ell=2, m=2$ term. This  aspect is explained by the fact that in the $\ell=2, m=0$ term both tidal force and centrifugal force are included while in $\ell=2, m=2$ only tidal force in present.  This indicates that the perturbative approach lacks precision when modelling tidal forces at low orbital separations, corroborating findings from \autoref{subsect_apidal_motion_theoritical_model} and  \autoref{fig.diff_pot_surface_all_models}.

\subsection{Higher order perturbative method}\label{subsect_higher_order_spherical_apsides}
An advantage of the perturbative approach lies in the possibility to extend the modelling to higher spherical orders.  In the previous section we showed that the $\ell=3$ component of the acceleration perturbation represents $\sim35\%$ of the discrepancies seen between our non-perturbative and the perturbative models.  In \autoref{annexe_l=3_perturbation}, we developed the equations necessary  to extend the perturbative approach to the $\ell=3$ term. In this section, we verify the relevance  
of adding the $\ell=3$ term in the computation of the apsidal motion in the perturbative approach.  As shown in \autoref{annexe_l=3_perturbation}, the apsidal motion arising from the $\ell=3$ component of the perturbed acceleration is given by 
\begin{equation}
\frac{\mathrm{d}\omega}{\mathrm{d}t}=\frac{56\pi}{P_{\mathrm{orb}}} v(e)  \left( k_{2,3}\frac{1}{q}\left(\dfrac{R_{\textrm{2}}}{a} \right)^{7} +  k_{1,3}q\left(\dfrac{R_{\textrm{1}}}{a} \right)^{7}\right),
\end{equation}
with
\begin{equation}
v(e)=\dfrac{1+\dfrac{15}{4}e^2+\dfrac{15}{8}e^4+\dfrac{5}{64}e^6}{(1-e^2)^7}.
\end{equation}
We took back the model of observed stars extensively presented in \autoref{subsect_application_observations}  and computed the apsidal motion using our method and the perturbative approach including the $\ell=3$ term.  The results from this modelling are presented in \autoref{table_apside_high_order_decomposition}.
\begin{table*}[h] \centering
\caption{Decomposition of the apsidal motion of the modelled observed systems including the $\ell=3$ term of the perturbative approach.  The relativistic component of the apsidal motion is noted $\dot{\omega}_{\mathrm{rel.}}$,  $\dot{\omega}_{\mathrm{pert.}}$ is the contribution from the pertubative approach and $\dot{\omega}_{\mathrm{non-pert.}}$ is the correction induced by the non-perturbative approach.  The latest contribution in then decomposed as spherical harmonics in the right region of the table noted $\dot{\omega}_{\mathrm{non-pert.}}$ contributions.  }\label{table_apside_high_order_decomposition}
\resizebox{17cm}{!}{
\begin{tabular}{llllllllll}
\hline\hline
                         &                                              &                                        &                                                                                                                              &                                                                                                                               &                                                                                                                                &                                                                                                                                     \multicolumn{4}{c}{\multirow{2}{*}{$\dot{\omega}_{N, \mathrm{non-pert.}}$ contributions {[}$\%${]}}}                                                          \\
                         &                                                                                                                                                                                                                    &                                                                                                                               &                                                                                                                                                                                                                                                                  &  & \multicolumn{5}{c}{}                                                                                                                                       \\ \cline{7-10} 
\multirow{2}{*}{Targets} &  \multicolumn{1}{c}{\multirow{2}{*}{\begin{tabular}[c]{@{}c@{}}$\dot{\omega}_{\mathrm{obs}}$\\ {[}$^\circ$cycle$^{-1}${]}\end{tabular}}} & \multicolumn{1}{c}{\multirow{2}{*}{\begin{tabular}[c]{@{}c@{}}$\dot{\omega}_{\mathrm{rel.}}$\\ {[}$^\circ$cycle$^{-1}${]}\end{tabular}}} & \multicolumn{1}{c}{\multirow{2}{*}{\begin{tabular}[c]{@{}c@{}}$\dot{\omega}_{N, \mathrm{pert.}}$\\ {[}$^\circ$cycle$^{-1}${]}\end{tabular}}} & \multicolumn{1}{c}{\multirow{2}{*}{\begin{tabular}[c]{@{}c@{}}$\dot{\omega}_{N, \mathrm{non-pert.}}$\\ {[}$^\circ$cycle$^{-1}${]}\end{tabular}}} &  & \multirow{2}{*}{$\ell=1$} & \multirow{2}{*}{$\ell=2$} & \multirow{2}{*}{$\ell=3$} & \multirow{2}{*}{$\ell>3$}  \\
                         & \multicolumn{1}{c}{}                         & \multicolumn{1}{c}{}                   & \multicolumn{1}{c}{}                                                                                                         & \multicolumn{1}{c}{}                                                                                                          & \multicolumn{1}{c}{}                                                                                                           & \multicolumn{1}{c}{}                                                                                                               &  &                                                                                 &                                                                       \\ \cline{1-5} \cline{7-10} 
                         &                                              &                                        &                                                                                                                              &                                                                                                                               &                                                                                                                                &                                                                                                                                    &  &                           &                                                                                                                             \\
PV Cas$^{1,2,3}$                                                          & 0.0212(2)                                                                                                                    &              0.0012 &                0.0194            &  0.0006 &   &               31        &          53                 &         14                  &             2               \\
IM Per$^{2,4}$                                              & 0.0146(4)                                                                                                                    &       0.0007                                                                                                                        &                                                                                                                               0.0133 &           0.0005                                                                                                                         &  &                         33 &     50                      &     14                    &     3                                                              \\
Y Cyg$^{5,2,3}$                                                    & 0.0618(3)                                                                                                                   &                                                   0.0029                                                                            &0.0570 &                                                                                                                                   0.0020 &  &        29                  &                          59 &            11              &           2                 \\
HD152248$^{6,7}$                                               & 0.0293(13)                                                                                                                   &          0.0026                                                                                                                     &                                                                                                                               0.0255 &                                0.0013                                                                                                    &  &                          30 &                      55     &       12                    &         3                                                         \\
                         &                                              &                                        &                                                                                                                              &                                                                                                                               &                                                                                                                                &                                                                                                                                    &  &                           &                                                                                                                           \\ \hline
\end{tabular}}
\begin{flushleft}
\footnotesize{\textbf{References.} (1) \cite{Torres2010};  (2) \cite{Claret2021}; (3) \cite{Marcussen2022}; (4) \cite{Lacy2015}; (5) \cite{Harmanec2014};  (6) \cite{Rosu2020}; (7) \cite{Rosu2020b}.}
\end{flushleft}
\end{table*}
By comparing Tables \ref{table_apside_low_order_decomposition} and \ref{table_apside_high_order_decomposition} we conclude that by including the $\ell=3$ component of the apsidal motion we corrected $\sim 20\%$ of the discrepancies seen in \autoref{subsect_low_order_spherical_apsides}.  However, the $\ell=3$ term is only partially corrected when included in the perturbative approach. Indeed, we found that for all the systems modelled this term stills represent $\sim 10\%$ of the observed model discrepancies. Due to the reduction of the contribution from the $\ell=3$ term, the quadrupolar term now represents $\sim 50 \%$ of the observed discrepancies while the dipolar term contributes to $\sim 30\%$ of the differences.  The exact repartition of these contributions can vary depending on the orbital separation and the stellar type, in particular, we saw a highly non-linear comportment of the dipolar term. Nonetheless, in the future, the $\ell=3$ contribution to the apsidal motion have to be included when modelling the apsidal motion of close binaries with the perturbative approach.

\section{Discussion}\label{sec_discussion_population} 
\begin{figure}[h]
\centering
\includegraphics[width=\hsize]{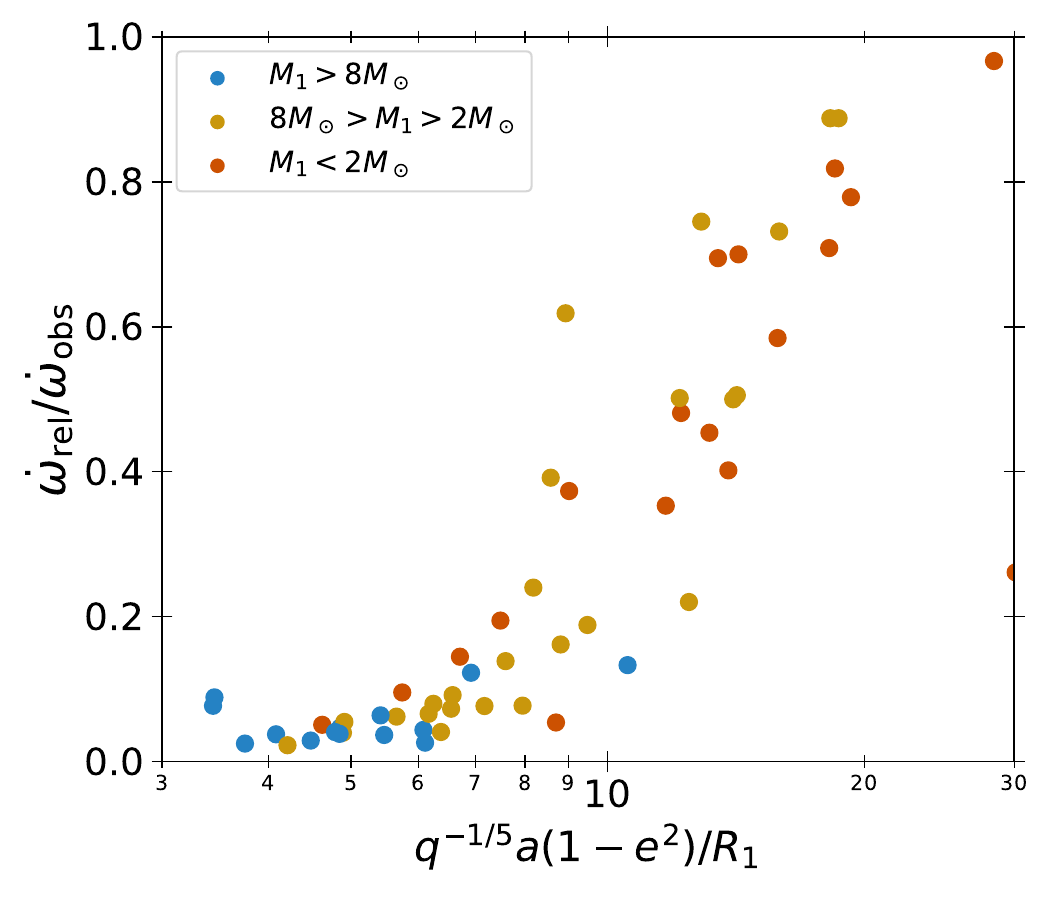}
\caption{Repartition of observed binary systems from our catalogue affected by our methodology.  Each point correspond to an observed system with a given orbital separation and a relativistic contribution to the apsidal motion.  The blue dots correspond to  massive binaries, the yellow points are intermediate mass binary stars,  and the red dots are low-mass stars.  For this diagram, the primary was chosen as the most massive star in the system.}
\label{fig._distribution binaries}
\end{figure}

In the previous sections, we showed that non-pertubative modelling is necessary for a precise interpretation of observations in close binary systems.  Aside from the orbital evolution considerations,  the detection of tight systems with apsidal motion is favoured due to the $(R/a)^5$ dependency of the tidal component.  The combination of high-precision observations and strong tidal effects on apsidal motion motivates their investigation as stellar laboratories and, consequently, the need to model them comprehensively.  Our modelling has conclusively shown that, for close binary systems, a non-perturbative treatment of deformations is necessary for accurately characterising stellar global properties and stellar structure.
\\~\\To assess the impact of our modelling method on observed binary systems with apsidal motion, we constructed a comprehensive catalogue of such systems. In total, our catalogue includes $61$ systems,  encompassing all systems with constraints on their stellar parameters and observed apsidal motion found in the literature. These systems were obtained by combining two recent catalogues of observed binary systems with apsidal motion \citep{Claret2021,  Marcussen2022} to systems from various literature sources where stellar parameter determinations were available \citep{Gimenez1987, Benvenuto2002,Wolf2006,Wolf2008,Wolf2010,Torres2010, Pablo2015, Rosu2020b, Baroch2021, Baroch2022,Rosu2022a, Rosu2022b, Rosu2023}.  With this catalogue,   \autoref{fig._distribution binaries} illustrates the fraction of observed apsidal motion arising from relativistic corrections against the parameter $q^{-1/5}a(1-e^2)/R_1$ of each observed system.
\\~\\The observed binary systems in our catalogue can be broadly categorised into two equal-sized groups: close binaries with most of the apsidal motion driven by tidal effects and wider binaries where apsidal motion is mostly due to relativistic corrections. The boundary between these two groups is located around $q^{-1/5}a(1-e^2)/R_1=8-9$.  Because of their low tidal contribution to the apsidal motion,  the group of systems with high relativistic contribution  is not favoured when wanting to constrain the stellar structure with apsidal motion.  For such systems,  as shown in \autoref{sect_comparaison_theoritical_models}, no  significant correction is expected from our non-perturbative method.  On the other hand,  for binary systems where apsidal motion is dominated by tidal effects (close systems),  which is the group preferred for drawing constraints on stellar structure with apsidal motion, our non-perturbative models show discrepancies compared to perturbative models.  In \autoref{subsect_theoritical_dependancy_k2} we showed that,  for a given orbital separation,  intermediate-high mass stars are the least impacted by our non-perturbative method,  independently from the $k_2$ value.  In \autoref{fig.apsidal_motion_diff_ecc} we also showed that for a theoretical massive star the apsidal motion relative difference reaches $5\%$ when $q^{-1/5}a(1-e^2)/R_1\lesssim 4.5$ and $2\%$ when $q^{-1/5}a(1-e^2)/R_1\lesssim 6.5$.  Therefore, we can estimate that in general when $q^{-1/5}a(1-e^2)/R_1\lesssim 4.5$ the relative difference of tidally induced apsidal motion reaches at least $5\%$ which corresponds to about $10\%$ of the stars in the catalogue presented in \autoref{fig._distribution binaries}.  Similarly,  for $42\%$ of the catalogue we expect at least a $2\%$ apsidal motion correction from our methodology,  corresponding to almost all the systems with the apsidal motion highly dominated by the tidal component.  Among the binaries impacted by our methodology,  a majority are high-mass or intermediate-mass binaries. In these catalogues and in general,  low-mass binaries are under-represented, and not commonly observed with apsidal motion probably due to observational biases.
\\~\\The discrepancies observed in the apsidal motion are a consequence of the underlying tidal force discrepancy.  However,  various aspects and phenomena of binary systems' evolutions are impacted by an underestimation of the tidal force.  One key quantity affected by this underestimation is tidal dissipation,  which plays a crucial role in the evolution of binary systems through orbital circularisation, transfer of angular momentum between bodies,  and orbital migrations.  Current models of the tidal dissipation rely on the perturbative approach \citep{Hut1981} combined to a free parameter that can be adjusted to control the dissipation.  Even if this parameter can be tuned to reproduce the dissipation obtained by our non-pertrubative models with enhanced tidal interactions at a given point of the evolution, it cannot capture the non-linearities that arise during the evolution of binary systems.  Consequently, our methodology predicts that tidal dissipation models used by the majority of binary stellar evolution codes underestimate the dissipation at low orbital separation.  This leads to imprecise predictions regarding the evolution of global orbital parameters, stellar rotation, and the overall outcomes of these models for systems with close orbits.  One specific area impacted by these findings is population synthesis, where the modelled population strongly depends on stellar interactions.  Therefore,  population models focused on the outcomes of close massive binary systems and the remnants of these systems face significant uncertainties, even before considering mass exchange and processes that rely on imprecise models with free parameters. These findings highlight the need for improved models that better account for the complex interplay of tidal forces in close binary systems.
\\~\\The discrepancies revealed by our modelling have wide-ranging implications that extend beyond the effects on global stellar and orbital parameters,  given that the entire structures of these distorted bodies are affected.  The enhanced stellar interaction obtained with our modelling drastically increases the stellar  surface deformations, as shown in \cite{Fellay2023}.  Furthermore, the internal structure of stars, which is typically modelled using the perturbative approach, is deeply influenced by our non-perturbative methodology.  To precisely capture the interior deformations, mass redistributions, and structural properties of deformed stars, a transition to 3D modelling becomes imperative.  Having highly accurate structural models is crucial for the precise study and modelling of the pulsation properties of deformed celestial bodies.  Among the oscillations propagating in deformed stars,  gravito-inertial waves, called dynamical tides,  significantly contribute to the orbital evolution of binary systems. However, the current dynamical tides models rely on structural modelling techniques based on the perturbative approach combined with 1D non-adiabatic spherically symmetric oscillations codes to treat this highly sensitive and non-linear problem.  As such, the dynamical tide models, as obtained in \cite{Ma2023}, for example,  are only estimations of the impact of dynamical tides on close binary systems. Without having appropriate 3D non-perturbative deformed structures and the associated 3D non-perturbative non-adiabatic oscillation codes this problem cannot be precisely solved.
\\~\\Our methodology,  similar to the perturbative approach, does have a notable limitation: it assumes solid body rotation within stars.  For single stars, this assumption is not fully justified because they often exhibit strong differential rotation, which varies over the course of stellar evolution.  However,  as we showed in \autoref{subsect_application_observations},  interactions between binary components induce significant extra mixing within deformed stars.  This additional mixing reduces the differential rotation in interacting bodies,  satisfying our assumption.  The subject of angular momentum transport processes in binary systems is a complex and extensive area of research that warrants further investigation.  MoBiDICT represents an important initial step in this direction. It allows for the first precise 3D computation of stellar interactions and the resulting forces at each point within the deformed stellar structure.
\\~\\Finally,  the modelling of mass transfer and phases of common envelopes in binary systems is still associated with significant uncertainties and often relies on the use of multiple free parameters.  In theory, MoBiDICT has the potential to be adapted for the static modelling of common envelope phases or, at the very least, to provide initial 3D deformed structures that can be used in conjunction with more sophisticated codes to simulate common envelope evolution.  Furthermore, our 3D non-perturbative models can be valuable in modelling mass transfer processes in binaries or providing initial 3D deformed structures for further studies.  In the near future, it would be useful to investigate the oscilation properties of binaries before they fill their Roche lobes as determining their configurations and properties before these phases play a crucial role in determining their ultimate fate.

\section{Conclusion}\label{sect_conclusion}
In this work, we investigated the impact of non-perturbative modelling on the theoretical computation of the apsidal motion and tidal force in eccentric binaries.  We developed formalisms for apsidal motion in our non-perturbative approach and the perturbative approach in respectively \autoref{sect_MoBIDICT} and \autoref{sect_Theoritical_apsides_pert}.  Then,  we compared the perturbed acceleration and apsidal motion for theoretical models, in \autoref{sect_comparaison_theoritical_models}, and observed binary systems, in \autoref{sect_application_observations}. Finally, we discussed the implications of our results, in \autoref{sec_discussion_population}, on the observed population of binaries and the general problem of binary modelling.
\\~\\All theoretical stellar models were significantly impacted at low orbital separation by our methodology.  For instance, we observed a maximum increase in tidal force of about $40 \%$ for our low-mass stars, $28\%$ for the RGB, $20\%$ for the solar types, and $12\%$ for massive stars. Consequently, the apsidal motion of these systems is impacted.  For an orbital eccentricity of $e=0.1$,  these discrepancies led to a respectively  increase in apsidal motion of $70\%$,  $45\%$,  $30\%$,  and $15\%$ for the low-mass, RGB,  solar type,  and massive twin binary systems.  The discrepancies  seen may vary depending on the exact structure of the stars or the architecture and the orbital parameters of the system. These results are consistent with \cite{Fellay2023}: the models with higher envelope mass experience greater impacts from our methodology.  From this study we noted that models with higher $k_2$ are more impacted by our methodology.  We attempted to establish an empirical relationship between tidal force discrepancy, the orbital separation, and $k_2$.  However, due to high non-linearities and the numerous free parameters of the problem we were not able to identify such a relationship.  We also explored the dependency of the apsidal motion modelling discrepancy on $q$ and found that this difference is proportional to $q$ except when stars have similar radii. In the latter case, the properties of the secondary are modifying  this relationship and the exact properties of the system have to be accounted for.
\\~\\In our analysis of observed systems,  we explored in more details the origins of these discrepancies.  Specifically,  we modelled four observed twin binary systems,  PV Cas,  IM Per,  Y Cyg,  and  HD152248, composed of respectively, intermediate mass, sub-giants, and massive stars for the two latests.  We started by modelling these systems using the pertubative model to reproduce the observed values of the constraints parameters.  When best fitting models were found,  we started from this solution to model stars using our non-perturbative method to obtain the theoretical apsidal motion.  In agreement with the results of \cite{Rosu2020, Rosu2022a, Rosu2022b, Rosu2023},  our results highlighted the necessity for significant extra mixing in the 1D stellar models to accurately reproduce the observational properties of these systems.  Moreover,  our methodology induced an additional increase of extra mixing due to the modification of apsidal motion,  reinforcing the results of \cite{Rosu2020, Rosu2022a, Rosu2022b, Rosu2023}.  
\\~\\For the systems we examined,  the discrepancies in apsidal motion exceeded the observed uncertainties, indicating their non-negligible impact. Our investigation into the origins of these discrepancies revealed that the $\ell=1$ term is responsible for $\sim23\%$ of the discrepancies, $\sim40\%$ for the $\ell=2$,  $\sim35\%$ for the $\ell=3$, and less than $2\%$ for the $\ell>3$ components. The exact distribution of these discrepancies may vary based on the model.  For the $\ell=1$ and $\ell=2$ terms the discrepancies cannot be recovered by the perturbative approach due to its formalism and approximations.  We were able to correct $\sim 20\%$ of the $\ell=3$ discrepancies by employing a higher spherical order perturbative approach. In opposition to the results of \cite{Rosu2020b}, the $\ell=3$ contribution to the apsidal motion cannot be neglected in close binaires.  
\\~\\Finally,  we constructed a catalogue of observed binary systems with apsidal motion from the literature to assess the fraction of the observed binaries with apsidal motion effectively impacted by our modelling method.  Our analysis indicated that differences in tidal force, and consequently apsidal motion,  are non-negligible for systems where $q^{-1/5}a(1-e^2)/R_1\lesssim 6.5$.  As a result, a significant fraction  of the observed binaries ($42\%$) with apsidal motion are indeed affected by our modelling approach.
\\~\\ When modelling the dynamics and deformations of binaries the perturbative approach is the most sophisticated method employed.  In this article,  we demonstrated that this approach loses precision for binaries with $q^{-1/5}a(1-e^2)/R_1\lesssim 6.5$.  The most immediate observable quantity affected is the apsidal motion.  While,  in this work, we only considered the contribution from the equilibrium tides and the general relativistic contributions, additional phenomenon could impact the determination of the apsidal motion.  Furthermore,  including the dynamical tides  can highly modify the theoretical determination of the apsidal motion. However, a proper non-perturbative  non-addiabatic 3D astreroseimic modelling is required to study the propagation of gravito-inertial waves in the stellar interior and their impact on the dynamics of binary systems.  In the future,  our aim is to develop such a tool.   In addition,  in the future,  the inclination of binary systems have to be considered when modelling the tidal contribution to the apsidal motion to account for all the parameters of the orbits.  An extension of MoBiDICT for non-aligned systems can be developed; however, all the system's symmetries exploited in this work will disappear, consequently our methodology will require significantly more resources to run. 
\\~\\Importantly, the underestimations seen extends beyond the apsidal motion to encompass tidal interactions between the binary components, thereby impacting tidal dissipation processes.  Notably,  a vast majority of binary stellar evoltion codes are also subject to these imprecisions at low orbital period,  which directly influences the predictive accuracy of binary system outcomes.  Given the significant tidal force discrepancies of our models at low orbital separations,  it is reasonable to anticipate an underestimation of tidal dissipations in such conditions.  Therefore,  with our methodology,  we expect to amplify the interactions and exchanges of angular momentum in close binary systems, ultimately leading to shorter dynamical evolution timescales. To solidify these observations,  the next step is to integrate our method into existing stellar binary evolution codes. This implementation will offer a more comprehensive and accurate understanding of the dynamics and evolution of close binary systems.
\begin{acknowledgements} The authors are thanking the anonymous referee for their comments. 
L.F was supported by the Fonds de la Recherche Scientifique F.R.S.-FNRS as a Research Fellow.  
\end{acknowledgements}

\bibliographystyle{aa}
\bibliography{Potentiel.bib}

\begin{thebibliography}{47}
\expandafter\ifx\csname natexlab\endcsname\relax\def\natexlab#1{#1}\fi

\bibitem[{{Adelberger} {et~al.}(2011){Adelberger}, {Garc{\'\i}a}, {Robertson},
  {Snover}, {Balantekin}, {Heeger}, {Ramsey-Musolf}, {Bemmerer}, {Junghans},
  {Bertulani}, {Chen}, {Costantini}, {Prati}, {Couder}, {Uberseder},
  {Wiescher}, {Cyburt}, {Davids}, {Freedman}, {Gai}, {Gazit}, {Gialanella},
  {Imbriani}, {Greife}, {Hass}, {Haxton}, {Itahashi}, {Kubodera}, {Langanke},
  {Leitner}, {Leitner}, {Vetter}, {Winslow}, {Marcucci}, {Motobayashi},
  {Mukhamedzhanov}, {Tribble}, {Nollett}, {Nunes}, {Park}, {Parker},
  {Schiavilla}, {Simpson}, {Spitaleri}, {Strieder}, {Trautvetter}, {Suemmerer},
  \& {Typel}}]{Reaction2011}
{Adelberger}, E.~G., {Garc{\'\i}a}, A., {Robertson}, R.~G.~H., {et~al.} 2011,
  Reviews of Modern Physics, 83, 195

\bibitem[{{Asplund} {et~al.}(2009){Asplund}, {Grevesse}, {Sauval}, \&
  {Scott}}]{Asplund2009}
{Asplund}, M., {Grevesse}, N., {Sauval}, A.~J., \& {Scott}, P. 2009, ARA\&A,
  47, 481

\bibitem[{{Baroch} {et~al.}(2022){Baroch}, {Gim{\'e}nez}, {Morales}, {Ribas},
  {Herrero}, {Perdelwitz}, {Jordi}, {Granzer}, \& {Allende
  Prieto}}]{Baroch2022}
{Baroch}, D., {Gim{\'e}nez}, A., {Morales}, J.~C., {et~al.} 2022, \aap, 665,
  A13

\bibitem[{{Baroch} {et~al.}(2021){Baroch}, {Gim{\'e}nez}, {Ribas}, {Morales},
  {Anglada-Escud{\'e}}, \& {Claret}}]{Baroch2021}
{Baroch}, D., {Gim{\'e}nez}, A., {Ribas}, I., {et~al.} 2021, \aap, 649, A64

\bibitem[{{Benvenuto} {et~al.}(2002){Benvenuto}, {Serenelli}, {Althaus},
  {Barb{\'a}}, \& {Morrell}}]{Benvenuto2002}
{Benvenuto}, O.~G., {Serenelli}, A.~M., {Althaus}, L.~G., {Barb{\'a}}, R.~H.,
  \& {Morrell}, N.~I. 2002, \mnras, 330, 435

\bibitem[{{Claret}(2023)}]{Claret2023}
{Claret}, A. 2023, \aap, 674, A67

\bibitem[{{Claret} {et~al.}(2021){Claret}, {Gim{\'e}nez}, {Baroch}, {Ribas},
  {Morales}, \& {Anglada-Escud{\'e}}}]{Claret2021}
{Claret}, A., {Gim{\'e}nez}, A., {Baroch}, D., {et~al.} 2021, \aap, 654, A17

\bibitem[{{Cox} \& {Giuli}(1968)}]{Cox1968}
{Cox}, J.~P. \& {Giuli}, R.~T. 1968, {Principles of stellar structure }

\bibitem[{{Fellay} \& {Dupret}(2023)}]{Fellay2023}
{Fellay}, L. \& {Dupret}, M.~A. 2023, arXiv e-prints, arXiv:2305.14139

\bibitem[{{Fitzpatrick}(2012)}]{Fitzpatrick2012}
{Fitzpatrick}, R. 2012, {An Introduction to Celestial Mechanics}

\bibitem[{{Gimenez} {et~al.}(1987){Gimenez}, {Kim}, \& {Nha}}]{Gimenez1987}
{Gimenez}, A., {Kim}, C.-H., \& {Nha}, I.-S. 1987, \mnras, 224, 543

\bibitem[{{Gimenez} \& {Margrave}(1985)}]{Gimenez1985}
{Gimenez}, A. \& {Margrave}, T.~E. 1985, \aj, 90, 358

\bibitem[{{Harmanec} {et~al.}(2014){Harmanec}, {Holmgren}, {Wolf},
  {Bo{\v{z}}i{\'c}}, {Guinan}, {Kang}, {Mayer}, {McCook}, {Nemravov{\'a}},
  {Yang}, {{\v{S}}lechta}, {Ru{\v{z}}djak}, {Sudar}, \&
  {Svoboda}}]{Harmanec2014}
{Harmanec}, P., {Holmgren}, D.~E., {Wolf}, M., {et~al.} 2014, \aap, 563, A120

\bibitem[{{Hut}(1981)}]{Hut1981}
{Hut}, P. 1981, \aap, 99, 126

\bibitem[{{Iglesias} \& {Rogers}(1996)}]{Iglesias1996}
{Iglesias}, C.~A. \& {Rogers}, F.~J. 1996, ApJ, 464, 943

\bibitem[{{Irwin}(2012)}]{Irwin2012}
{Irwin}, A.~W. 2012, {FreeEOS: Equation of State for stellar interiors
  calculations}

\bibitem[{{Jeans}(1929)}]{Jeans1929}
{Jeans}, J.~H. 1929, {The universe around us}

\bibitem[{{Kopal}(1959)}]{Kopal1959}
{Kopal}, Z. 1959, {Close binary systems}

\bibitem[{{Kopal}(1978)}]{Kopal1978}
{Kopal}, Z. 1978, {Dynamics of close binary systems}

\bibitem[{{Lacy} {et~al.}(2015){Lacy}, {Torres}, {Fekel}, {Muterspaugh}, \&
  {Southworth}}]{Lacy2015}
{Lacy}, C. H.~S., {Torres}, G., {Fekel}, F.~C., {Muterspaugh}, M.~W., \&
  {Southworth}, J. 2015, \aj, 149, 34

\bibitem[{{Ma} \& {Fuller}(2023)}]{Ma2023}
{Ma}, L. \& {Fuller}, J. 2023, \apj, 952, 53

\bibitem[{{Marcussen} \& {Albrecht}(2022)}]{Marcussen2022}
{Marcussen}, M.~L. \& {Albrecht}, S.~H. 2022, \apj, 933, 227

\bibitem[{{Naoz} {et~al.}(2013){Naoz}, {Farr}, {Lithwick}, {Rasio}, \&
  {Teyssandier}}]{Naoz2013}
{Naoz}, S., {Farr}, W.~M., {Lithwick}, Y., {Rasio}, F.~A., \& {Teyssandier}, J.
  2013, \mnras, 431, 2155

\bibitem[{{Pablo} {et~al.}(2015){Pablo}, {Richardson}, {Moffat}, {Corcoran},
  {Shenar}, {Benvenuto}, {Fuller}, {Naz{\'e}}, {Hoffman}, {Miroshnichenko},
  {Ma{\'\i}z Apell{\'a}niz}, {Evans}, {Eversberg}, {Gayley}, {Gull},
  {Hamaguchi}, {Hamann}, {Henrichs}, {Hole}, {Ignace}, {Iping}, {Lauer},
  {Leutenegger}, {Lomax}, {Nichols}, {Oskinova}, {Owocki}, {Pollock},
  {Russell}, {Waldron}, {Buil}, {Garrel}, {Graham}, {Heathcote}, {Lemoult},
  {Li}, {Mauclaire}, {Potter}, {Ribeiro}, {Matthews}, {Cameron}, {Guenther},
  {Kuschnig}, {Rowe}, {Rucinski}, {Sasselov}, \& {Weiss}}]{Pablo2015}
{Pablo}, H., {Richardson}, N.~D., {Moffat}, A. F.~J., {et~al.} 2015, \apj, 809,
  134

\bibitem[{{Packet}(1981)}]{Packet1981}
{Packet}, W. 1981, \aap, 102, 17

\bibitem[{{Paxton} {et~al.}(2011){Paxton}, {Bildsten}, {Dotter}, {Herwig},
  {Lesaffre}, \& {Timmes}}]{MESA2011}
{Paxton}, B., {Bildsten}, L., {Dotter}, A., {et~al.} 2011, \apjs, 192, 3

\bibitem[{{Paxton} {et~al.}(2013){Paxton}, {Cantiello}, {Arras}, {Bildsten},
  {Brown}, {Dotter}, {Mankovich}, {Montgomery}, {Stello}, {Timmes}, \&
  {Townsend}}]{MESA2013}
{Paxton}, B., {Cantiello}, M., {Arras}, P., {et~al.} 2013, \apjs, 208, 4

\bibitem[{{Paxton} {et~al.}(2015){Paxton}, {Marchant}, {Schwab}, {Bauer},
  {Bildsten}, {Cantiello}, {Dessart}, {Farmer}, {Hu}, {Langer}, {Townsend},
  {Townsley}, \& {Timmes}}]{MESA2015}
{Paxton}, B., {Marchant}, P., {Schwab}, J., {et~al.} 2015, \apjs, 220, 15

\bibitem[{{Paxton} {et~al.}(2018){Paxton}, {Schwab}, {Bauer}, {Bildsten},
  {Blinnikov}, {Duffell}, {Farmer}, {Goldberg}, {Marchant}, {Sorokina},
  {Thoul}, {Townsend}, \& {Timmes}}]{MESA2018}
{Paxton}, B., {Schwab}, J., {Bauer}, E.~B., {et~al.} 2018, \apjs, 234, 34

\bibitem[{{Paxton} {et~al.}(2019){Paxton}, {Smolec}, {Schwab}, {Gautschy},
  {Bildsten}, {Cantiello}, {Dotter}, {Farmer}, {Goldberg}, {Jermyn}, {Kanbur},
  {Marchant}, {Thoul}, {Townsend}, {Wolf}, {Zhang}, \& {Timmes}}]{MESA2019}
{Paxton}, B., {Smolec}, R., {Schwab}, J., {et~al.} 2019, \apjs, 243, 10

\bibitem[{{Rosu} {et~al.}(2020{\natexlab{a}}){Rosu}, {Noels}, {Dupret}, {Rauw},
  {Farnir}, \& {Ekstr{\"o}m}}]{Rosu2020}
{Rosu}, S., {Noels}, A., {Dupret}, M.~A., {et~al.} 2020{\natexlab{a}}, \aap,
  642, A221

\bibitem[{{Rosu} {et~al.}(2023){Rosu}, {Quintero}, {Rauw}, \&
  {Eenens}}]{Rosu2023}
{Rosu}, S., {Quintero}, E.~A., {Rauw}, G., \& {Eenens}, P. 2023, \mnras, 521,
  2988

\bibitem[{{Rosu} {et~al.}(2020{\natexlab{b}}){Rosu}, {Rauw}, {Conroy},
  {Gosset}, {Manfroid}, \& {Royer}}]{Rosu2020b}
{Rosu}, S., {Rauw}, G., {Conroy}, K.~E., {et~al.} 2020{\natexlab{b}}, \aap,
  635, A145

\bibitem[{{Rosu} {et~al.}(2022{\natexlab{a}}){Rosu}, {Rauw}, {Farnir},
  {Dupret}, \& {Noels}}]{Rosu2022a}
{Rosu}, S., {Rauw}, G., {Farnir}, M., {Dupret}, M.~A., \& {Noels}, A.
  2022{\natexlab{a}}, \aap, 660, A120

\bibitem[{{Rosu} {et~al.}(2022{\natexlab{b}}){Rosu}, {Rauw}, {Naz{\'e}},
  {Gosset}, \& {Sterken}}]{Rosu2022b}
{Rosu}, S., {Rauw}, G., {Naz{\'e}}, Y., {Gosset}, E., \& {Sterken}, C.
  2022{\natexlab{b}}, \aap, 664, A98

\bibitem[{{Scuflaire} {et~al.}(2008){Scuflaire}, {Th{\'e}ado}, {Montalb{\'a}n},
  {Miglio}, {Bourge}, {Godart}, {Thoul}, \& {Noels}}]{Scuflaire2008a}
{Scuflaire}, R., {Th{\'e}ado}, S., {Montalb{\'a}n}, J., {et~al.} 2008, ApSS,
  316, 83

\bibitem[{{Siess} {et~al.}(2013){Siess}, {Izzard}, {Davis}, \&
  {Deschamps}}]{Siess2013}
{Siess}, L., {Izzard}, R.~G., {Davis}, P.~J., \& {Deschamps}, R. 2013, \aap,
  550, A100

\bibitem[{{Sterne}(1939)}]{Sterne1939}
{Sterne}, T.~E. 1939, \mnras, 99, 451

\bibitem[{{Torres} {et~al.}(2010){Torres}, {Andersen}, \&
  {Gim{\'e}nez}}]{Torres2010}
{Torres}, G., {Andersen}, J., \& {Gim{\'e}nez}, A. 2010, \aapr, 18, 67

\bibitem[{{Vernazza} {et~al.}(1981){Vernazza}, {Avrett}, \&
  {Loeser}}]{Atmosphere1981}
{Vernazza}, J.~E., {Avrett}, E.~H., \& {Loeser}, R. 1981, \apjs, 45, 635

\bibitem[{{Vink} {et~al.}(2001){Vink}, {de Koter}, \& {Lamers}}]{Vink2001}
{Vink}, J.~S., {de Koter}, A., \& {Lamers}, H.~J.~G.~L.~M. 2001, \aap, 369, 574

\bibitem[{{Willems} \& {Claret}(2002)}]{Willems2002}
{Willems}, B. \& {Claret}, A. 2002, \aap, 382, 1009

\bibitem[{{Wolf} {et~al.}(2010){Wolf}, {Claret}, {Kotkov{\'a}},
  {Ku{\v{c}}{\'a}kov{\'a}}, {Koci{\'a}n}, {Br{\'a}t}, {Svoboda}, \&
  {{\v{S}}melcer}}]{Wolf2010}
{Wolf}, M., {Claret}, A., {Kotkov{\'a}}, L., {et~al.} 2010, \aap, 509, A18

\bibitem[{{Wolf} {et~al.}(2006){Wolf}, {Ku{\v{c}}{\'a}kov{\'a}}, {Kolasa},
  {{\v{S}}tastn{\'y}}, {Bozkurt}, {Harmanec}, {Zejda}, {Br{\'a}t}, \&
  {Hornoch}}]{Wolf2006}
{Wolf}, M., {Ku{\v{c}}{\'a}kov{\'a}}, H., {Kolasa}, M., {et~al.} 2006, \aap,
  456, 1077

\bibitem[{{Wolf} {et~al.}(2008){Wolf}, {Zejda}, \& {de Villiers}}]{Wolf2008}
{Wolf}, M., {Zejda}, M., \& {de Villiers}, S.~N. 2008, \mnras, 388, 1836

\bibitem[{{Zahn}(1975)}]{Zahn1975}
{Zahn}, J.~P. 1975, \aap, 41, 329

\bibitem[{{Zahn}(1977)}]{Zahn1977}
{Zahn}, J.~P. 1977, \aap, 57, 383

\end{thebibliography}

\clearpage
\appendix

\section{The Clairaut-Radau equation}\label{annexe_Clairaut-Radau}
\begin{figure}[h]
\centering
\includegraphics[width=\hsize]{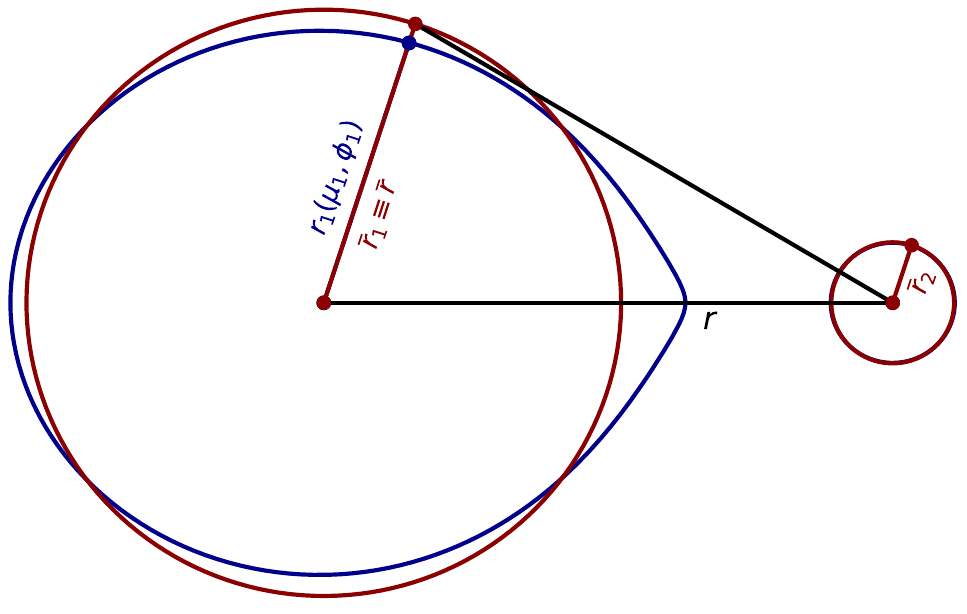}
\caption{Scheme of a binary system separated by a distance $r$. The different quantities used to model the deformations in the perturbative framework are detailed and illustrated here. The blue curve is the surface of the deformed star while the red curves are the spheres of equivalent averaged radii. As the surface is represented here, $\overline{r}_i(\mu_i,\phi_i)= R_i$, $R_i$ being the radial coordinate of the stellar surface in the spherically symmetric case. }
\label{fig.scheme_perturbative}
\end{figure}

A partial derivation of the Clairaut-Radau equation can already be found in \cite{Sterne1939,Kopal1959, Fitzpatrick2012}. In this section we provide a detailed proper derivation of this equation and focus on the implication  of the assumptions of the perturbative approach. 
Let us express the radial coordinate of a given equipotential in the direction $r_1(\theta, \phi)$  with its spherical expansion:
\begin{equation}\label{eq_perturbative_1}
r_1(\overline{r},\theta, \phi)= \overline{r}\left(1+ \sum^{L}_{\ell=1} \sum^{\ell}_{m=-\ell} r_{\ell}^m(\overline{r}) Y_{\ell m}(\theta,\phi) \right),
\end{equation}
where  $\overline{r}$ denotes the averaged radial coordinate of the considered equipotential and $r_\ell^m(r) $ the spectral terms of the equipotential radii that can be obtained with
\begin{equation}
r_{\ell}^m(r_1)=\frac{1}{\overline{r}}\int_0^1\int_0^\pi r_1(\overline{r},\mu, \phi) Y_{\ell m}(\theta,\phi) d\phi d\mu,
\end{equation}
where the spherical harmonics were appropriately normalised. Throughout this  entire section, \autoref{fig.scheme_perturbative} illustrates the different variables and orbital quantities used to describe the deformations of a binary system. 
In the perturbative approach, $\overline{r}$ can be obtained by expanding \autoref{eq_perturbative_1} to the first order ($r_{\ell}^m(r_1)\ll 1$):
\begin{align}
\overline{r}(r_1,\theta, \phi)&=\dfrac{r_1}{1+ \sum^{L}_{\ell=1} \sum^{\ell}_{m=-\ell} r_{\ell}^m(r_1)Y_{\ell m}(\theta,\phi) } \\ \ &\simeq r_1\left(1- \sum^{L}_{\ell=1} \sum^{\ell}_{m=-\ell} r_{\ell}^m(r_1)Y_{\ell m}(\theta,\phi)\right). \nonumber
\end{align}
The density of an equipotential passing by $(r_1,\theta, \phi)$, denoted $\rho(\overline{r}(r_1,\theta, \phi))$, can be approximated to the first order using a Taylor expansion of $\overline{r}$ around $r_1$:
\begin{align}
\rho(\overline{r}(r_1,\theta, \phi))&=\rho(r_1)+\frac{\textrm{d}\rho}{\textrm{d}r_1}(\overline{r}-r_1)\\ \ &=\rho(r_1)-r_1\frac{\textrm{d}\rho}{\textrm{d}r_1}(r_1)\sum^{L}_{\ell=1} \sum^{\ell}_{m=-\ell} r_{\ell}^{m}(r_1) Y_{\ell m}(\theta,\phi), \nonumber
\end{align}
where $\rho(r_1)$ is the density of the equipotential with $\overline{r}=r_1$.
The densities of a sphere of radius $r_1$ projected on the spherical harmonics basis is obtained through 
\begin{equation}
\rho_\ell^m(r_1)= \int_0^{\pi}  \int_0^{2\pi} \rho(r_1, \theta,\phi)Y_{\ell m}(\theta,\phi)\sin(\theta)  \textrm{d}\phi \textrm{d}\theta 
\end{equation}
that can be rewritten with
\begin{align}
\rho_\ell^m(r_1)&= \int_0^{\pi}  \int_0^{2\pi}  \left[  \rho(r_1)-  r_1\frac{\textrm{d}\rho}{\textrm{d}r_1}(r_1)\sum^{L^\prime}_{\ell^\prime=1}\right. \\ \ & \left. \sum^{\ell^\prime}_{m^\prime=-\ell^\prime} r_{\ell^\prime}^{m^\prime}(r_1) Y_{\ell^\prime m^\prime}(\theta,\phi)   \right] Y_{\ell m}(\theta,\phi)\sin(\theta)  \textrm{d}\phi \textrm{d}\theta. \nonumber
\end{align}
Using the orthogonality properties of the spherical harmonics, at the first order  the projected densities are 
\begin{align}\label{eq_spectral_developpement_Rho}
\rho_\ell^m (r_1)=
    \begin{cases}
       \rho(r_1) \qquad  &\ell,m=0 \\
      -r_1\dfrac{\textrm{d}\rho}{\textrm{d}r_1}(r_1)r_{\ell}^{m}(r_1) &\ell>0
    \end{cases}.
\end{align}
The spectral development of the gravitational potential on a sphere of radius $r_1$ is given by \autoref{eq_solution_poisson}. The gravitational potential on a sphere of radius $r_1$ will be noted in the future $\Psi_r(r_1,\theta, \phi)$ while the gravitational potential on an equipotential is noted $\Psi_{\overline{r}}(\overline{r},\theta, \phi)$ and can be expressed as
\begin{align}
\Psi_{\overline{r}}(\overline{r},\theta, \phi)&=\Psi^\prime_0(\overline{r})+\sum^{L}_{\ell=1} \sum^{\ell}_{m=-\ell} \Psi_{\ell}^{\prime m}(\overline{r})Y_{\ell m}(\theta,\phi)\\ \ &=\Psi_r(r_1(\overline{r},\theta, \phi),\theta, \phi)\nonumber 
\end{align}
that can be developed to the first order using the Taylor approximation around $r_1=\overline{r}$:
\begin{align}
&\Psi_r(r_1(\overline{r},\theta, \phi),\theta, \phi)=\\ \nonumber &\Psi_0(r_1(\overline{r},\theta, \phi))+\sum^{L}_{\ell=1} \sum^{\ell}_{m=-\ell} \Psi_{\ell}^{m}(r_1(\overline{r}))Y_{\ell m}(\theta,\phi) \\ &=\nonumber \Psi_0(\overline{r})+\frac{\textrm{d}\Psi_0}{\textrm{d}r_1}(\overline{r}-r_1)+\sum^{L}_{\ell=1} \sum^{\ell}_{m=-\ell} \Psi_{\ell}^{m}(\overline{r})Y_{\ell m}(\theta,\phi) \\ &=\nonumber  \Psi_0(\overline{r})+\sum^{L}_{\ell=1} \sum^{\ell}_{m=-\ell} \left( \Psi_{\ell}^{m}(\overline{r} )+\overline{r} \frac{\textrm{d}\Psi_0}{\textrm{d}\overline{r}}r_{\ell}^{m}(\overline{r})\right)Y_{\ell m}(\theta,\phi),\nonumber 
\end{align}
and therefore
\begin{align}
    \begin{cases}
      \Psi^\prime_0(\overline{r})= \Psi_0(\overline{r})\\
      \Psi_{\ell}^{\prime m}(\overline{r})=\Psi_{\ell}^{m}(\overline{r})+\overline{r} \dfrac{\textrm{d}\Psi_0}{\textrm{d}\overline{r}}r_{\ell}^{m}(\overline{r})
    \end{cases}.
\end{align}
The spectral components of the gravitational potential (\autoref{eq_solution_poisson}) on an equipotential  can be rewritten using \autoref{eq_spectral_developpement_Rho} before  integrating by parts:
\begin{align}
\nonumber & \Psi_\ell^m(\overline{r})\ =\ - \frac{4\pi G}{(2\ell+1)\overline{r}^{\ell+1}} \left(- \int_0^{\overline{r}} \overline{r}^{\prime \ell+3} \dfrac{\textrm{d}\rho}{\textrm{d}\overline{r}^\prime}(\overline{r}^\prime)r_{\ell}^{m}(\overline{r}^\prime)\textrm{d}\overline{r}^\prime \right)\\ &-   \frac{4\pi G \overline{r}^\ell}{2\ell+1}   \left(-\int_{\overline{r}}^{\infty}\overline{r}^{\prime 2-\ell}   \dfrac{\textrm{d}\rho}{\textrm{d}\overline{r}^\prime}(\overline{r}^\prime)r_{\ell}^{m}(\overline{r}^\prime)  \textrm{d}\overline{r}^\prime\right) \\ &=\  \nonumber - \frac{4\pi G}{(2\ell+1)\overline{r}^{\ell+1}} \left(-\cancel{\left[ \overline{r}^{\prime \ell+3} r_{\ell}^{m} \rho  \right]_0^{\overline{r}}} + \int_0^{\overline{r}} \rho(\overline{r}^\prime) \textrm{d}(r_{\ell}^{m}\overline{r}^{\prime \ell+3}) \right) \\ &-  \frac{4\pi G \overline{r}^\ell}{2\ell+1} \left(-\cancel{\left[ \overline{r}^{\prime 2-\ell} r_{\ell}^{m} \rho  \right]_{\overline{r}}^{R_{\textrm{s}}}} + \int_{\overline{r}}^{R_{\textrm{s}}}  \rho(\overline{r}^\prime) \textrm{d}(r_{\ell}^{m}\overline{r}^{\prime 2-\ell}) \right) \\ &=\  \nonumber - \frac{4\pi G}{(2\ell+1)\overline{r}^{\ell+1}}  \int_0^{\overline{r}} \rho(\overline{r}^\prime) \textrm{d}(r_{\ell}^{m}\overline{r}^{\prime \ell+3}) \\ &-  \frac{4\pi G \overline{r}^\ell}{2\ell+1}  \int_{\overline{r}}^{R}  \rho(\overline{r}^\prime) \textrm{d}(r_{\ell}^{m}\overline{r}^{\prime 2-\ell})\label{eq_perturbative_3},
\end{align}
where at the second step the two crossed terms are cancelling each others out.
\\As the total potential is constant on an equipotential, one can write that:
\begin{equation}
\Psi_{\textrm{tot}}=\Psi_1+\Psi_2+\Psi_{\textrm{c}}=\textrm{const},
\end{equation}
implying that for all the $\ell$  and $m$ different from $0$,
\begin{equation}
\Psi^{\prime m}_{1\ell}+\Psi^{\prime m}_{2,\ell}+\Psi^{\prime m}_{\textrm{c},\ell}=0.
\end{equation}
\autoref{eq_perturbative_3} can be developed and multiplied by $\overline{r}^{\ell+1}/4\pi G$:
\begin{align}
\nonumber& \frac{-1}{(2\ell+1)}  \int_0^{\overline{r}} \rho(\overline{r}^\prime) \textrm{d}(r_{\ell}^{m}\overline{r}^{\prime \ell+3}) -  \frac{\overline{r}^{2\ell+1}}{2\ell+1}  \int_{\overline{r}}^{R}  \rho(\overline{r}^\prime) \textrm{d}(r_{\ell}^{m}\overline{r}^{\prime 2-\ell})\\ \ &+ \label{eq_perturbative_2}\frac{ \overline{r} m( \overline{r})}{4\pi}  r_{\ell}^{m}(\overline{r})+ \frac{\overline{r}^{\ell+1}}{4\pi G}\left( \Psi^{\prime}_{2\ell m}+\Psi^{\prime}_{\textrm{c},\ell m} \right)=0.
\end{align}
At first order $\Psi_{2,l}^{\prime,m} = \Psi_{2,l}^m$  and $\Psi_{c,l}^{\prime,m} = \Psi_{c,l}^m$, By differentiating \autoref{eq_perturbative_2}, we obtain
\begin{align}
&  - \overline{r}^{2\ell} \int_{\overline{r}}^{R}  \rho(\overline{r}^\prime) \textrm{d}(r_{\ell}^{m}\overline{r}^{\prime 2-\ell}) + \frac{ m( \overline{r})}{4\pi} \overline{r}^{\ell-1}\left(\overline{r} \dfrac{\textrm{d}r_{\ell}^{m}}{\textrm{d}\overline{r}}+\ell r_{\ell}^{m} \right)\nonumber \\ \ &+  \frac{1}{4\pi G}  \dfrac{\textrm{d}}{\textrm{d}\overline{r}}\left(  (\Psi_{2,\ell}^m+\Psi_{\textrm{c},\ell}^m)\overline{r}^{\ell+1}  \right) =0 \label{eq_perturbative_4}.
\end{align}
This expression can then be divided by $\overline{r}^{2\ell}$ and again differentiated by $\overline{r}$ to obtain
\begin{align}
& \nonumber \overline{r}^2 \frac{\textrm{d}^2r_{\ell}^{m}}{\textrm{d}\overline{r}^2} + \frac{6\rho(\overline{r})}{\overline{\rho}(\overline{r})}\left(\overline{r}\frac{\textrm{d}r_{\ell}^{m}}{\textrm{d}\overline{r}}+ r_{\ell}^{m}(\overline{r})\right) -\ell(\ell+1)r_{\ell}^{m}(\overline{r}) \\ \ &+\frac{3 \overline{r}^{\ell-1}}{4\pi G \overline{\rho}(\overline{r})}\dfrac{\textrm{d}}{\textrm{d}\overline{r}}\left( \frac{1}{\overline{r}^{2\ell}} \dfrac{\textrm{d}}{\textrm{d}\overline{r}}\left(  (\Psi_{2,\ell}^m+\Psi_{\textrm{c},\ell}^m)\overline{r}^{\ell+1}  \right)   \right)=0,
\end{align}
where we used the definition of the mean density under an isobar:
\begin{equation}
\overline{\rho}= \dfrac{m(\overline{r})}{\dfrac{4}{3}\pi \overline{r}^3}.
\end{equation}
To the first order $(\Psi_{2,\ell}^m+\Psi_{\textrm{c},\ell}^m)\propto \overline{r}^{\ell}$ meaning that $ \dfrac{\textrm{d}}{\textrm{d}\overline{r}}\left(  (\Psi_{2,\ell}^m+\Psi_{\textrm{c},\ell}^m)\overline{r}^{\ell+1}  \right) \propto \overline{r}^{2\ell}$ therefore the last term of the previous equation is cancelling out during the differentiation. 
Finally by defining 
\begin{equation}\label{eq_def_eta_ell}
\eta_{\ell}^{m} \equiv \frac{\overline{r}}{r_{\ell}^{m}} \frac{\textrm{d}r_{\ell}^{m}}{\textrm{d}\overline{r}},
\end{equation}
and developing the second derivative of $r_{\ell}^{m}$:
\begin{equation}
\dfrac{\mathrm{d}^2r_{\ell}^{m}}{\mathrm{d}\overline{r}^2}=\dfrac{\mathrm{d}}{\mathrm{d}\overline{r}} \left( \dfrac{r_{\ell}^{m}\eta_{\ell}^{m}}{\overline{r}}\right)=\eta_{\ell}^{m}\left( \dfrac{1}{\overline{r}}\dfrac{\mathrm{d}\eta_{\ell}^{m}}{\mathrm{d}\overline{r}}+\dfrac{\eta_{\ell}^{m}(\eta_{\ell}^{m}-1)}{\overline{r}^2}\right),
\end{equation}
we obtain the well known Clairaut-Radau equation expressed as
\begin{equation}\label{eq_clairaut-radau}
\overline{r}\frac{d \eta_{\ell}}{d\overline{r}}+6\frac{\rho(\overline{r})}{\overline{\rho}(\overline{r})}(\eta_{\ell}+1)+\eta_{\ell}(\eta_{\ell}-1)=\ell(\ell+1).
\end{equation}

\section{Perturbation of the surface total potential }\label{annexe_perturbation_surface}
In this section our aim is to look at how the perturbative model is modifying the surface gravitational potential of each star composing the system.  Let us go back to \autoref{eq_perturbative_2} and divide it by $\overline{r}^{2\ell+1}$ and next differentiate it to make $\int_{\overline{r}}^{R_{\textrm{s}}} ...$ fall:
\begin{align}
& \nonumber\frac{1}{\overline{r}^{2\ell+2}}  \int_0^{\overline{r}} \rho (\overline{r}^\prime) \textrm{d}(r_{\ell}^{m}\overline{r}^{\prime \ell+3})+\frac{1}{4\pi G} \dfrac{\textrm{d}}{\textrm{d}\overline{r}}\left(\frac{\Psi_{2,\ell}^m+\Psi_{\textrm{c},\ell}^m}{\overline{r}^{\ell}}  \right)\\ \ &+\frac{r_{\ell}^{m}(\overline{r})}{\overline{r}^{\ell+2}}  \int_0^{\overline{r}} \rho (\overline{r}^\prime)\overline{r}^{\prime 2} \textrm{d}\overline{r}^\prime \left( \eta_{\ell}^{m}  -1-\ell\right) =0.
\end{align}
The previous equation is then multiplied by $\dfrac{\overline{r}^{\ell+1}4\pi G}{2\ell+1} $  before being injected in  \autoref{eq_perturbative_3} to eliminate $\int_0^{\overline{r}} \rho (\overline{r}^\prime) \textrm{d}(r_{\ell}^{m}\overline{r}^{\prime \ell+3})$ and obtain the spectral projections of the gravitational potentials expressed as
\begin{align}
& \nonumber \Psi_{1\ell}^m(\overline{r}) = \frac{4\pi G}{2\ell+1} \left( \dfrac{r_{\ell}^{m}(\overline{r}) m(\overline{r})}{4\pi \overline{r}}  \left( \eta_{1\ell}^{m}  -1-\ell\right)\right. \\ \ &- \left.\overline{r}^{\ell} \int_{\overline{r}}^{R_{\textrm{s}}} \rho (\overline{r}^\prime) \textrm{d}(r_{\ell}^{m}\overline{r}^{\prime 2-\ell}) \right) +\dfrac{\overline{r}^{\ell+1}}{2\ell+1}\dfrac{\textrm{d}}{\textrm{d}\overline{r}}\left(\frac{\Psi_{2,\ell}^m+\Psi_{\textrm{c},\ell}^m}{\overline{r}^{\ell}} \right).
\end{align}
The integral term $\int_{\overline{r}}^{R_{\textrm{s}}}...$ is negligible at the surface. For an arbitrary chosen primary with a radius $R_1$ the surface gravitational potential is now expressed as 
\begin{equation}\label{eq_perturbative_5}
 \Psi_{1\ell}^m(R_{\textrm{1}})\simeq \dfrac{r_{\ell}^{m}(R_{\textrm{1}}) G M}{(2\ell+1)R_{\textrm{1}} }   \left( \eta_{1\ell}^{m}(R_{\textrm{1}})  -1-\ell\right).
\end{equation}
\autoref{eq_perturbative_4} is then evaluated at the stellar surface and to the first order to obtain 
\begin{equation}
r_{\ell}^{m}(R_{\textrm{1}}) G M=-R_{\textrm{1}}(2\ell+1)\dfrac{\Psi_{2,\ell}^m(R_{\textrm{1}}) +\Psi_{\textrm{c},\ell}^m(R_{\textrm{1}}) }{\eta_{1\ell}^{m}(R_{\textrm{1}})+\ell},
\end{equation}
that we inject in \autoref{eq_perturbative_5} to express the perturbation of the spectral surface gravitational potential 
\begin{equation}\label{eq_anx_Psi_lm_pert}
 \Psi_{1\ell}^m(R_{\textrm{1}})=\dfrac{1+\ell-\eta_{1\ell}^{m}(R_{\textrm{1}})}{\eta_{1\ell}^{m}(R_{\textrm{1}})+\ell}\left(\Psi_{2,\ell}^m(R_{\textrm{1}}) +\Psi_{\textrm{c},\ell}^m(R_{\textrm{1}}) \right).
\end{equation}
With the definition of the Love numbers given in \autoref{eq_k2}, we have finally:
\begin{equation}\label{anx_eq_perturbation_surface}
 \Psi_{1\ell}^m(R_{\textrm{1}})=2k_{\ell, 1}\left(\Psi_{2,\ell}^m(R_{\textrm{1}}) +\Psi_{\textrm{c},\ell}^m(R_{\textrm{1}}) \right).
\end{equation}

\section{First terms of the centrifugal,  gravitational,  and tidal potentials}\label{annexe_first_termes}

\subsection{Centrifugal potential in unsynchronised systems}
In a non-synchronised system, the centrifugal potential expressed with the Cartesian coordinates centred on the primary is given by 
\begin{equation}
\Psi_{\textrm{c}}(x,y,z)=-\tilde{n}^2x_{\rm{CM}} x -\frac{\Omega_{\star}^2}{2}(x^2+y^2),
\end{equation}
where $(x,y,z)$ are the Cartesian coordinates centred on the star studied, $x$ pointing towards the centre of mass of the system,  $\tilde{n}$ is the orbital rotation rate of the binary system with a separation $r$ and $\Omega_{\star}$ is the rotation rate of this star. The quantity $x^2+y^2$ can be written when passing in spherical coordinates as $r_1^2-z^2$ with $z=r_1\cos\theta$ that can be expressed with the spherical harmonics
\begin{equation}
x^2+y^2=\dfrac{2}{3}r_1^2-Y^0_2(\cos\theta).
\end{equation}
The centrifugal potential can therefore be expressed as 
\begin{align}\label{anx_eq_Psi_centri}
\nonumber \Psi_{\textrm{c}}(x,y,z)=&\tilde{n}^2x_{\rm{CM}}r_1Y^1_1(\mu)  -\frac{\Omega_{\star}^2}{3} r_1^2  \left(1-  Y^0_2(\mu)\right),
\end{align}
and decomposed as
\begin{align}
& \Psi_{\textrm{c},0}^0= -  \dfrac{\Omega_{\star}^2}{3} r_1^2  ,
\\& \Psi_{\textrm{c},1}^1= -  \tilde{n}^2x_{\rm{CM}}r_1,
\\& \Psi_{\textrm{c},2}^0=   \dfrac{\Omega_{\star}^2}{3} r_1^2,
\\& \Psi_{\textrm{c},\ell}^m= 0 \textrm{ for all other combinaisons of $\ell$ and $m$}\nonumber.
\end{align}


\subsection{Gravitational and tidal potential}
Treating the secondary as a point source, the spectral component of the gravitational potential exerted by the secondary star at the surface of the primary can be expressed as
\begin{equation}\label{eq_anx_tidal_force}
\Psi_{2}(R_{\textrm{1}},\theta_1, \phi_1)= -\dfrac{GM_2}{R_{\textrm{1}}}\sum_\ell^L\left(\dfrac{R_{\textrm{1}}}{r} \right)^{\ell+1}   P_{\ell}(\sin\theta_1 \cos\phi_1),
\end{equation}
where $r$ is the instantaneous orbital separation of the stars and $P_{\ell}(\sin\theta_1 \cos\phi_1)$ is the classical Legendre polynomial that can be decomposed using spherical harmonics as
\begin{equation}
 P_{\ell}(\sin\theta_1 \cos\phi_1)=\sum_{m=-\ell}^\ell Y^m_{\ell}(\theta_1, \phi_1)d^m_{\ell},
\end{equation}
where $d^m_{\ell}$ are constant coefficients linked to the normalisation of the spherical harmonics.  In the following we define and use
\begin{equation}
\lambda_1\equiv\sin\theta_1 \cos\phi_1.
\end{equation}
Using this notation:
\begin{equation}
P_1(\lambda_1)=\lambda_1=\sin\theta_1 \cos\phi_1=\frac{x}{r_1}= -Y^1_1(\theta_1,\phi_1),
\end{equation}
yield  $d_1^0=0, \ d_1^1=-1$.  Similarly,
\begin{equation}
P_2(\lambda_1)=\dfrac{1}{4}Y^2_2(\theta_1,\phi_1)- \dfrac{1}{2}Y^0_2(\theta_1,\phi_1),
\end{equation}
yields $d_2^0=- \dfrac{1}{2}\textrm{ and }  d_2^2=\dfrac{1}{4}$.Therefore,
\begin{align}
& \Psi_{2,1}^1(R_{\textrm{1}})=   \dfrac{GM_2}{R_{\textrm{1}}} \left(\dfrac{R_{\textrm{1}}}{r} \right)^{2},\label{anx_eq_Psi_11_tides}
\\& \Psi_{2,2}^0(R_{\textrm{1}})= \dfrac{1}{2}\dfrac{GM_2}{R_{\textrm{1}}} \left(\dfrac{R_{\textrm{1}}}{r} \right)^{3},
\\& \Psi_{2,2}^2(R_{\textrm{1}})= -\dfrac{1}{4}\dfrac{GM_2}{R_{\textrm{1}}} \left(\dfrac{R_{\textrm{1}}}{r} \right)^{3}.
\end{align}
Summing Eqs. (\ref{anx_eq_Psi_centri}),  (\ref{anx_eq_Psi_11_tides})  and using the definition of $\tilde{n}$ (\autoref{eq_def_n_tilde}), we have: 
\begin{equation}\resizebox{\hsize}{!}{$
\Psi_{\textrm{c},1}^1 (R_{\textrm{1}})+\Psi_{2,1}^1(R_{\textrm{1}})= \dfrac{GM_2}{R_{\textrm{1}}} \left(\dfrac{R_{\textrm{1}}}{r} \right)^{2} -\dfrac{GM_2}{R_{\textrm{1}}} \left(\dfrac{R_{\textrm{1}}}{r} \right)^{2}=0.$}
\end{equation}
Substituting this result in the right hand side of \autoref{anx_eq_perturbation_surface}, we see that the perturbative theory predicts that the dipolar component of the gravitational potential generated by a star is equal to zero when its companion is treated as a point source. This is an important result explaining why the dipolar component was systematically neglected in previous studies.
\\The sums of the quadrupolar terms of the centrifugal and secondary tidal forces at the surface of the primary are finally given by 
\begin{align}
\Psi_{\textrm{c},2}^0(R_{\textrm{1}})+\Psi_{2,2}^0(R_{\textrm{1}})=\left( \dfrac{R_{\textrm{1}}}{r} \right)^2\left(\dfrac{\Omega_{\star}^2}{3} r^2 +  \dfrac{1}{2}\dfrac{GM_2}{r}\right)
\end{align}
and
\begin{equation}
\Psi_{\textrm{c},2}^2(R_{\textrm{1}})+\Psi_{2,2}^2(R_{\textrm{1}})=-\dfrac{1}{4}\dfrac{GM_2}{R_{\textrm{1}}} \left(\dfrac{R_{\textrm{1}}}{r} \right)^{3}.
\end{equation}

\section{Perturbation of the force}\label{annexe_perturbation of the force}
In this section, we develop the expression of the tidal force exerted by a deformed body on its companion, specifically $F_{2^{\prime}10}$ as introduced in \autoref{subsect_perturbation_gravity}. We limit ourselves to the dominant term, which arises from the quadrupolar deformation of the secondary. The system of coordinates used in this Section is described in \autoref{fig.scheme_perturbative}. Let us consider the secondary,  which is deformed due to the presence of an unperturbed primary.  In the section,  $r_i$ denotes the radial coordinate of the spherical coordinate frame centred on each star.  With this formalism the potential perturbation at the surface of the secondary is given by \autoref{eq_anx_Psi_lm_pert} applied to the $\ell=2$ terms:
\begin{equation}
\Psi_{2,2}(R_{\textrm{2}})=\Psi_{2,2}^0(R_{\textrm{2}})Y^2_0(\theta_2, \phi_2)+\Psi_{2,2}^2(R_{\textrm{2}})Y^2_2(\theta_2, \phi_2)
\end{equation}
that can be rewritten, using \autoref{eq_anx_tidal_force} as 
\begin{equation}\resizebox{\hsize}{!}{$
\Psi_{2,2}(R_{\textrm{2}}, \theta_2, \phi_2)=2k_{2,2}\left(\dfrac{R_{\textrm{2}}}{r} \right)^{2}\left[-\dfrac{GM_1}{r} P_2(\lambda_2) + \dfrac{\Omega_{\star, 2}^2}{3} r^2 P_2(\mu_2)\right],$}
\end{equation}
where $P_2$ is the Legendre polynomial of order two and $\lambda_i\equiv\sin\theta_i \cos\phi_i$ and $\mu_i = \cos\theta_i$. To study the impact of this potential perturbation on the primary, we can express the $\ell=2$ component of the potential outside of the secondary as
\begin{align}\label{eq_anx_perturbation_pot_1}
& \Psi_{2,2}(r_{2}, \lambda_2, \mu_2)=\Psi_{2,2}(R_2,\lambda_2,\mu_2) \left(\dfrac{R_{\textrm{2}}}{r_2} \right)^{3}\\ &= 2k_{2,2}\left(\dfrac{R_{\textrm{2}}}{r} \right)^{5}\left(\dfrac{r}{r_2} \right)^{3}\left[-\dfrac{GM_1}{r} P_2(\lambda_2) + \dfrac{\Omega_{\star,2}^2}{3} r^2 P_2(\mu_2)\right]\nonumber.
\end{align}
Following \cite{Kopal1959}, different coordinates related quantities from the secondary star can be approximated with series developments in the coordinates of the primary star:
\begin{equation}
\left(\dfrac{r}{r_2} \right)^{3}=1+3\dfrac{r_{\textrm{1}}}{r} P_1(\lambda_1)+\left(\dfrac{r_{\textrm{1}}}{r}\right)^2 (5P_2(\lambda_1) +1) +...
\end{equation}
\begin{align}
\nonumber P_2(\lambda_2)\left(\dfrac{r}{r_2} \right)^{3}&\simeq\left(1+\left(\dfrac{r_{\textrm{1}}}{r} \right)^{2} (P_2(\lambda_1)-1)  \right)\\ &\times\left(1+3\dfrac{r_{\textrm{1}}}{r} P_1(\lambda_1)  +...\right) \\ &\simeq 1+3\dfrac{r_{\textrm{1}}}{r} P_1(\lambda_1)+6\left(\dfrac{r_{\textrm{1}}}{r} \right)^{2} P_2(\lambda_1),\nonumber
\end{align}
\begin{align}
P_2(\mu_2)&=\nonumber \frac{1}{2}\left(3 \frac{z_1^2}{r_2^2}-1\right)=\frac{1}{2}\left(3 \frac{z_1^2}{r^2}-1\right) +\mathcal O\left(\left(\dfrac{r_1}{r} \right)^{3}\right)\\ &\simeq \frac{3}{2} \left(\frac{r_1}{r}\right)^2\mu_1^2 -\frac{1}{2}+\mathcal O\left(\left(\dfrac{r_1}{r} \right)^{3}\right)\\ &= \left(\frac{r_1}{r}\right)^2P_2(\mu_1)+\frac{1}{2}\left(\frac{r_1}{r}\right)^2 -\frac{1}{2} +\mathcal O\left(\left(\dfrac{r_1}{r} \right)^{3}\right),\nonumber
\end{align}
and
\begin{align}
P_2(\mu_2)\left(\dfrac{r}{r_2} \right)^{3}&=-\frac{1}{2}-\frac{3}{2}\dfrac{r_{\textrm{1}}}{r} P_1(\lambda_1)\\ &+\left(\dfrac{r_1}{r} \right)^{2}\left[ P_2(\mu_1)  -\frac{5}{2}P_2(\lambda_1)   \right]+\mathcal O\left(\left(\dfrac{r_1}{r} \right)^{3}\right).\nonumber
\end{align}
All those approximations can be inserted back in \autoref{eq_anx_perturbation_pot_1} to obtain the perturbation of the secondary gravitational potential in the coordinates of the primary:
\begin{align}
& \Psi_{2,2}(r_{1}, \lambda_1, \mu_1)=\Psi_{2,0}(r_1^2) \\ &- \nonumber 2k_{2,2}\left(\dfrac{R_{\textrm{2}}}{r} \right)^{5}\dfrac{r_1}{r}\left(3\dfrac{GM_1}{r}+\dfrac{\Omega_{\star,2}^2}{2} r^2  \right)P_1(\lambda_1)\nonumber\\ &-2k_{2,2}\left(\dfrac{R_{\textrm{2}}}{r} \right)^{5}\left(\dfrac{r_1}{r}\right)^2\left[\left(6\dfrac{GM_1}{r}+\dfrac{5}{6}\Omega_{\star,2}^2 r^2  \right)P_2(\lambda_1)\right.\nonumber \\&\left. - \dfrac{\Omega_{\star,2}^2}{3} r^2 P_2(\mu_1)\right],
\end{align}
that can be rewritten using the third Kepler law including the periods of the system as follows
\begin{align}\label{anx_perturbed_surface_grid_primary}
\Psi_{2,2}(r_{1}, \lambda_1, \mu_1)&=\Psi_{2,0}(r_1^2)-2k_{2,2}\dfrac{GM_1}{r}\left(\dfrac{R_{\textrm{2}}}{r} \right)^{5}\\&\left[\dfrac{r_1}{r}\left( 3+\dfrac{1}{2}\dfrac{M_{\mathrm{tot}}}{M_1} \left( \dfrac{\Omega_{\star, 2}}{\tilde{n}}\right)^2  \right)P_1(\lambda)\right.\nonumber\\&+\left.\left(\dfrac{r_1}{r}\right)^2\left( 6+\dfrac{5}{6}\dfrac{M_{\mathrm{tot}}}{M_1} \left( \dfrac{\Omega_{\star, 2}}{\tilde{n}}\right)^2  P_2(\lambda)\right.\right.\nonumber  \\&-\left.\left.\dfrac{1}{3}\dfrac{M_{\mathrm{tot}}}{M_1} \left( \dfrac{\Omega_{\star, 2}}{\tilde{n}}\right)^2 P_2(\mu)\right)\right].\nonumber
\end{align}
We can note that all the constants appearing in the expressions have been included in the $\Psi_{2,0}$  as they are not depending on $r_1/r$. 
\\Let us now compute the force exerted by the perturbation of the secondary on the unperturbed primary,  $F_{2^{\prime}10}$, that is given by
\begin{align}
F_{2^{\prime}10}&= -\int \rho_{10}\frac{\partial \Psi_{2,2}}{\partial x_1}\mathrm{d} V_1\simeq -M_1\frac{\partial \Psi_{2,2}}{\partial x_1}(r_1=0),
\end{align}
when treating the primary as point like ($r_1=0$). In this case, only the gradient of the $\ell=1$ component is non-null at the centre of the star ($r_1=0$) as 
\begin{equation}
\frac{r_1}{r}P_1(\lambda)=\frac{x_1}{r}.
\end{equation}
Therefore, by only taking the $\ell=1$ component,  the force can be expressed as 
\begin{align}
F_{2^{\prime}10}&=-M_1\frac{\partial\Psi_{2,2}}{\partial x_1}(r_1=0)\nonumber
\\&=2k_{2,2}\dfrac{GM_1^2}{r^2}\left(\dfrac{R_{\textrm{2}}}{r} \right)^{5}\left( 3+\dfrac{1}{2}\dfrac{M_{\mathrm{tot}}}{M_1} \left( \dfrac{\Omega_{\star, 2}}{\tilde{n}}\right)^2  \right).
\end{align}
\section{Higher order perturbations}\label{annexe_l=3_perturbation}
In this section,  we apply the developments done in  \autoref{annexe_perturbation of the force}  and \autoref{subsec_pert_apsidal_motion_theory} to obtain the apsidal motion resulting from the $\ell=3$ perturbation of the potential.  By combining \autoref{anx_eq_perturbation_surface} and  \autoref{eq_anx_tidal_force}, the surface potential perturbation $\ell $ of the secondary originating from the presence of the primary is given by 
\begin{equation}
\Psi_{2,3}(R_2, \theta_2, \phi_2)= -2 k_{2,3} \frac{GM_1}{r} \left(\frac{R_2}{r}\right)^3 P_3(\lambda_2).
\end{equation}
Beyond the surface of the secondary, at a distance $r_2$ from its centre, the potential perturbation from the $\ell=3$ perturbed potential of the secondary is expressed as 
\begin{align}
\Psi_{2,3}(r_2, \theta_2, \phi_2)&= -2 k_{2,3} \frac{GM_1}{r} \left(\frac{R_2}{r}\right)^3 P_3(\lambda_2) \left(\frac{R_2}{r_2}\right)^4\\ &=-2 k_{2,3} \frac{GM_1}{r} \left(\frac{R_2}{r}\right)^7 P_3(\lambda_2) \left(\frac{r}{r_2}\right)^4.\label{anx_eq_perturbed_l3_surface}
\end{align}
Based on \cite{Kopal1959}, $\lambda_2$ can be expended as
\begin{equation}
\lambda_2 = 1+ \dfrac{1}{3} (P_2(\lambda_1)+1)\left(\dfrac{r_{\textrm{1}}}{r} \right)^{2} + ... ,
\end{equation}
meaning that 
\begin{equation}
P_3(\lambda_2) =  \dfrac{1}{2}\left(5\lambda_2^2-3\lambda_2\right)= 1 -\dfrac{1}{2}(P_2(\lambda_1)+1)\left(\dfrac{r_{\textrm{1}}}{r} \right)^{2} +..., 
\end{equation}
and
\begin{equation}
P_3(\lambda_2)\left(\dfrac{r}{r_2} \right)^{4}\simeq 1+4\dfrac{r_{\textrm{1}}}{r} P_1(\lambda_1)+ ...,
\end{equation}
which can be inserted back in \autoref{anx_eq_perturbed_l3_surface} to obtain the perturbation originating from the $\ell=3$ perturbation of the secondary in the coordinate frame of the primary:
\begin{equation}
\Psi_{2,3}(r_{1}, \lambda_1)=\Psi_{2,0}(r_1) -8k_{2,3}\dfrac{GM_1}{r}\left(\dfrac{R_{\textrm{2}}}{r} \right)^{7}\dfrac{r_1}{r}P_1(\lambda_1)+... \ .
\end{equation}
By using the same methodology as in the  \autoref{annexe_perturbation of the force}, the force perturbation originating from the $\ell=3$ deformation of the secondary on the primary is expressed as
\begin{equation}
F_{2^{\prime}10}= - 8k_{2,3}\dfrac{GM_1^2}{r^2}\left(\dfrac{R_{\textrm{2}}}{r} \right)^{7}
\end{equation}
Following the \autoref{subsect_perturbation_gravity}, the acceleration perturbation undergone by the system due to the tidal perturbation is expressed as
\begin{align}
F_{\mathrm{R}}&=-8 \dfrac{GM_{\mathrm{tot}}}{a^2} \left( k_{2,3}\frac{1}{q}\left(\dfrac{R_{\textrm{2}}}{a} \right)^{7} +  k_{1,3}q\left(\dfrac{R_{\textrm{1}}}{a} \right)^{7}\right) \left( \dfrac{a}{r}\right)^9 \\&\equiv   F_{R,\mathrm{tides}, \ell=3}  \left( \dfrac{a}{r}\right)^9, 
\end{align}
when defining $a$ as the semi-major axis of the system.  Following the same approach as in \autoref{subsec_pert_apsidal_motion_theory}, the time orbital average of the perturbation is formally expressed as
\begin{equation}
\langle F_{\mathrm{R}} \cos{\varphi}\rangle=\frac{1}{P_{\mathrm{orb}}}F_{R,\mathrm{tides}, \ell=3} \int_0^{P_{\mathrm{orb}}} \left( \dfrac{a}{r}\right)^9 \cos{\varphi}\ \mathrm{d}t,
\end{equation}
which can be developed into polynomial functions of the eccentricity using Hansen's functions:
\begin{equation}\label{anx_eq_Rcos_l3}
\langle F_{\mathrm{R}} \cos{\varphi}\rangle=F_{R,\mathrm{tides}, \ell=3}  X_0^{-9,1}(e),
\end{equation}
with 
\begin{equation}
X_0^{-9,1}(e)=\dfrac{7}{2}e(1-e^2)^{-15/2}\left(1+\dfrac{15}{4}e^2+\dfrac{15}{8}e^4+\dfrac{5}{64}e^6\right).
\end{equation}
The apsidal motion caused by the $\ell=3$ acceleration perturbation can finally be obtained by inserting \autoref{anx_eq_Rcos_l3} in \autoref{eq_domega_base} and simplifying:
\begin{equation}
\frac{\mathrm{d}\omega}{\mathrm{d}t}=\frac{56\pi}{P_{\mathrm{orb}}} v(e)  \left( k_{2,3}\frac{1}{q}\left(\dfrac{R_{\textrm{2}}}{a} \right)^{7} +  k_{1,3}q\left(\dfrac{R_{\textrm{1}}}{a} \right)^{7}\right),
\end{equation}
with,
\begin{equation}
v(e)=\dfrac{1+\dfrac{15}{4}e^2+\dfrac{15}{8}e^4+\dfrac{5}{64}e^6}{(1-e^2)^7}.
\end{equation}

\end{document}